\documentclass[fleqn,usenatbib,twocolumn]{mnras}
\usepackage{graphicx} 
\usepackage{xcolor} 
\usepackage{hyperref}
\makeatletter\let\expandableinput\@@input\makeatother
\title{Quasar absorption lines}
\usepackage[normalem]{ulem}

\usepackage[T1]{fontenc}
\usepackage{ae,aecompl}
\usepackage{float,comment}
\usepackage{longtable}
\usepackage{array}
\newcolumntype{H}{>{\setbox0=\hbox\bgroup}c<{\egroup}@{}}

\definecolor{DG}{rgb}{0.09, 0.45, 0.27}

\def\kms{km~s$^{-1}$}
\def\kmsMpc{km~s$^{-1}$~Mpc$^{-1}$}

\def\Ob{$\Omega_b$}

\def\dof{$\mathrm{d.o.f.}$}

\def\hst{\it HST\rm}
\def\fuse{\it FUSE\rm}
\def\rosat{\emph{ROSAT}}
\def\rxte{\emph{RXTE}}
\def\nodata{-}
\def\lya{Lyman-$\alpha$}
\def\h{$H$}
\def\he{$He$}

\def\oi{O~I}

\def\oiv{O~IV}
\def\ov{O~V}
\def\ovi{O~VI}
\def\ovii{O~VII}
\def\oviii{O~VIII}
\def\neviii{Ne~VIII}
\def\neix{Ne~IX}

\def\hi{H~I}

\def\bla{BLA}

\def\chandra{\it Chandra\rm}
\def\xmm{\it XMM-Newton\rm}
\def\xmmshort{\it XMM\rm}

\def\es{1ES~1553+113}
\def\kms{km~s$^{-1}$}

\def\gof{goodness--of--fit}
\def\cmin{$C_\mathrm{min}$}
\def\cmineq{C_\text{min}}
\def\lcdm{$\Lambda$CDM}

\def\nSources{51}
\def\nsystems{1,224}
\def\nDet{33}
\def\nEm{29}

\usepackage{graphicx}   
\usepackage{amsmath}    
\usepackage{scalerel}
\title[X--ray absorption lines in FUV--detected quasars I]{X--ray absorption lines in FUV--detected quasars:\\
I. Sample and analysis of the XMM--Newton and Chandra data}

\author[D.~Spence et al.]{
David~Spence$^{1}$, Massimiliano Bonamente$^{1}$\thanks{E-mail: bonamem@uah.edu}, Jussi~Ahoranta$^{2}$,  Nastasha~Wijers$^{3}$
\newauthor{Toni~Tuominen$^{2}$ and Jelle~de~Plaa$^{4}$}\\
$^{1}$Department of Physics and Astronomy, University of Alabama in Huntsville, Huntsville, AL \\
$^{2}$ Department of Physics, University of Helsinki, PO Box 64, 00014 Helsinki, Finland \\
$^{3}$ Center for Interdisciplinary Exploration and Research in Astrophysics (CIERA) and Department of Physics and Astronomy,\\ Northwestern University, 1800 Sherman Ave, Evanston, IL 60201, USA\\
$^{4}$ SRON, Netherlands Institute for Space Research Astrophysics, Niels Bohrweg 4, 2333CA Leiden, Netherlands\\
}
\date{}

\begin{document}

\maketitle

\begin{abstract}
   This paper presents initial results of a systematic search for  resonance X--ray absorption lines from \h--like \oviii\ and 
   \he--like \ovii\ caused by the intervening warm--hot intergalactic medium (WHIM).
   The search is based on far ultra--violet redshift  priors from \ovi\ and \hi\ broad \lya\ lines that were previously detected by \hst\ and \fuse\ in a sample of 51~sources with either \xmm\ or \chandra\ data, for a total X--ray redshift path of $\Delta z=10.9$. 
   Of the \nsystems\ absorption--line systems with FUV priors that were analyzed, \nDet\ 
   systems feature an absorption--line feature detected with $\geq 99$~\% confidence at the same redshift of the FUV prior, some coincident with previously reported absorption line detections. The ultimate goal of this search is to test the hypothesis that  X--ray absorbing WHIM gas is the repository of the missing baryons in the local universe.  Further
   results and the cosmological implications of this analysis are presented in a companion paper.
\end{abstract}

\section{Introduction: missing baryons in the low--redshift universe}
\label{sec:introduction}
For the standard Friedmann--Robertson--Walker \lcdm\ cosmological model of the universe, the
density of baryons has been accurately measured by a variety of probes, with a
broad consensus that \Ob$\simeq 0.05$ \citep[e.g.][]{planck2020}, in accordance
with big--bang nucleosynthesis predictions \citep[e.g.][]{kirkman2003}. 
At high redshift, baryons are accounted for primarily via 
detection of the \lya\ forest \citep[e.g.][]{weinberg1997, rauch1998}.
At lower redshift, far ultra--violet surveys have successfully 
detected only a fraction of these baryons, primarily in the
warm--hot intergalactic medium \citep[WHIM, e.g., ][]{tilton2012, danforth2016}.

For over two decades, numerical simulations have suggested that the
missing baryons may be located in filamentary structures of galaxies 
that host a warm--hot
intergalactic medium at temperatures of approximately $\log T(K)=5-7$ \citep[the WHIM, e.g.][]{cen1999,dave2001,bertone2008,cautun2014}.
There is growing evidence that the missing baryons are in the high--temperature
($\log T(K)\geq 6$) portion of the WHIM, and therefore primarily accessible
only via X--ray observations of such ions as \ovii\ and \oviii, and others \citep[see, e.g.][]{martizzi2019,tuominen2021}.
The higher range of WHIM temperatures have been more challenging to 
probe than  lower temperatures, primarily because the available spectrometers on board
\xmm\ and \chandra\ are far less  sensitive than the FUV instruments.
One of cosmology's open questions is therefore the location of these \textit{missing
baryons}, which is the main goal of this project.

To date, the search for X--ray absorption lines in the spectrum of background quasars has yielded 
a few possible detections that we review in Sec.~\ref{sec:previousDetections}. However, all claimed detections have
limited signal--to--noise, and may have sometimes been the result of confusion
with intrinsic absorption, as is likely in the case for a high--redshift system in  \es\ 
that was claimed to resolve the missing baryons problem \citep[e.g.][]{nicastro2018}.
In particular, the recent survey by \cite{gatuzz2023} failed to detect significant absorption in a sample
of six sources, which are all included in the present analysis. A more comprehensive review of the literature on X--ray absorption lines in provided in Sec.~\ref{sec:previousDetections}.

Given the ongoing challenges in the detection of X--ray absorption lines with the current--generation
spectrometers, this project focuses on using all
available \chandra\ and \xmm\ data of sources with prior \ovi\ or \hi\ broad Lyman--$\alpha$ (\bla) 
absorption line detections \citep{tilton2012, danforth2016}, to search for X--ray absorption lines and to set upper limits to their non--detection. 
The search focuses
on the two most prominent X--ray ions of the most abundant element at $Z>2$, namely \ovii\ and \oviii,
and the FUV detections provide the needed priors that make the search manageable given the
resolution of the data \citep[see, e.g.][for a similar method on individual sources]{bonamente2016, ahoranta2020}.

In this paper we present the sample of sources used for this project, the
X--ray data from \xmm\ and \chandra\  and the main results for the detection
of the \ovii\ and \oviii\ absorption lines in the entire sample.
Given the large number of sightlines under consideration, the results of the analysis  therefore provide the means to estimate the  cosmological density associated to the X--ray WHIM that is more representative of the low--redshift universe than that of just one or few sources, as we provided for \es\ \citep{spence2023}. The cosmological 
interpretations of the results are presented in a companion paper.

This paper is structured as follows. Section~\ref{sec:sample} describes the sample of sources
used in this study and the methods of analysis. Section~\ref{sec:results} presents the results of the data analysis, including possible
new detections. 
Section~\ref{sec:discussion} contains a discussion of the results and the systematics associated with this search, and  conclusions are presented in Sec.~\ref{sec:conclusions}. 

\section{Sample and data analysis methods}
\label{sec:sample}

\subsection{Search at fixed FUV prior redshifts}
\label{sec:searchPrior}
Previous searches for X--ray absorption from the WHIM have focused primarily on
the few sightlines towards X--ray bright quasars and other compact sources,~\footnote{In this paper,
we simply refer to point--like background sources as quasars, although some sources would
be more aptly referred to as either blazars (such as \es) or AGN. Since the intrinsic properties of
the X--ray sources are not of interest to this study, we do not discuss their nature beyond their
X--ray flux and redshift.} 
such as \es\ \citep[e.g.][]{nicastro2018, spence2023},
Markarian~421 \citep[e.g.][]{nicastro2005, rasmussen2007, yao2012}, Markarian~501 \citep{ren2014},
H~2356-309 \citep[][]{fang2010, buote2009}, 3C~273 \citep[e.g.][]{ahoranta2020}
or H~1821+643 \citep{kovacs2019}, among others.
In certain cases, the searches were conducted serendipitously at all redshifts, as in the case of 
\cite{nicastro2018} or \cite{gatuzz2023}. In other cases, the searches were guided by prior detections
of far ultra--violet (FUV) lines such as \ovi\ or \hi\ BLA, as in \cite{bonamente2016}, \cite{ahoranta2021} and \cite{kovacs2019}, among others. 

This project focuses on X--ray searcher at fixed redshifts that were previously identified
by the FUV detections of \cite{tilton2012} and \cite{danforth2016}. 
There is a two--fold reason for carrying out a to search for X--ray absorption lines at a fixed
redshift that is set by a prior FUV detection, rather than to conduct a blind search. 
First, the significance of detection of a line is higher when
the prior redshift is known \citep[see, e.g., the statistical treatment of
\emph{redshift trials} by ][and Sec.~\ref{sec:redshiftTrials} below]{bonamente2019, nicastro2013}. Second, and more relevant to this study, is the lower likelihood
of line confusion, especially for the lower--resolution data that are present in some of the fainter
sources. 
Therefore, this paper focuses exclusively on the study of X--ray absorption lines
at redshifts with prior detections of FUV absorption that act as signpost of possible X--ray absorbing WHIM, following the methods of \cite{spence2023}.
Blind searches \emph{à la} \cite{gatuzz2023} or \cite{nicastro2018} are certainly valuable, in that the search may indeed reveal absorption lines that are unassociated
with FUV priors, e.g., as one would expect for warmer absorbers. The FUV--prior method and the blind method are therefore complementary, and this project focuses exclusively on the former.

\subsection{Selection of sightlines and FUV redshift priors}

We used the \cite{tilton2012}, hereafter T12, and \cite{danforth2016}, hereafter D16, to
identify all available sightlines with FUV detection of either 
\ovi\ ($1032, 1038$~\AA) or \hi\ \lya\ (1216~\AA). Of all the FUV systems in these two papers, we
only consider those with positive \ovi\ detection of at least one component of the doublet, and 
those sources with a \hi\ absorption with Doppler parameter $b\geq 60$~\kms, indicative of gas at $\log T(\mathrm{K}) \geq 5$. These are the same criteria we used in \cite{spence2023}, to which we refer the
reader for additional considerations concerning the sample selection; we only loosened
the criteria for use of \ovi\ FUV data to systems with at least one of the two lines of the
doublet, instead of requiring a detection of both, for consideration in this work.  This choice was made as an effort to provide as broad a search for X--ray absortion lines as possible, while still being guided by FUV priors with reliable line identification.

We then searched the the \chandra\ and \xmm\
archive for X--ray grating observations; for \chandra, we only considered the LETG spectrometer.
Table~\ref{tab:sample} contains the list of \nSources\ sources that we analyze for this study. 
We chose to analyze all the available X--ray data, instead of focusing only on the brightest sources
(e.g., PKS~2155-304 or \es) that were most likely to yield positive detections. This choice
was dictated by the goal to set upper limits to the non--detections of selected X--ray absorption
lines for the largest possible
redshift path, in addition to seeking possible line detections.  Upper limits
to non--detections can then be used to set global limits to the cosmological density of X--ray absorbing
baryons, with the method that was presented in \cite{spence2023}, even if no significant detections were to
be found.
Cosmological conclusions from this project, using both possible detections
and upper limits to non--detections, are presented in a companion paper.

\begin{table} 
\caption{List of X--ray sources with \ovi\ and/or \hi\ BLA FUV priors. "FUV sys." is the number of \ovi\ and \hi\ BLA FUV priors, "Obs." is the number of pointed observations and the total exposure time in ks and $z$ is redshift of the source.}
\label{tab:sample} 
\setlength\extrarowheight{-1pt}
\begin{tabular}{lp{2cm}p{0.6cm}p{0.15cm}p{0.15cm}p{0.15cm}p{0.4cm}p{0.15cm}p{0.2cm}}
\hline \hline
\multicolumn{2}{c}{Source} & $z$ & \multicolumn{2}{c}{FUV sys.} & \multicolumn{4}{c}{Obs. (number, ks)} \\
& & & OVI & HI & \multicolumn{2}{c}{\xmmshort} & \multicolumn{2}{c}{\chandra}\\
\hline\\[0pt]
\footnotesize
1 & 1ES1028+511 & 0.3604 & 3 & 3 & 5 & 312 & 1 & 149   \\
2 & 1ES1553+113 & 0.4140 & 7 & 4 & 22 & 1998 & 3 & 496   \\
3 & 3C249 & 0.3115 & 4 & 2 & 1 & 37 & 0 & 0 \\
4 & 3C273 & 0.1583 & 10 & 4 & 35 & 1282 & 2 & 70   \\
5 & 3C66A & 0.3347 & 2 & 1 & 2 & 27 & 0 & 0   \\
6 & H1821+643 & 0.2968 & 15 & 7 & 15 & 130 & 4 & 470   \\
7 & H2356-309 & 0.1651 & 2 & 4 & 9 & 704 & 11 & 587   \\
8 & HE0056-3622 & 0.1641 & 2 & 0 & 4 & 186 & 0 & 0   \\
9 & HE0226-4110 & 0.4934 & 61 & 3 & 1 & 33 & 0 & 0   \\
10 & IRASF22456-5125 & 0.1000 & 8 & 2 & 4 & 108 & 0 & 0   \\
11 & MR2251-178 & 0.0640 & 1 & 7 & 10 & 610 & 1 & 78   \\
12 & Mrk421 & 0.0300 & 1 & 0 & 102 & 2876 & 38 & 639  \\
13 & Mrk478 & 0.0791 & 0 & 3 & 5 & 252 & 1 & 80   \\
14 & Mrk876 & 0.1290 & 6 & 6 & 2 & 21 & 0 & 0   \\
15 & NGC7469 & 0.0163 & 3 & 1 & 11 & 857 & 0 & 0  \\
16 & PG0157+001 & 0.1631 & 1 & 0 & 1 & 15 & 0 & 0   \\
17 & PG0003+158 & 0.4509 & 16 & 5 & 1 & 26 & 0 & 0   \\
18 & PG0804+761 & 0.1000 & 4 & 3 & 3 & 103 & 0 & 0   \\
19 & PG0832+251 & 0.3298 & 5 & 8 & 1 & 23 & 0 & 0   \\
20 & PG0838+770 & 0.1310 & 0 & 3 & 1 & 24 & 0 & 0   \\
21 & PG0953+414 & 0.2341 & 10 & 2 & 1 & 16 & 0 & 0   \\
22 & PG1048+342 & 0.1671 & 0 & 2 & 1 & 33 & 0 & 0   \\
23 & PG1115+407 & 0.1546 & 4 & 3 & 1 & 21 & 0 & 0   \\
24 & PG1116+215 & 0.1763 & 9 & 7 & 6 & 395 & 11 & 356   \\
25 & PG1211+143 & 0.0809 & 4 & 2 & 12 & 883 & 3 & 134  \\
26 & PG1216+069 & 0.3313 & 8 & 8 & 1 & 17 & 0 & 0   \\
27 & PG1229+204 & 0.0630 & 0 & 1 & 1 & 26 & 0 & 0   \\
28 & PG1259+593 & 0.4778 & 14 & 10 & 1 & 34 & 0 & 0   \\
29 & PG1307+085 & 0.1550 & 2 & 1 & 1 & 14 & 0 & 0   \\
30 & PG1309+355 & 0.1829 & 11 & 5 & 1 & 30 & 0 & 0   \\
31 & PG1444+407 & 0.2673 & 6 & 1 & 1 & 22 & 0 & 0  \\
32 & PG1626+554 & 0.1330 & 1 & 1 & 1 & 11 & 0 & 0   \\
33 & PHL1811 & 0.1920 & 13 & 6 & 2 & 92 & 0 & 0   \\
34 & PKS0312-770 & 0.2230 & 2 & 1 & 4 & 153 & 0 & 0\\
35 & PKS0405-123 & 0.5740 & 25 & 12 & 2 & 170 & 4 & 376   \\
36 & PKS0558-504 & 0.1372 & 0 & 1 & 20 & 1096 & 0 & 0   \\
37 & PKS1302-102 & 0.2784 & 11 & 8 & 1 & 16 & 0 & 0   \\
38 & PKS2005-489 & 0.0710 & 1 & 0 & 3 & 62 & 4 & 282   \\
39 & PKS2155-304 & 0.1165 & 2 & 3 & 36 & 2191 & 15 & 319   \\
40 & PMNJ1103-2329 & 0.1860 & 5 & 0 & 2 & 27 & 0 & 0   \\
41 & PMNJ2345-1555 & 0.6210 & 12 & 5 & 1 & 32 & 0 & 0   \\
42 & Q1230+0115 & 0.1170 & 15 & 13 & 1 & 71 & 0 & 0   \\
43 & QSO0045+3926 & 0.1340 & 2 & 2 & 1 & 17 & 0 & 0   \\
44 & RBS1892 & 0.2000 & 3 & 6 & 1 & 22 & 0 & 0   \\
45 & RBS542 & 0.1040 & 2 & 2 & 11 & 369 & 0 & 0   \\
46 & RXJ0439.6-5311 & 0.2430 & 6 & 2 & 2 & 162 & 0 & 0   \\
47 & S50716+714 & 0.2315 & 1 & 4 & 5 & 180 & 0 & 0   \\
48 & TonS210 & 0.1160 & 0 & 2 & 1 & 9 & 0 & 0   \\
49 & Ton28 & 0.3297 & 7 & 1 & 3 & 108 & 0 & 0  \\
50 & Ton580 & 0.2902 & 2 & 0 & 2 & 48 & 0 & 0   \\
51 & TonS180 & 0.0620 & 3 & 1 & 4 & 223 & 1 & 77  \\
\hline
   &         &        & \multicolumn{6}{c}{Totals}\\
   &         &  10.94      & 332 & 178 & 365 & 16174 & 99 & 4113 
\\
\hline
\hline
\end{tabular}
\end{table}

To ensure an accurate X--ray search, the following criteria were adopted for
the T12 and D16 FUV detections for the sources in Table~\ref{tab:sample}:
(a) The D16 paper reports individual detections for the two components of
the $1032, 1038 \AA$ doublet, whereas the T12 has a combined column density for the
entire \ovi\ absorption line. Accordingly, when both components of the doublet
were detected by D12, they were averaged into a single detection, with corresponding
redshift and column density given by the weighted average of the two lines.
(2) When the redshift of D16 and T12 lines differed by $\Delta z \leq 0.0001$,
the two lines were deemed to be the same, and only the D16 was considered.
The sample of \ovi\ and \hi\ BLA absorption lines for the combined T12/D16 sample considered
in this search are reported respectively in Tables~\ref{tab:oviPaper} and \ref{tab:hiPaper} 
(full lists are reported in the on--line version of the paper). 

\subsection{Survey design}
\label{sec:surveyDesign}
A data analysis project of this magnitude required a number of
choices in its design, as a compromise between accuracy and overall feasibility and ease of interpretation of the results. The main choices that were made are discussed in this section.

First, we opted for a simple power--law model for the continuum plus a simple absorption--line model (\texttt{line} in SPEX), as described in detail in Sec.~\ref{sec:spectralModel}. This is in contrast with more accurate or physically--motivated models, such as the ones used in some of our previous analyses \citep[e.g.][]{nevalainen2019, spence2023}, or by \cite{gatuzz2023}. We found that this simple model was a good compromise between accuracy of the results and the ability to manage the automation of the data analysis. A similar power--law model was in fact successfully applied to the \es\ data in \cite{spence2023} to find results that were consistent with a more complex \texttt{spline} model for the continuum. Moreover, a simple
\texttt{line} model for the absorption line features at the \ovii\ and \oviii\ wavelengths lets us use the corresponding ion column densities to constrain the cosmological density of the ions \cite[e.g., following the methods of][]{nicastro2018, spence2023}.

Second, we limited the analysis to the \ovii\ and \oviii\ \lya\ lines of oxygen, the most abundant "metal" in the cosmos \citep[e.g.][]{anders1989, asplund2009}. Given that in collisional equilibrium  these ions have a substantial ionization fraction  in the range $\log T(K)=5.5-6.5$ \citep[e.g.][]{mazzotta1998}, these two ions sample effectively temperatures that are immediately adjacent to the one probed by the FUV lines. In principle, other ions such as \neviii\ or \neix\ could be searched for, but their lower expected abundances make them less interesting than \ovii\ and \oviii.

Third, we chose a simple search at the \emph{exact} FUV redshift priors, instead of making allowances for possible redshift differences between the lower--temperature FUV systems and the sought--after higher--temperature X--ray systems. The motivation for this choice is both for the sake of simplicity of the analysis, and to avoid complications in the statistical interpretation of the results, i.e., the issue of redshift trials. This choice, however,  may cause this analysis to miss 
or underestimate
the significance of certain absorption line systems. Systematics associated with this choice are further discusses in Sec.~\ref{sec:systematics}.

\begin{table*}
    \centering
    \begin{tabular}{lllccccc}
\# & Name & Redshift & $\log N$ (cm$^{-2}$) & $b$ (km~s$^{-1}$) & Source\\ 
\hline
1 & 1es1028 & 0.12314 & 14.3$\pm$0.1 & 41.37$\pm$5.57 & Danforth$^{M}$ \\ 
2 & 1es1028 & 0.13706 & 13.6$\pm$0.1 & 14.50$\pm$6.30 & Danforth$^{1032}$ \\ 
3 & 1es1028 & 0.33735 & 13.9$\pm$0.1 & 72.87$\pm$15.73 & Danforth$^{M}$ \\ 
\hline
4 & 1es1553 & 0.18759 & 13.8$\pm$0.1 & 12.10$\pm$3.00 & Danforth$^{M}$ \\ 
5 & 1es1553 & 0.18775 & 13.9$\pm$0.1 & 23.40$\pm$4.50 & Danforth$^{M}$ \\ 
6 & 1es1553 & 0.18984 & 13.4$\pm$1.5 & 23.37$\pm$6.57 & Danforth$^{M}$ \\ 
7 & 1es1553 & 0.21631 & 13.3$\pm$0.1 & 19.80$\pm$5.80 & Danforth$^{1032}$ \\ 
8 & 1es1553 & 0.31130 & 13.4$\pm$0.1 & 31.90$\pm$7.50 & Danforth$^{1032}$ \\ 
9 & 1es1553 & 0.37868 & 12.9$\pm$0.2 & 17.30$\pm$3.70 & Danforth$^{1032}$ \\ 
10 & 1es1553 & 0.39497 & 13.9$\pm$0.1 & 44.07$\pm$3.33 & Danforth$^{M}$ \\ 
\hline
\dotfill \\
    \end{tabular}
    \caption{List of \ovi\ absorption line systems from the T12/D16 samples. A '$^M$' indicates that there was an OVI 1032 and 1038 doublet present. These 1032 and 1038 doublet values were averaged into one value on a weighted system of 2:1 respectively. A '*' indicates a duplicate redshift in both Danforth and Tilton. The corresponding Tilton duplicates are omitted from the table.
    A complete list of T12/D16 absorption line systems is available in the on--line version of the paper, which contains additional
    sources to those of Table~\ref{tab:sample}.}
    \label{tab:oviPaper}
\end{table*}


\begin{table*}
\begin{tabular}{lllccccc}
    \# & Name & Redshift & $\log N$ (cm$^{-2}$) & $b$ (km~s$^{-1}$) & Source\\ 
\hline
1 & 1es1028 & 0.13714 & 13.3$\pm$0.1 & 91.60$\pm$24.00 & Danforth \\ 
2 & 1es1028 & 0.20383 & 13.2$\pm$0.1 & 63.60$\pm$14.20 & Danforth \\ 
3 & 1es1028 & 0.22121 & 13.4$\pm$0.8 & 68.60$\pm$71.40 & Danforth \\ 
\hline
4 & 1es1553 & 0.03466 & 13.1$\pm$0.0 & 72.00$\pm$9.40 & Danforth \\ 
5 & 1es1553 & 0.04273 & 13.4$\pm$0.0 & 63.00$\pm$4.50 & Danforth \\ 
6 & 1es1553 & 0.06364 & 13.0$\pm$0.2 & 76.30$\pm$19.90 & Danforth \\ 
7 & 1es1553 & 0.21869 & 12.7$\pm$0.1 & 62.60$\pm$21.80 & Danforth \\ 
\hline\\
\dotfill \\
\end{tabular}
\caption{List of \hi\ BLA absorption line systems from the T12/D16 samples. A '*' indicates a duplicate redshift in both Danforth and Tilton. The corresponding T12 duplicates are omitted from the table.
    A complete list is available in the on--line version of the paper.}
    \label{tab:hiPaper}
\end{table*}

\subsection{X--ray Data Reduction}

Details of the data reduction were provided in \cite{spence2023}, which was a pilot study for this project, 
and the key aspects are briefly described here.
The X--ray data from \xmm\ and \chandra\ were processed using
standard processing tools available from the \texttt{SAS} and the \texttt{CIAO}
packages, using the standard calibration files from \texttt{CALDB}.
The \xmm\ data were processed with \texttt{rgsproc}, and first--order grating
spectra were extracted separately for RGS1 and RGS2 with \texttt{rgscombine}, and then 
combined for multiple observations of a given source with \texttt{trafo}.
The spectra were binned to a 20~m\AA\ bin size for the analysis.
The \chandra\ data were reduced with the \texttt{chandra\_repro} pipeline, and the
$\pm1$ order LETG spectra were extracted in a 50~m\AA\ bin size and
then combined
for all observations with \texttt{trafo}, same as the \xmm\ spectra.  More details 
of the data reduction are available in \cite{spence2023}, \cite{ahoranta2020}
and \cite{bonamente2016}, who followed the same methods of data reduction.

\def\nXMM{51} 
\def\nChandra{14} 
\def\nXMMObs{365} 
\def\nXMMTime{16.2} 
\def\nChandraObs{99} 
\def\nChandraTime{4.1} 

In summary, the X--ray data analyzed for this paper therefore consists of
\nXMM\ \xmm\ sources for a total of \nXMMObs\ observations and
\nXMMTime~Ms of exposure time; and \nChandra\ \chandra\ LETG sources
for a total of \nChandraObs\ observations and
\nChandraTime~Ms of exposure time (Table~\ref{tab:sample}).

\subsection{X--ray data analysis}
\label{sec:dataAnalysis}
 As discussed in Sec.~\ref{sec:surveyDesign}, the large number of individual
spectra to analyze, and that of the FUV priors to search, 
are such that a number of choices were made to
render the task manageable, while ensuring accuracy for the task of 
measuring the \ovii\ and \oviii\ absorption lines at the
fixed redshifts of Tables~\ref{tab:oviPaper} and \ref{tab:hiPaper}.
The spectral analysis of the X--ray data was performed in \texttt{SPEX} \citep{kaastra1996},
same as in \cite{spence2023}.
The data analysis choices are explained in detail this section.

\subsubsection{Spectral model}
\label{sec:spectralModel}
Given that the main
goal of the search is to detect the presence of a faint absorption line
at a fixed wavelength, we fit a 1--\AA\ segment of the spectra around the target
wavelength with a simple power--law model, same as in \cite{spence2023}. Such wavelength baseline was deemed sufficient to determine the continuum around the line with a simple model for all sources, in place of using
a more complicated model such as the spline that was used for the entire $\sim 20$~\AA\
span of the
spectrum of \es, which is one of the brightest sources
in the sample \citep{spence2023}.  Appendix~\ref{app:NumericalTests} presents numerical simulations
that address the accuracy of this method to measure typical WHIM column densities.

The X--ray spectra were fit to a \texttt{power--law} for the continuum plus a \texttt{line} model.
The power--law model was used with its index fixed at $\alpha=1$, which was found to be an acceptable fit for
the majority of the systems given the restricted wavelength range used, and therefore the continuum
has only one free parameter. For certain sources with higher counting statistics for which the fixed $\alpha$ was not sufficient, the data were fit to a model with free index.
The \texttt{line} component models an absorption or emission line, using a simplified Gaussian line shape with a Gaussian line width parameter (\texttt{awhg}) fixed at a fiducial value of 10~m\AA, as was done in \cite{spence2023}.  We point out that the choice of fixing this model parameter is due to the limited S/N of the data, and it does not affect the ability to provide an unbiased measurement of the equivalent width of the line.~\footnote{Details on the measurements of the equivalent width of the lines, and the
associated constraints on column densities, will be provided in a companion paper that provides the cosmological implications of these measurements.}
This
model component has the central wavelength of the line fixed at the FUV prior, and it therefore features
just one additional free parameter $\tau_0$, which corresponds to the optical depth at line center when $\tau_0>0$, which indicates absorption.
The \texttt{SPEX} \texttt{line} model also returns the
associated line equivalent width, which can be then related
to the column density of the line assuming an optically--thin line, e.g., using Eq.~1 of \cite{bonamente2016}, or by performing the appropriate curve--of--growth analysis in the case
of line saturation \citep[e.g.][]{bonamente2016,ahoranta2021}. Column densities will be discussed in a companion paper.


 The choice of using a fixed wavelength for the absorption line model, as obtained by the FUV prior, is again dictated by the limited S/N of the data. The effects of  using redshift uncertainties are discussed in Sec.~\ref{sec:redshiftTrials}, where we discuss the
concept of `redshift trials', and in Sec.~\ref{sec:systematics}, where we discuss our use of a fixed redshift as a  possible source of systematic error.

\subsubsection{Exclusion of wavelengths with detector artifacts} The \xmm\
RGS camera is especially plagued by the presence of multiple observation--dependent pixels with
reduced efficiency \citep{denherder2001, detmers2010}, which we excluded from the analysis following the same 
methods described in \cite{spence2023}, including the entire 20-24~\AA\
band for RGS2 which is unavailable due to the failure of one of the CCD chips.
After a first analysis, each spectrum was visually inspected for obviously miscalibrated pixels, e.g., identified as pixel with a sudden $\geq 30$~\% drop in their efficiency relative to neighboring pixels, as discussed in e.g. \cite{ahoranta2020, spence2023}. These pixels
were removed from the analysis. 

While such `bad bands' may change from observation to observation, the region $21.5-21.9$~\AA\ is routinely excluded in all
\xmm\ fits, because of the combination of possible Galactic \ovii\ near $\lambda=21.6$, and the poorly calibrated 21.7-21.9~\AA\ band. This means that \xmm\ data are unavailable to detect \ovii\ at $z\leq 0.014$, and \oviii\ at  $0.133 \leq z \leq 0.155$.
Lines with \ovi\ priors falling in these redshift ranges are listed in the tables with a `$\nodata$' symbol.

\subsubsection{Wavelengths with known Galactic spectral features and source confusion}
The wavelength band between the 21.6~\AA\ \ovii\ resonance line and the 
oxygen edge at approximately 23.5~\AA\ has a number of zero--redshift 
oxygen resonance lines, including from \oi\ \citep[e.g.][]{gatuzz2015, nicastro2016}, that could be mistaken for redshifted \ovii\ or
other lines (see Table~3 of \citealt{spence2023}). 
First, a region of $\pm0.1$~\AA\ around $\lambda=21.6$~\AA\ and was excluded in each fit,
to avoid possible contamination from Galactic \ovii, which is sometimes detected
in extragalactic sources \citep[e.g.][]{das2019}. 
Second,  we are cognizant that it would be difficult to
ascertain whether \emph{any} absorption feature in the 21.6-23.5~\AA\ is
a redshifted line from the WHIM or a Galactic line, due to the presence of multiple
Galactic lines in that band, as discussed above. For the sake of completeness of the analysis,
we did retain absorption--line systems in this wavelength range, but the interpretation
of possible detections in this wavelength range was subject to further analysis for
possible contamination. This issue is further discussed for all relevant sources
in Sec.~\ref{sec:detections}.

There are several \ovi\ systems and a few \hi\ BLA systems that are indistinguishable amongst themselves
at the resolution of these X--ray data. FUV redshift priors that are separated by $\Delta z \leq  10^{-3}$ correspond to a wavelength difference $\Delta \lambda \leq 20$~m\AA\
at $\lambda_0=20$~\AA. For a typical resolution of order $50$~m\AA\
for the RGS and LETG spectra, our X--ray data cannot resolve systems that have a redshift difference of order $\Delta z \leq 2-3 \times 10^{-3}$. The effect of such close 
FUV priors for the purpose of cosmological interpretation will be will
discussed in a follow--up paper.

\begin{table}
    \centering
    \begin{tabular}{l|cccc}
    \hline
    \hline
    Sample & \multicolumn{4}{c}{Number}\\
           & Systems & Poor fits & Abs. & Em.\\
           \hline
             XMM OVII/OVI &   308 &    13 &     6 &     3 \\
    Chandra OVII/OVI &    83 &     0 &     1 &     3 \\
       XMM OVIII/OVI &   308 &    10 &    10 &     9 \\
   Chandra OVIII/OVI &    83 &     1 &     1 &     3 \\
         XMM OVII/HI &   164 &     7 &     3 &     5 \\
     Chandra OVII/HI &    57 &     0 &     4 &     1 \\
        XMM OVIII/HI &   164 &     8 &     7 &     5 \\
    Chandra OVIII/HI &    57 &     0 &     1 &     0 \\
    \hline
                     &  1224 &    39 &    33 &    29 \\
\hline
\hline

    \end{tabular}
    \caption{Number of X--ray systems for each instrument and FUV prior. Poor fits are those characterized by \cmin/\dof$\geq 3$. "Abs." denotes fits with $\Delta C\geq 6.6$, $\tau_0 \geq 0$
    for possible absorption lines, and "Em." those with  $\Delta C\geq 6.6$, $\tau_0 \leq 0$ for emission line features.
    \label{tab:statistics}}
\end{table}

\subsubsection{Special treatment of selected spectra} 
Certain high--resolution data, including sources with
long exposure such as PKS~2155-304 or MRK~421, are fit with the same \texttt{power--law} plus
\texttt{line} model, where the index of the power--law model is also free, for a total of three adjustable parameters in the fits. Moreover, the \chandra\ spectrum of Mrk~421 around the \oviii\ $z=0.01$ FUV prior ($\lambda=19.15$) has a sharp behavior in its continuum,
and therefore the spectrum is fit to the 19.0-19.3~\AA\ range only. While this range is probably insufficient for a proper line
detection, it is considered sufficient to set an upper limit. In all these cases, the special allowances do not
affect the
distribution of the $\Delta C$ statistic that is used to assess the significance of detection
of the \texttt{line} component (see Sec.~\ref{sec:basicStatistics}).

This simple two--parameter model in a narrow wavelength range around the
expected line center was fit to all the X--ray sources, separately for
\chandra\ and for \xmm, and for all the \ovi\ and \hi\ BLA FUV priors
for both the 21.60~\AA\ \ovii\ line and the 18.97~\AA\ \oviii\ line. 
The results of the analysis are reported in Tables~\ref{tab:OVIIOVIXMM} through \ref{tab:OVIIIHIChandra}, where
uncertainties are according to the $\Delta C=1$ criterion \citep[e.g.][]{cash1979}, corresponding to a 68.3\% confidence interval
for each interesting parameter. 

\begin{table*}
    \centering
    \begin{tabular}{lllllllllllH|l}
\hline
\hline
 \multicolumn{2}{c}{Target line} & \cmin & \multicolumn{2}{c}{RGS1} & \multicolumn{2}{c}{RGS2} & \multicolumn{2}{c}{power--law} & \multicolumn{3}{c}{line component} & $\Delta C$\\
 Name & z & (d.o.f.)& avg exp (s) & \%  & avg exp (s) & \%       & \text{norm.} & \text{index} & $\lambda$ (\AA) & $\tau_0$ & $\log N (\text{cm}^{-2})$ & \\
 \hline 
1es1028 & $0.12314$ (\#1)  & 95.40(79) & 144787 & 10 & 147275 & 7 & 1102$\pm^{ 30 }_{ 30 }$ & $1.00$ & 24.2598 & $ 0.55\pm^{ 1.74 }_{ 0.88 }$ & $ 15.24\pm^{ 0.44 }_{ nan }$ & 0.40\\ 
1es1028 & $0.13706$ (\#2)  & 92.82(90) & 143804 & 9 & 146370 & 7 & 1121$\pm^{ 33 }_{ 32 }$ & $1.00$ & 24.5605 & $ -0.31\pm^{ 0.85 }_{ 0.56 }$ & $ nan\pm^{ nan }_{ nan }$ & 0.22\\ 
1es1028 & $0.33735$ (\#3)  & 86.46(91) & 145450 & 17 & 146420 & 22 & 1324$\pm^{ 48 }_{ 46 }$ & $1.00$ & 28.8868 & $ -0.78\pm^{ 0.84 }_{ 0.57 }$ & $ nan\pm^{ nan }_{ nan }$ & 1.08\\ 
1es1553 & $0.18759$ (\#4)  & 102.10(88) & 1793737 & 7 & 1757303 & 7 & 1796$\pm^{ 13 }_{ 12 }$ & $1.00$ & 25.6519 & $ 0.60\pm^{ 0.32 }_{ 0.26 }$ & $ 15.27\pm^{ 0.15 }_{ 0.21 }$ & 7.82\\ 
1es1553 & $0.18775$ (\#5)  & 102.34(88) & 1793737 & 7 & 1757303 & 7 & 1796$\pm^{ 13 }_{ 12 }$ & $1.00$ & 25.6554 & $ 0.57\pm^{ 0.30 }_{ 0.25 }$ & $ 15.26\pm^{ 0.15 }_{ 0.22 }$ & 7.53\\ 
1es1553 & $0.18984$ (\#6)  & 105.72(86) & 1794800 & 7 & 1753128 & 7 & 1778$\pm^{ 12 }_{ 12 }$ & $1.00$ & 25.7005 & $ -0.05\pm^{ 0.17 }_{ 0.23 }$ & $ nan\pm^{ nan }_{ nan }$ & 0.16\\ 
1es1553 & $0.21631$ (\#7)  & 84.95(81) & 1756247 & 7 & 1757203 & 8 & 1778$\pm^{ 14 }_{ 14 }$ & $1.00$ & 26.2723 & $ 0.19\pm^{ 0.25 }_{ 0.22 }$ & $ 14.83\pm^{ 0.33 }_{ nan }$ & 0.95\\ 
1es1553 & $0.31130$ (\#8)  & 113.79(90) & 1740303 & 11 & 1677696 & 14 & 1807$\pm^{ 15 }_{ 15 }$ & $1.00$ & 28.3241 & $ -0.15\pm^{ 0.26 }_{ 0.23 }$ & $ nan\pm^{ nan }_{ nan }$ & 0.45\\ 
1es1553 & $0.37868$ (\#9)  & 119.25(98) & 1809416 & 21 & 1817188 & 12 & 1834$\pm^{ 17 }_{ 17 }$ & $1.00$ & 29.7795 & $ 0.00\pm^{ 0.09 }_{ 0.80 }$ & $ 13.08\pm^{ 1.43 }_{ nan }$ & -0.04\\ 
1es1553 & $0.39497$ (\#10)  & 90.13(98) & 1751580 & 21 & 1813451 & 12 & 1810$\pm^{ 18 }_{ 18 }$ & $1.00$ & 30.1314 & $ 0.02\pm^{ 0.38 }_{ 0.24 }$ & $ 13.80\pm^{ 1.32 }_{ nan }$ & 0.02\\ 
\dotfill\\
\hline
\hline
    \end{tabular}
    \caption{\ovii\ measurements with \xmm\ data, at the prior redshift from the \ovi\ lines from Table~\ref{tab:oviPaper}. Column "\%" reports the percent background level, ${B}/{S+B}$, where $B$ is the background count rate, and $S$ the source count rate. Entries with an asterisk sign indicate poor fits; entries that have a `$\nodata$' sign indicate lines that fall in a region of reduced efficiency, and therefore they could not be constrained (see Sec.~\ref{sec:dataAnalysis}). Full table is provided in the on--line version of the paper.}
    \label{tab:OVIIOVIXMM}
\end{table*}

\begin{table*} 
\begin{tabular}{cc|lllllllllH|l}
 \hline
 \hline
 \multicolumn{2}{c}{Target line} & \cmin & \multicolumn{2}{c}{RGS1} & \multicolumn{2}{c}{RGS2} & \multicolumn{2}{c}{power--law} & \multicolumn{3}{c}{line component} & $\Delta C$\\
 Name & z & (d.o.f.)& avg exp (s) & \% & avg exp (s) & \%   & \text{norm.} & \text{index} & $\lambda$ (\AA) & $\tau_0$ & $\log N (\text{cm}^{-2})$ & \\
 \hline 
1es1028 & $0.13714$ (\#1)  & 92.78(90) & 143804 & 9 & 146370 & 7 & 1121$\pm^{ 32 }_{ 32 }$ & $1.00$ & 24.5622 & $ -0.33\pm^{ 0.82 }_{ 0.55 }$ & $ nan\pm^{ nan }_{ nan }$ & 0.27\\ 
1es1028 & $0.20383$ (\#2)  & 114.22(94) & 146421 & 9 & 145107 & 9 & 1207$\pm^{ 32 }_{ 35 }$ & $1.00$ & 26.0027 & $ 0.03\pm^{ 0.97 }_{ 0.83 }$ & $ 14.01\pm^{ 1.43 }_{ nan }$ & -0.01\\ 
1es1028 & $0.22121$ (\#3)  & 101.58(92) & 144278 & 9 & 146704 & 9 & 1198$\pm^{ 35 }_{ 35 }$ & $1.00$ & 26.3781 & $ -0.70\pm^{ 0.67 }_{ 0.48 }$ & $ nan\pm^{ nan }_{ nan }$ & 1.33\\ 
1es1553 & $0.03466$ (\#4)  & 38.25(27) & 1791095 & 9 & nan & nan & 747$\pm^{ 397 }_{ 326 }$ & $1.97$ & 22.3487 & $ -0.18\pm^{ 0.33 }_{ 0.25 }$ & $ nan\pm^{ nan }_{ nan }$ & 0.53\\ 
1es1553 & $0.04273$ (\#5)  & 30.79(21) & 1784759 & 9 & nan & nan & 489$\pm^{ 628 }_{ 294 }$ & $2.72$ & 22.5230 & $ 0.51\pm^{ 0.64 }_{ 0.44 }$ & $ 15.21\pm^{ 0.28 }_{ 0.85 }$ & 2.85\\ 
1es1553 & $0.06364$ (\#6*)  & 64.48(28) & 1661330 & 9 & nan & nan & 27$\pm^{ 12 }_{ 9 }$ & $7.49$ & 22.9746 & $ 0.53\pm^{ 0.63 }_{ 0.44 }$ & $ 15.22\pm^{ 0.27 }_{ 0.73 }$ & 1.84\\ 
1es1553 & $0.21869$ (\#7)  & 108.81(87) & 1691684 & 7 & 1788288 & 8 & 1010$\pm^{ 415 }_{ 304 }$ & $1.74$ & 26.3237 & $ -0.01\pm^{ 0.65 }_{ 0.08 }$ & $ nan\pm^{ nan }_{ nan }$ & -0.07\\ 
\dotfill\\
\hline
\hline
    \end{tabular}
    \caption{\ovii\ measurements with \xmm\ data, at the prior redshift from the \hi\ lines from Table~\ref{tab:hiPaper}.}
    \label{tab:OVIIHIXMM}
\end{table*}

\begin{table*}
\begin{tabular}{cc|lllllllllH|l}
 \hline
 \hline
 \multicolumn{2}{c}{Target line} & \cmin & \multicolumn{2}{c}{RGS1} & \multicolumn{2}{c}{RGS2} & \multicolumn{2}{c}{power--law} & \multicolumn{3}{c}{line component} & $\Delta C$\\
 Name & z & (d.o.f.)& avg exp (s) & \% & avg exp (s) & \%   & \text{norm.} & \text{index} & $\lambda$ (\AA) & $\tau_0$ & $\log N (\text{cm}^{-2})$ &  \\
 \hline 
1es1028 & $0.12314$ (\#1)  & 33.31(31) & 146792 & 8 & nan & nan & 963$\pm^{ 40 }_{ 26 }$ & $1.00$ & 21.2947 & $ 0.02\pm^{ 3.99 }_{ 0.61 }$ & $ 14.13\pm^{ 2.00 }_{ nan }$ & 0.01\\ 
1es1028 & $0.13706$ (\#2)  & 37.48(28) & 147949 & 9 & nan & nan & 956$\pm^{ 28 }_{ 28 }$ & $1.00$ & 21.5587 & $ 1.00E03\pm^{ 1.00E20 }_{ 921.62 }$ & $ 16.44\pm^{ 0.30 }_{ 0.09 }$ & 0.55\\
1es1028 & $0.33735$ (\#3)  & 92.62(96) & 147900 & 8 & 144798 & 8 & 1196$\pm^{ 31 }_{ 31 }$ & $1.00$ & 25.3562 & $ 5.01\pm^{ 15.34 }_{ 3.36 }$ & $ 16.16\pm^{ 0.15 }_{ 0.24 }$ & 6.58\\ 
1es1553 & $0.18759$ (\#4)  & 29.26(27) & 1729540 & 9 & nan & nan & 540$\pm^{ 318 }_{ 194 }$ & $2.55$ & 22.5167 & $ 0.37\pm^{ 0.44 }_{ 0.33 }$ & $ 15.43\pm^{ 0.29 }_{ 0.90 }$ & 1.81\\ 
1es1553 & $0.18775$ (\#5)  & 28.91(27) & 1729540 & 9 & nan & nan & 550$\pm^{ 295 }_{ 211 }$ & $2.52$ & 22.5197 & $ 0.39\pm^{ 0.49 }_{ 0.31 }$ & $ 15.45\pm^{ 0.29 }_{ 0.62 }$ & 2.20\\ 
1es1553 & $0.18984$ (\#6)  & 25.62(27) & 1729542 & 9 & nan & nan & 406$\pm^{ 219 }_{ 140 }$ & $3.05$ & 22.5594 & $ 0.98\pm^{ 0.60 }_{ 0.48 }$ & $ 15.77\pm^{ 0.14 }_{ 0.23 }$ & 8.20\\ 
1es1553 & $0.21631$ (\#7)  & 33.20(25) & 1723969 & 9 & nan & nan & 60$\pm^{ 36 }_{ 21 }$ & $6.19$ & 23.0612 & $ -0.08\pm^{ 0.43 }_{ 0.30 }$ & $ nan\pm^{ nan }_{ nan }$ & 0.03\\ 
1es1553 & $0.31130$ (\#8)  & 113.12(91) & 1744157 & 7 & 1748313 & 7 & 2268$\pm^{ 757 }_{ 440 }$ & $0.67$ & 24.8622 & $ 1.01\pm^{ 0.37 }_{ 0.33 }$ & $ 15.80\pm^{ 0.10 }_{ 0.14 }$ & 17.08\\ 
1es1553 & $0.37868$ (\#9)  & 113.07(88) & 1784762 & 7 & 1713272 & 8 & 1213$\pm^{ 467 }_{ 336 }$ & $1.52$ & 26.1398 & $ 1.03\pm^{ 0.44 }_{ 0.35 }$ & $ 15.79\pm^{ 0.11 }_{ 0.14 }$ & 15.34\\ 
1es1553 & $0.39497$ (\#10)  & 118.64(92) & 1704141 & 7 & 1756581 & 9 & 481$\pm^{ 192 }_{ 138 }$ & $2.74$ & 26.4486 & $ -0.20\pm^{ 0.19 }_{ 0.18 }$ & $ nan\pm^{ nan }_{ nan }$ & 1.31\\ 
\dotfill\\
    \hline
\hline
\end{tabular}
    \caption{\oviii\ measurements with \xmm\ data, at the prior redshift from the \ovi\ lines from Table~\ref{tab:oviPaper}.}
    \label{tab:OVIIIOVIXMM}
\end{table*}

\begin{table*}
    \begin{tabular}{cc|lllllllllH|l}
 \hline
 \hline
\multicolumn{2}{c}{Target line} & \cmin & \multicolumn{2}{c}{RGS1} & \multicolumn{2}{c}{RGS2} & \multicolumn{2}{c}{power--law} & \multicolumn{3}{c}{line component} & $\Delta C$\\
 Name & z & (d.o.f.)& avg exp (s) & \% & avg exp (s) & \%   & \text{norm.} & \text{index} & $\lambda$ (\AA) & $\tau_0$ & $\log N (\text{cm}^{-2})$ & \\
 \hline 
1es1028 & $0.13714$ (\#1)  & 37.48(28) & 147949 & 9 & nan & nan & 956$\pm^{ 35 }_{ 28 }$ & $1.00$ & 21.5602 & $ 2.91E+03\pm^{ 1.00E+20 }_{ 2.63E+03 }$ & $ 16.45\pm^{ 0.30 }_{ 0.06 }$ & 0.55\\ 
1es1028 & $0.20383$ (\#2)  & 62.35(44) & 146640 & 9 & nan & nan & 1006$\pm^{ 29 }_{ 29 }$ & $1.00$ & 22.8246 & $ 0.05\pm^{ 2.21 }_{ 0.94 }$ & $ 14.58\pm^{ 1.44 }_{ nan }$ & 0.01\\ 
1es1028 & $0.22121$ (\#3)  & 57.62(42) & 144940 & 9 & nan & nan & 1031$\pm^{ 32 }_{ 31 }$ & $1.00$ & 23.1541 & $ -1.47\pm^{ 0.70 }_{ 0.53 }$ & $ nan\pm^{ nan }_{ nan }$ & 3.89\\ 
1es1553 & $0.03466$ (\#4)  & 90.75(81) & 1722815 & 7 & 1780022 & 7 & 752$\pm^{ 116 }_{ 96 }$ & $2.09$ & 19.6172 & $ -0.03\pm^{ 0.65 }_{ 0.06 }$ & $ nan\pm^{ nan }_{ nan }$ & -0.41\\ 
1es1553 & $0.04273$ (\#5)  & 89.06(72) & 1720386 & 8 & 1774803 & 7 & 776$\pm^{ 127 }_{ 108 }$ & $2.01$ & 19.7702 & $ 0.08\pm^{ 0.18 }_{ 0.21 }$ & $ 14.82\pm^{ 0.47 }_{ nan }$ & 0.14\\ 
1es1553 & $0.06364$ (\#6)  & 46.18(56) & 1785421 & 8 & 1740261 & 7 & 1161$\pm^{ 253 }_{ 220 }$ & $1.18$ & 20.1666 & $ 0.07\pm^{ 0.22 }_{ 0.32 }$ & $ 14.72\pm^{ 0.61 }_{ nan }$ & -0.04\\ 
1es1553 & $0.21869$ (\#7)  & 36.89(28) & 1666450 & 9 & nan & nan & 36$\pm^{ 20 }_{ 13 }$ & $7.01$ & 23.1064 & $ -0.43\pm^{ 0.30 }_{ 0.26 }$ & $ nan\pm^{ nan }_{ nan }$ & 2.52\\ 
\dotfill\\
\hline
\hline
\end{tabular}
    \caption{\oviii\ measurements with \xmm\ data, at the prior redshift from the \hi\ lines from Table~\ref{tab:hiPaper}.}
    \label{tab:OVIIIHIXMM}
\end{table*}

\begin{table*}
    \begin{tabular}{cc|lllllllH|l}
 \hline
 \hline
 \multicolumn{2}{c}{Target line} & \cmin & \multicolumn{2}{c}{LETG} & \multicolumn{2}{c}{power--law} & \multicolumn{3}{c}{line component} & $\Delta C$\\
 Name & z & (d.o.f.)& avg exp (s) & \%  & \text{norm.} & \text{index} & $\lambda$ (\AA) & $\tau_0$ & $\log N (\text{cm}^{-2})$ & \\
 \hline 
1es1028 & $0.12314$ (\#1)  & 32.34(39) & 148933 & 24 & 1757$\pm^{ 63 }_{ 62 }$ & $1$ & 24.2598 & $ -0.56\pm^{ 0.97 }_{ 0.68 }$ & $ nan\pm^{ nan }_{ nan }$ & 0.44\\ 
1es1028 & $0.13706$ (\#2)  & 35.52(39) & 148933 & 24 & 1830$\pm^{ 64 }_{ 63 }$ & $1$ & 24.5605 & $ 1.29\pm^{ 3.09 }_{ 1.34 }$ & $ 15.56\pm^{ 0.30 }_{ nan }$ & 0.99\\ 
1es1028 & $0.33735$ (\#3)  & 37.26(39) & 148933 & 25 & 2138$\pm^{ 81 }_{ 79 }$ & $1$ & 28.8868 & $ -0.66\pm^{ 0.98 }_{ 0.74 }$ & $ nan\pm^{ nan }_{ nan }$ & 0.56\\ 
1es1553 & $0.18759$ (\#4)  & 44.14(38) & 495645 & 22 & 3135$\pm^{ 48 }_{ 48 }$ & $1$ & 25.6519 & $ 0.34\pm^{ 0.57 }_{ 0.46 }$ & $ 15.06\pm^{ 0.37 }_{ nan }$ & 0.58\\ 
1es1553 & $0.18775$ (\#5)  & 44.13(38) & 495645 & 22 & 3135$\pm^{ 48 }_{ 48 }$ & $1$ & 25.6554 & $ 0.35\pm^{ 0.62 }_{ 0.47 }$ & $ 15.07\pm^{ 0.37 }_{ nan }$ & 0.58\\ 
1es1553 & $0.18984$ (\#6)  & 42.53(38) & 495645 & 22 & 3106$\pm^{ 48 }_{ 48 }$ & $1$ & 25.7005 & $ -0.38\pm^{ 0.40 }_{ 0.36 }$ & $ nan\pm^{ nan }_{ nan }$ & 1.00\\ 
1es1553 & $0.21631$ (\#7)  & 51.98(38) & 495645 & 23 & 3137$\pm^{ 49 }_{ 49 }$ & $1$ & 26.2723 & $ 0.29\pm^{ 0.58 }_{ 0.46 }$ & $ 15.00\pm^{ 0.42 }_{ nan }$ & 0.42\\ 
1es1553 & $0.31130$ (\#8)  & 36.35(38) & 495645 & 25 & 3202$\pm^{ 53 }_{ 54 }$ & $1$ & 28.3241 & $ -0.04\pm^{ 0.53 }_{ 0.44 }$ & $ nan\pm^{ nan }_{ nan }$ & 0.01\\ 
1es1553 & $0.37868$ (\#9)  & 48.60(38) & 495645 & 28 & 3322$\pm^{ 62 }_{ 61 }$ & $1$ & 29.7795 & $ -0.69\pm^{ 0.39 }_{ 0.33 }$ & $ nan\pm^{ nan }_{ nan }$ & 2.89\\ 
1es1553 & $0.39497$ (\#10)  & 39.90(38) & 495645 & 28 & 3374$\pm^{ 65 }_{ 64 }$ & $1$ & 30.1314 & $ -0.48\pm^{ 0.47 }_{ 0.37 }$ & $ nan\pm^{ nan }_{ nan }$ & 1.16\\ 
\dotfill\\
\hline
\hline
\end{tabular}
    \caption{\ovii\ measurements with \chandra\ data, at the prior redshift from the \ovi\ lines from Table~\ref{tab:oviPaper}.}
    \label{tab:OVIIOVIChandra}
\end{table*}

\begin{table*}
    \begin{tabular}{cc|lllllllH|l}
 \hline
 \hline
 \multicolumn{2}{c}{Target line} & \cmin & \multicolumn{2}{c}{RGS1} & \multicolumn{2}{c}{power--law} & \multicolumn{3}{c}{line component} & $\Delta C$\\
 Name & z & (d.o.f.)& avg exp (s) & \% & \text{norm.} & \text{index} & $\lambda$ (\AA) & $\tau_0$ & $\log N (\text{cm}^{-2})$ & \\
 \hline 
1es1028 & $0.13714$ (\#1)  & 35.38(39) & 148933 & 24 & 1831$\pm^{ 64 }_{ 63 }$ & $1$ & 24.5622 & $ 1.34\pm^{ 2.73 }_{ 1.32 }$ & $ 15.58\pm^{ 0.28 }_{ 1.74 }$ & 1.14\\ 
1es1028 & $0.20383$ (\#2)  & 38.65(38) & 148933 & 24 & 1963$\pm^{ 69 }_{ 68 }$ & $1$ & 26.0027 & $ 1.18\pm^{ 2.92 }_{ 1.29 }$ & $ 15.52\pm^{ 0.31 }_{ nan }$ & 0.87\\ 
1es1028 & $0.22121$ (\#3)  & 35.87(38) & 148933 & 23 & 2058$\pm^{ 71 }_{ 70 }$ & $1$ & 26.3781 & $ 0.80\pm^{ 2.27 }_{ 1.13 }$ & $ 15.39\pm^{ 0.38 }_{ nan }$ & 0.48\\ 
1es1553 & $0.03466$ (\#4)  & 38.20(38) & 495645 & 25 & 2408$\pm^{ 40 }_{ 39 }$ & $1$ & 22.3487 & $ -0.03\pm^{ 0.50 }_{ 0.43 }$ & $ nan\pm^{ nan }_{ nan }$ & 0.00\\ 
1es1553 & $0.04273$ (\#5)  & 54.53(38) & 495645 & 25 & 2461$\pm^{ 41 }_{ 41 }$ & $1$ & 22.5230 & $ 0.31\pm^{ 0.61 }_{ 0.49 }$ & $ 15.03\pm^{ 0.41 }_{ nan }$ & 0.42\\ 
1es1553 & $0.06364$ (\#6)  & 48.63(38) & 495645 & 24 & 2557$\pm^{ 43 }_{ 42 }$ & $1$ & 22.9746 & $ -1.05\pm^{ 0.36 }_{ 0.32 }$ & $ nan\pm^{ nan }_{ nan }$ & 7.56\\ 
1es1553 & $0.21869$ (\#7)  & 44.59(38) & 495645 & 23 & 3121$\pm^{ 49 }_{ 48 }$ & $1$ & 26.3237 & $ -0.47\pm^{ 0.39 }_{ 0.35 }$ & $ nan\pm^{ nan }_{ nan }$ & 1.52\\ 
\dotfill\\
\hline
\hline
\end{tabular}
    \caption{\ovii\ measurements with \chandra\ data, at the prior redshift from the \hi\ lines from Table~\ref{tab:hiPaper}.}
    \label{tab:OVIIHIChandra}
\end{table*}

\begin{table*}
    \begin{tabular}{cc|lllllllH|l}
 \hline
 \hline
 \multicolumn{2}{c}{Target line} & \cmin & \multicolumn{2}{c}{RGS1} & \multicolumn{2}{c}{power--law} & \multicolumn{3}{c}{line component} & $\Delta C$\\
 Name & z & (d.o.f.)& avg exp (s) & \%  & \text{norm.} & \text{index} & $\lambda$ (\AA) & $\tau_0$ & $\log N (\text{cm}^{-2})$ & \\
 \hline 
1es1028 & $0.12314$ (\#1)  & 35.82(30) & 148933 & 23 & 1585$\pm^{ 62 }_{ 61 }$ & $1$ & 21.2947 & $ -0.79\pm^{ 0.77 }_{ 0.55 }$ & $ nan\pm^{ nan }_{ nan }$ & 1.19\\ 
1es1028 & $0.13706$ (\#2)  & 32.10(31) & 148933 & 22 & 1747$\pm^{ 65 }_{ 79 }$ & $1$ & 21.5587 & $ 0.22\pm^{ 6.36E14 }_{ 6.35 }$ & $ 15.23\pm^{ 1.55 }_{ nan }$ & -0.01\\
1es1028 & $0.33735$ (\#3)  & 34.59(38) & 148933 & 24 & 1851$\pm^{ 66 }_{ 65 }$ & $1$ & 25.3562 & $ -0.24\pm^{ 1.24 }_{ 0.72 }$ & $ nan\pm^{ nan }_{ nan }$ & 0.07\\ 
1es1553 & $0.18759$ (\#4)  & 43.76(33) & 495645 & 25 & 2462$\pm^{ 44 }_{ 44 }$ & $1$ & 22.5167 & $ 0.23\pm^{ 0.63 }_{ 0.47 }$ & $ 15.23\pm^{ 0.50 }_{ nan }$ & 0.23\\ 
1es1553 & $0.18775$ (\#5)  & 43.78(32) & 495645 & 25 & 2464$\pm^{ 44 }_{ 44 }$ & $1$ & 22.5197 & $ 0.24\pm^{ 0.63 }_{ 0.48 }$ & $ 15.25\pm^{ 0.48 }_{ nan }$ & 0.23\\ 
1es1553 & $0.18984$ (\#6)  & 40.55(33) & 495645 & 25 & 2490$\pm^{ 43 }_{ 43 }$ & $1$ & 22.5594 & $ 0.20\pm^{ 1.23 }_{ 0.76 }$ & $ 15.18\pm^{ 0.73 }_{ nan }$ & 0.06\\ 
1es1553 & $0.21631$ (\#7)  & 30.16(26) & 495645 & 24 & 2584$\pm^{ 53 }_{ 52 }$ & $1$ & 23.0612 & $ -0.45\pm^{ 0.44 }_{ 0.38 }$ & $ nan\pm^{ nan }_{ nan }$ & 1.20\\ 
1es1553 & $0.31130$ (\#8)  & 50.56(39) & 495645 & 22 & 2998$\pm^{ 46 }_{ 44 }$ & $1$ & 24.8622 & $ -0.05\pm^{ 0.58 }_{ 0.33 }$ & $ nan\pm^{ nan }_{ nan }$ & -0.03\\ 
1es1553 & $0.37868$ (\#9)  & 48.06(39) & 495645 & 23 & 3124$\pm^{ 49 }_{ 48 }$ & $1$ & 26.1398 & $ -0.37\pm^{ 0.41 }_{ 0.35 }$ & $ nan\pm^{ nan }_{ nan }$ & 0.89\\ 
1es1553 & $0.39497$ (\#10)  & 41.84(38) & 495645 & 23 & 3095$\pm^{ 49 }_{ 49 }$ & $1$ & 26.4486 & $ -0.63\pm^{ 0.38 }_{ 0.33 }$ & $ nan\pm^{ nan }_{ nan }$ & 2.79\\
\dotfill\\
\hline
\hline
\end{tabular}
    \caption{\oviii\ measurements with \chandra\ data, at the prior redshift from the \ovi\ lines from Table~\ref{tab:oviPaper}.}
    \label{tab:OVIIIOVIChandra}
\end{table*}

\begin{table*}
    \begin{tabular}{cc|lllllllH|l}
 \hline
 \hline
 \multicolumn{2}{c}{Target line} & \cmin & \multicolumn{2}{c}{RGS1} & \multicolumn{2}{c}{power--law} & \multicolumn{3}{c}{line component} & $\Delta C$\\
 Name & z & (d.o.f.)& avg exp (s) & \% & \text{norm.} & \text{index} & $\lambda$ (\AA) & $\tau_0$ & $\log N (\text{cm}^{-2})$ &\\
 \hline 
1es1028 & $0.13714$ (\#1)  & 32.10(31) & 148933 & 22 & 1747$\pm^{ 63 }_{ 70 }$ & $1$ & 21.5602 & $ -0.06\pm^{ 9.42E15 }_{ 6.69 }$ & $ nan\pm^{ nan }_{ nan }$ & 0.00\\
1es1028 & $0.20383$ (\#2)  & 24.47(38) & 148933 & 25 & 1653$\pm^{ 61 }_{ 59 }$ & $1$ & 22.8246 & $ -1.22\pm^{ 0.69 }_{ 0.52 }$ & $ nan\pm^{ nan }_{ nan }$ & 2.76\\ 
1es1028 & $0.22121$ (\#3)  & 26.80(38) & 148933 & 25 & 1757$\pm^{ 63 }_{ 61 }$ & $1$ & 23.1541 & $ 3.82\pm^{ 13.01 }_{ 2.75 }$ & $ 16.13\pm^{ 0.19 }_{ 0.32 }$ & 3.68\\ 
1es1553 & $0.03466$ (\#4)  & 51.93(39) & 495645 & 21 & 2174$\pm^{ 31 }_{ 31 }$ & $1$ & 19.6172 & $ 0.21\pm^{ 0.53 }_{ 0.42 }$ & $ 15.19\pm^{ 0.49 }_{ nan }$ & 0.23\\ 
1es1553 & $0.04273$ (\#5)  & 42.22(38) & 495645 & 22 & 2213$\pm^{ 32 }_{ 32 }$ & $1$ & 19.7702 & $ 0.00\pm^{ 0.58 }_{ 0.35 }$ & $ 13.57\pm^{ 2.03 }_{ nan }$ & 0.00\\ 
1es1553 & $0.06364$ (\#6)  & 42.31(39) & 495645 & 23 & 2255$\pm^{ 35 }_{ 35 }$ & $1$ & 20.1666 & $ 0.86\pm^{ 0.85 }_{ 0.59 }$ & $ 15.74\pm^{ 0.21 }_{ 0.44 }$ & 2.66\\ 
1es1553 & $0.21869$ (\#7)  & 36.69(30) & 495645 & 24 & 2585$\pm^{ 49 }_{ 48 }$ & $1$ & 23.1064 & $ -0.35\pm^{ 0.43 }_{ 0.35 }$ & $ nan\pm^{ nan }_{ nan }$ & 0.78\\ 
\dotfill\\
\hline
\hline
\end{tabular}
    \caption{\oviii\ measurements with \chandra\ data, at the prior redshift from the \hi\ lines from Table~\ref{tab:hiPaper}.}
    \label{tab:OVIIIHIChandra}
\end{table*}

\section{Results of the X--ray search}
\label{sec:results}

\begin{figure*}
    \centering
    \includegraphics[width=6.5in]{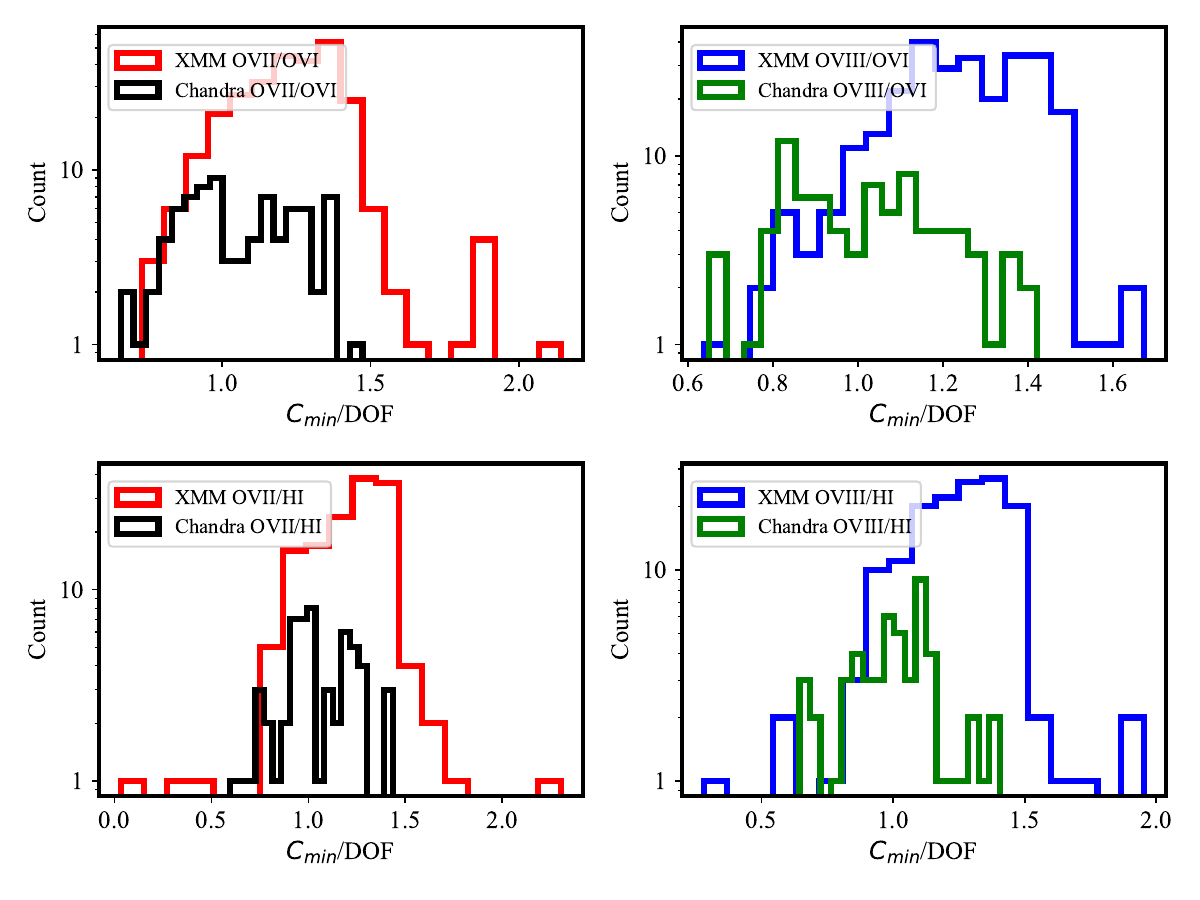}
    \caption{Distribution of number of systems ($y$ axis) as a function of \gof\ \cmin\ per degree of freedom ($x$ axis) for all fits with \cmin/DOF$\leq 3$. All regressions with \cmin/DOF$\geq 2$ (i.e., substantially poor fits) are indicated in the tables with an asterisk.}
    \label{fig:Cmin}
\end{figure*}

\subsection{Basic Statistics}
\label{sec:basicStatistics}
This project uses the  \cmin\ statistic as the \gof\ statistic of choice for the type 
of integer--count Poisson data used in the spectral analysis,  as implemented by the \texttt{SPEX} software package \citep{kaastra1996}. In the large--count limit, the \cmin\ statistic is asymptotically distributed like a chi--squared distribution with a number of degrees of freedom equal to $N-m$,
where $N$ is the number of datapoints, and $m$ is the number of free parameters \citep[e.g.][]{cash1979,bonamente2023}. Figure~\ref{fig:Cmin} shows the distribution of the \gof\ statistic \cmin\  for the spectral fits performed for this sample.

Several of the short exposure observations are in the low--count regime (i.e, $\leq 20$ counts per bin), where the asymptotic distribution no longer applies accurately \citep[e.g.][]{kaastra2017,bonamente2020,li2024}. Given the inhomogeneity of the sample, which features sources 
in a range from few  counts to several hundred counts per bin, we do not enforce a uniform 'cut' in the acceptable
\gof, which could be done by using the $p$--value associated with each regression. Rather, we remove the worst fits (\cmin/\dof $\geq 3$; these are 39 out of \nsystems\ fits) from future consideration, 
and indicate all poor fits (reduced \cmin$\geq 2$) with an asterisk in the tables.~\footnote{The figures in this section
 report fits with \cmin/\dof $\leq 3$, and poor fits (\cmin/\dof $\geq 2$) that are present in the tables are indicated with an asterisk. We opted to report the poor fits for the sake of completeness, so as to illustrate the difficulties in the regression in a limited number of spectra.} 
 This choice permits the analyst to see the largest number possible of fit results, and maintains the completeness of the sample to the largest degree possible.
Fig.~\ref{fig:Cmin} show that the vast majority of fits have reasonable values for the \gof\ (i.e, \cmin/\dof $\simeq 1$), which was the goal in the analysis of such large and heterogeneous sample of sources.

\begin{figure*}
    \centering
    \includegraphics[width=6.5in]{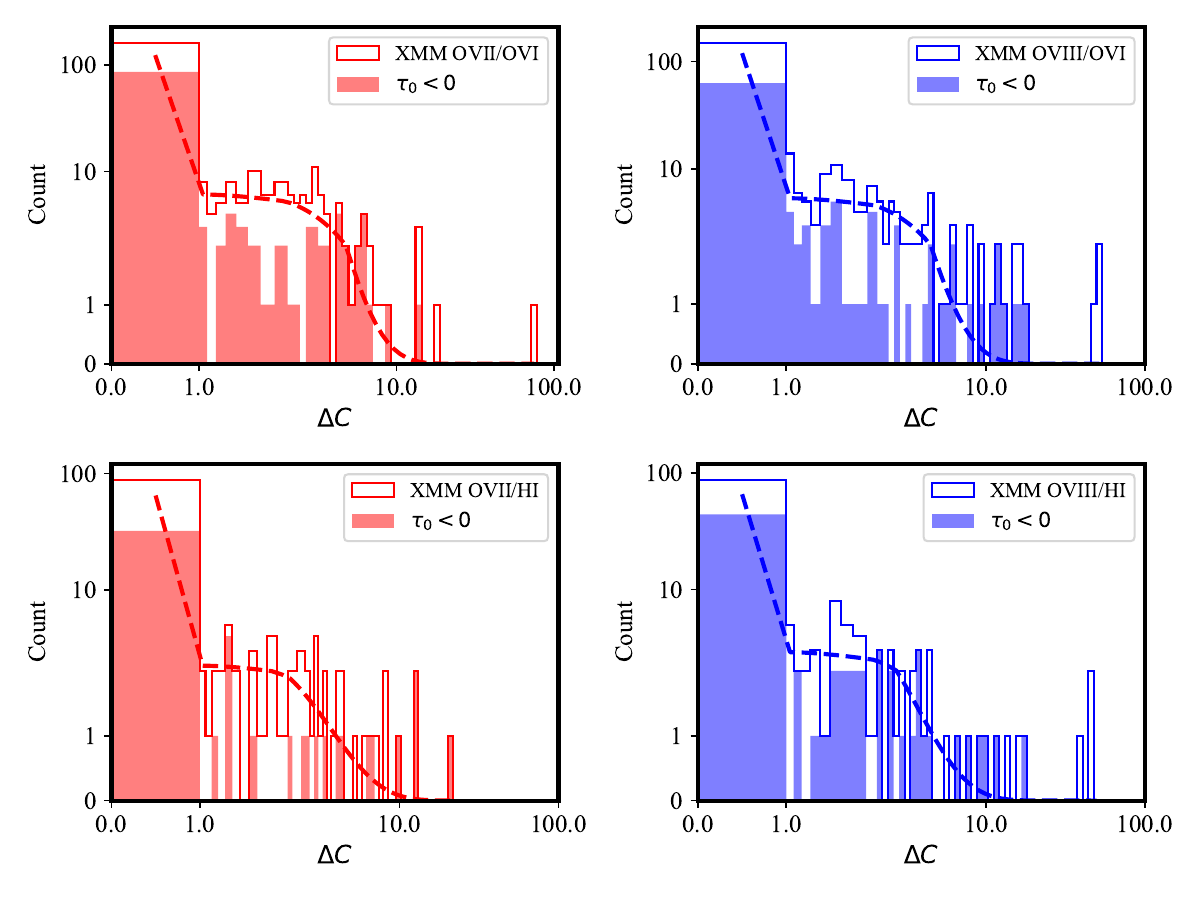}
    \caption{Distribution of $\Delta C$ statistics for the \xmm\ fits; open histogram plots are all the fits, and filled histogram a sub--set with negative best--fit $\tau_0$. The dashed line represent the expected cumulative distribution function (CDF) of a 
    $\chi^2(1)$ variable, representing the null hypothesis that there are no emission or absorption lines in the sample. Note that the \texttt{symlog} scale and the logarithmic binning contribute to the shape of the CDF. For this figure, 
    we only report statistics for fits with \cmin/\dof $\leq 3$, same as in Fig.~\ref{fig:Cmin} To further aid in the identification of possible absorption line detection, the sub--set of fits with negative best--fit parameter $\tau_0$, and therefore with \emph{positive} excess flux above the continuum, where highlighted with a solid histogram plot. Possible absorption line detections are therefore for large values of the $\Delta C$ statistic, and with an open histogram.}
    \label{fig:DeltaCXMM}
\end{figure*}

\begin{figure*}
    \centering
    \includegraphics[width=6.5in]{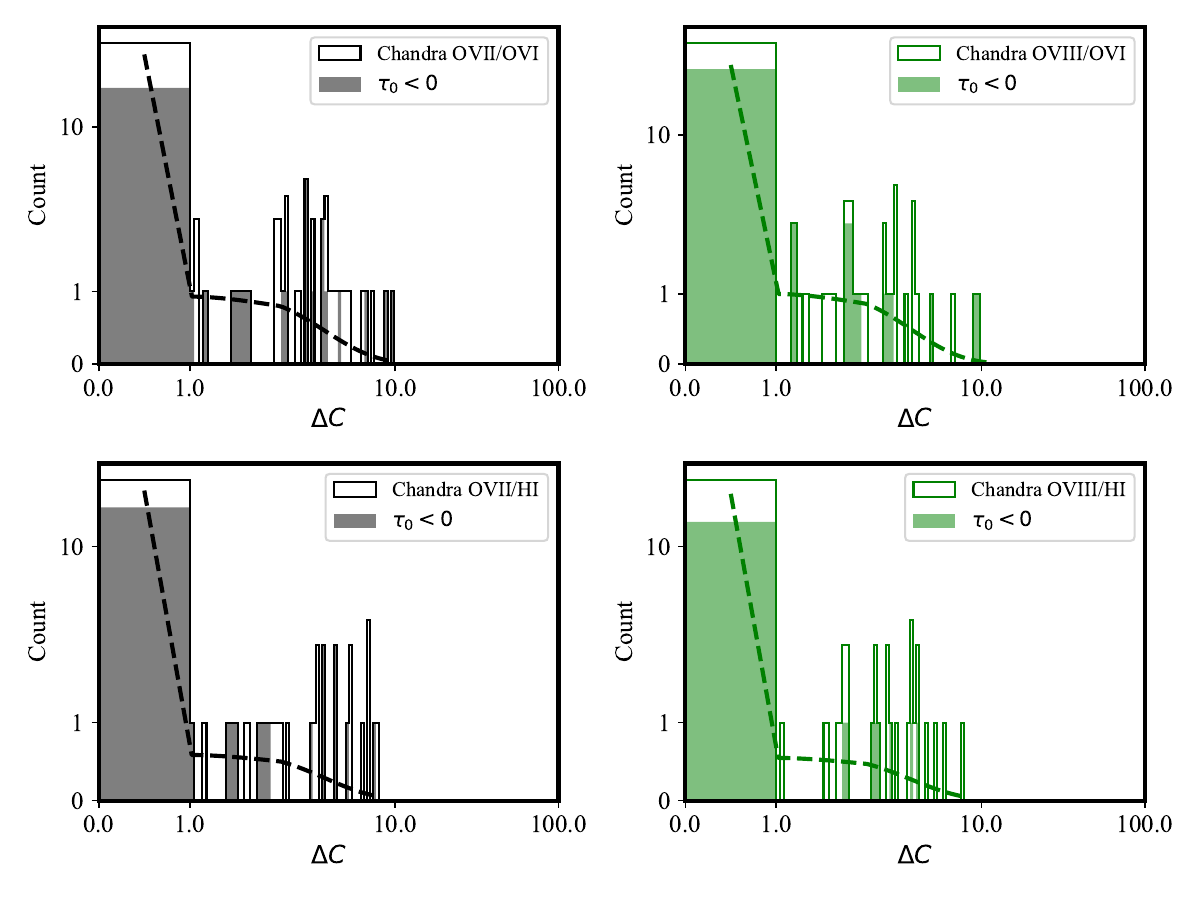}
    \caption{Distribution of $\Delta C$ statistics for the detection of possible absorption lines for the \chandra\ fits. See caption of Fig.~\ref{fig:DeltaCXMM} for additional comments.}
    \label{fig:DeltaCChandra}
\end{figure*}

We use the $\Delta C$ statistic \citep{cash1976, cash1979} as a measure of the significance of detection of a possible
    absorption/emission line. This statistic is defined as
    \begin{equation}
        \Delta C = \cmineq(\tau_0=0) - \cmineq \geq 0
        \label{eq:DeltaC}
    \end{equation}
    where \cmin\ is the global \gof\ statistic for the \texttt{power--law} plus \texttt{line} model, and $\cmineq(\tau_0=0)$ 
    is the statistic obtained by fixing the \texttt{line} parameter to zero, i.e., for the fit to the  \texttt{power--law} model alone.
    The $\Delta C$ statistic is positive--definite, and it is distributed as a chi--squared random variable with one degree of freedom that is introduced by the single additional adjustable parameter $\tau_0$ of the \texttt{line}
model relative to the null--hypothesis or baseline model with $\tau_0=0$ \citep[e.g.][]{cash1979}. Therefore, the $1-\sigma$ or 68\%, 90 and 99\% confidence levels for the detection of a $\tau_0 \geq 0$
parameter correspond respectively to critical values of $\Delta C=1.0, 2.7, 6.6$. A large value of
the $\Delta C$ statistic, together with a positive value for the best--fit $\tau_0$ parameter, therefore suggest the possible detection of an absorption line. Distribution of the $\Delta C$ statistics are illustrated in Fig.~\ref{fig:DeltaCXMM} and \ref{fig:DeltaCChandra}, along with the expected $\chi^2(1)$ distribution that represents the null hypothesis of no detection. Further discussion
of the applicability of the $\Delta C$ statistic to this method of analysis was provided in
\cite{spence2023}, especially with regards to the requirement that 
the null--hypothesis parameter $\tau_0=0$ falls in the interior of the allowed parameter space
\citep[e.g., as discussed in ][]{protassov2002}.

A summary of basic statistics for the \nsystems\ fits is provided in Table~\ref{tab:statistics}. The main result is  that there is a total of \nDet\ possible absorption line features detected at a 99\% significance
according to the $\Delta C$ statistic ($\Delta C \geq 6.6$  and positive values of $\tau_0$). These are the
systems that will be analyzed in more detail in Sec.~\ref{sec:detections}.
The choice of $\Delta C \geq 6.6$ as a threshold for consideration (see Sec.~\ref{sec:detections} below) as a possible detection is somewhat arbitrary. It exceeds the $p=0.05$ value
often used as discriminant for significance \citep{asa2016} in an effort to
be conservative, and also to account for possible systematic errors that would have a net effect to reduce the significance of detection
\citep[e.g.][]{bonamente2024}.  To facilitate the identification of possible absorption line detections, the distribution of fits that result in negative $\tau_0$ parameters (i.e., emission--line features) are represented as a filled--style histogram in Figs.~\ref{fig:DeltaCXMM} and \ref{fig:DeltaCChandra}. 

\begin{figure*}
    \centering
    \includegraphics[width=6.5in]{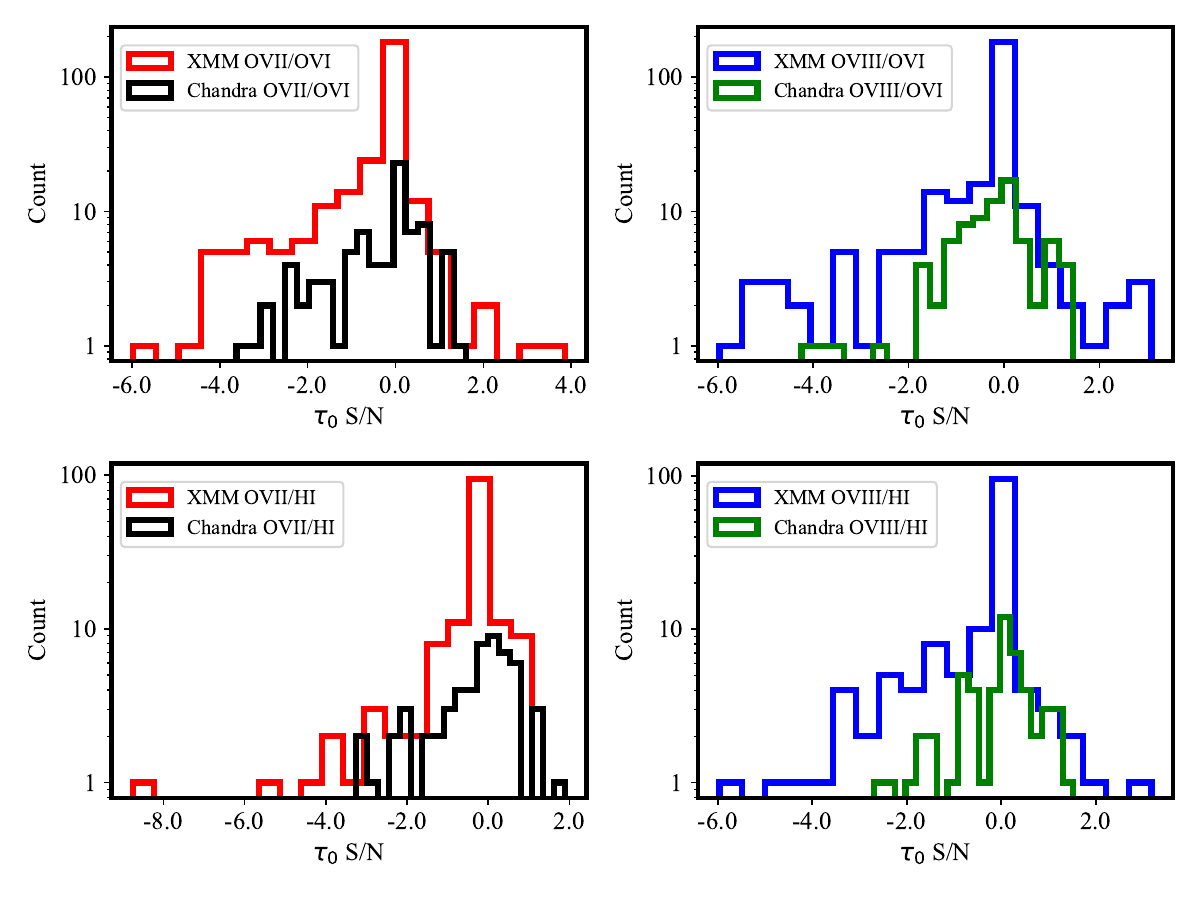}
    \caption{Distribution of the signal--to--noise ratio for the $\tau_0$
    parameter, $\hat{\tau}_0/\sigma_{\tau}$, where $\hat{\tau}_0$ is the best--fit value and $\sigma_{\tau}$ is the standard deviation of the parameter obtained by averaging the two uncertainties obtained with the usual $\Delta C=1$ criterion.}
    \label{fig:tau0}
\end{figure*}

Moreover, Fig.~\ref{fig:tau0} shows the distribution of the approximate signal--to--noise ratios of the $\tau_0$ parameter, obtained as a ratio of best--fit value to the average error. The distributions illustrates that there is a tail of \emph{negative} values, which is indicative of positive fluctuations or emission--line features. Figure~\ref{fig:tau0}, however, does not quantify
the significance of detection of a feature --- which is measured instead by
the $  \Delta C$ statistic --- but only the relative uncertainty of the estimated
$\tau_0$ parameter. It is necessary to point out that there is a nearly identical number of statistically significant
($\Delta C \geq 6.6$) positive and negative fluctuations (\nDet\ absorption--like features vs. \nEm\ emission--like features at the 99\% confidence), as shown in Tab.~\ref{tab:statistics}.
 Given that emission line features are not the focus of this paper, such features are not discussed further. 

\subsection{Redshift trials}
\label{sec:redshiftTrials}

\cite{nicastro2013} first introduced the concept of `redshift trials' to indicate the number of independent opportunities to detect a given feature, in this case an absorption--line feature. The issue is especially relevant to the type of blind searches performed by, e.g., \cite{nicastro2018} or \cite{gatuzz2023}. For the type of search conducted in this paper or by others at fixed redshift \citep[e.g.][]{ahoranta2020,ahoranta2021}, each FUV absorption line system has just one opportunity of being detected in X--rays, given that the search is performed at a fixed redshift. In this case, therefore, the statistical significance of detection of an individual feature, as performed via the $\Delta C$ statistic, is the formal significance of detection
according to the parent distribution of the statistic (i.e., the $\chi^2(1)$ distribution). The number of redshift trials, therefore, plays no role in determining whether any given FUV system has evidence for X--ray absorption.

One could seek to answer the associated question of whether the sample of \nsystems\ systems has evidence for the detection of at least one (or more) WHIM features, at a given level of significance. This is clearly a different question from that of assessing the significance of absorption in a \emph{given} system, in that this new question addresses the aggregate statistical behavior of the sample. To answer this question, we can use the framework laid out in \cite{bonamente2019}, which makes use of the binomial distribution to find the significance of detection of at least $m$ features, at a given $p$--level, given $N$ independent tries. In this application, $N=\nsystems$, and the $p$--value is obtained by the parent distribution of $\Delta C$: for example, features at $\Delta C=6.6$ correspond to $p=0.01$. According to this framework, which makes use of the binomial
distribution and the concept of redshift trials,
a simple way to determine whether the sample as a whole has
statistically significant evidence for absorption line features
is to test whether the $r=33$ features detected at $\geq 99$~\% significance can happen by random chance in $N=\nsystems$ trials.
Using the binomial distribution method, this null--hypothesis probability is calculated as
\[
P= \sum_{i=r}^N f(i,p) =  \sum_{i=r}^N \binom{N}{i} p^i (1-p)^{N-i}
\]
where $f(i,p)$ is the binomial probability of having outcome $i$ when the probability of a single success is $p$ (see Eq.~7 in \citealt{bonamente2019}). This probability evaluates to $P=5.7 \times 10^{-7}$, meaning that there is a negligible probability of a chance occurrence of so many significant features. 

Based on this statistical analysis, we therefore conclude that there is highly--significant evidence for absorption lines in these data also from the aggregate analysis of the sample. This 
aggregate probability can also be interpreted as the probability to exceed $\pm 5.0$ in a standard
normal distribution, e.g., the probability $P$ is equivalent to what is normally referred to as a $5 \sigma$ detection of absorption lines in this sample.~\footnote{This null--hypothesis probability is strictly conservative, since many of the features have substantially large values of $\Delta C$.}

\subsection{Summary of redshifted \ovii\ and \oviii\ absorption lines}
\label{sec:resultsRedshifted}

Tables~\ref{tab:OVIIOVIXMM} through \ref{tab:OVIIIHIChandra} report the
results of our search for \ovii\ and \oviii\ absorption lines at
the fixed redshifts provided by the FUV detection of \ovi\ and \hi\ BLA by 
\cite{tilton2012} and \cite{danforth2016}. Each line corresponds to a fit
with the \texttt{line} model for the possible emission or absorption line at the fixed FUV redshift. Each regression also shows the percent of background level above the source's signal, e.g., a value of 10\% means that the background has 10\% of the source's flux. As discussed in Sec.~\ref{sec:dataAnalysis}, most of the fits are with a fixed value of the power--law index of 1.0, and in some cases the index was left free, and the best--fit value is reported in the tables.

Given our choice of using fixed FUV priors, the significance of the line component is assessed via the $\Delta C$ statistic (see Eq.~\ref{eq:DeltaC}), as was discussed in Sec.~\ref{sec:redshiftTrials}. 
The statistic is positive--definite, and occasional (slightly) negative numbers for $\Delta C$ in the tables  are simply a result of errors in the numerical implementation of the procedure, and should be regarded as a value of zero.
Certain regressions have large uncertainties in the $\tau_0$ parameter and in the corresponding equivalent width,
signifying the inability of the data to provide meaningful constraints. Although the best--fit values and uncertainties are reported for all fits, those with the largest values should not be considered as physically meaningful, but simply a result of the poor quality of the data.


\subsection{Possible absorption line detections}
\label{sec:detections}
The \nDet\ systems in Tables~\ref{tab:OVIIOVIXMM} through \ref{tab:OVIIIHIChandra} with
best--fit parameter $\tau_0 >0$ and $\Delta C \geq 6.6$,
are discussed as candidates for the possible detection of absorption
lines at the redshift of the FUV priors with $\geq$~99\% confidence.
Spectra for the sources with possible WHIM absorption line detection are shown in Figs.~\ref{fig:detections} through \ref{fig:detections3}; systems that have a substantial wavelength overlap with others are not plotted. We also add to the discussion 
the \ovii\ system in 3C~273 (Sec.~\ref{sec:3c273ovii}), despite its \gof\ value that is slightly higher than the threshold, given the prior history with this system. A summary of key statistics for these
possible detections is provided in Table~\ref{tab:detections}.

\begin{table}
    \centering
    \begin{tabular}{lllllr}
    \hline
    \hline
    Source & ID & $z_{\text{abs}}$ & \cmin\ & \dof\ & $\Delta C$ \\
    \hline
\multicolumn{5}{c}{XMM OVII/OVI}\\ 
pks0405 & 261 & 0.4089 & 110.2 & 86 & 7.0 \\ 
1es1553 & 5 & 0.1878 & 102.3 & 88 & 7.5 \\ 
1es1553 & 4 & 0.1876 & 102.1 & 88 & 7.8 \\ 
pg1211 & 180 & 0.0512 & 45.2 & 34 & 13.5 \\ 
tons180 & 308 & 0.0456 & 39.9 & 45 & 14.4 \\ 
3c273 & 23 & 0.1466 & 91.0 & 79 & 17.9 \\ 
3c273 ($\star$)& 21 & 0.0902 & 100.6 & 47 & 78.3 \\ 
\hline 
\multicolumn{5}{c}{Chandra OVII/OVI}\\ 
pks0405 & 45 & 0.1657 & 45.8 & 38 & 7.2 \\ 
\hline 
\multicolumn{5}{c}{XMM OVIII/OVI}\\ 
1es1028 & 3 & 0.3373 & 92.6 & 96 & 6.6 \\ 
pks2155 & 267 & 0.0571 & 76.8 & 59 & 7.3 \\ 
pks0405 & 259 & 0.3633 & 110.1 & 94 & 8.0 \\ 
1es1553 & 6 & 0.1898 & 25.6 & 27 & 8.2 \\ 
mrk421 & 114 & 0.0101 & 95.4 & 57 & 9.7 \\ 
1es1553 & 9 & 0.3787 & 113.1 & 88 & 15.3 \\ 
1es1553 & 8 & 0.3113 & 113.1 & 91 & 17.1 \\ 
ngc7469 & 121 & 0.0096 & 84.3 & 74 & 45.8 \\ 
ngc7469 & 123 & 0.0115 & 85.8 & 74 & 53.4 \\ 
ngc7469 & 122 & 0.0099 & 94.8 & 74 & 53.8 \\ 
\hline 
\multicolumn{5}{c}{Chandra OVIII/OVI}\\ 
3c273 & 77 & 0.0902 & 29.9 & 38 & 6.6 \\ 
\hline 
\multicolumn{5}{c}{XMM OVII/HI}\\ 
pg0804 & 52 & 0.0502 & 46.3 & 37 & 7.4 \\ 
s50716 & 159 & 0.0883 & 48.9 & 39 & 8.0 \\ 
pg1116 & 75 & 0.0838 & 46.6 & 32 & 8.2 \\ 
\hline 
\multicolumn{5}{c}{Chandra OVII/HI}\\ 
mr2251 & 28 & 0.0633 & 50.5 & 39 & 6.8 \\ 
h1821 & 16 & 0.1982 & 30.7 & 38 & 6.9 \\ 
pg1116 & 38 & 0.1337 & 41.7 & 34 & 7.0 \\ 
mr2251 & 29 & 0.0638 & 48.0 & 38 & 8.0 \\ 
\hline 
\multicolumn{5}{c}{XMM OVIII/HI}\\ 
pks2155 & 129 & 0.0571 & 51.9 & 51 & 7.0 \\ 
pks0405 & 121 & 0.1946 & 47.4 & 42 & 13.4 \\ 
mr2251 & 33 & 0.0619 & 63.0 & 47 & 16.5 \\ 
mr2251 & 34 & 0.0628 & 50.9 & 49 & 38.9 \\ 
mr2251 & 36 & 0.0638 & 45.7 & 49 & 45.5 \\ 
mr2251 & 35 & 0.0633 & 44.9 & 49 & 45.7 \\ 
ngc7469 & 46 & 0.0098 & 85.7 & 65 & 51.6 \\ 
\hline 
\multicolumn{5}{c}{Chandra OVIII/HI}\\ 
h1821 & 13 & 0.0678 & 39.4 & 38 & 7.8 \\ 
\hline 
\hline
    \end{tabular}
    \caption{List of \nDet\ systems with \cmin/\dof $\leq 2$ and $\geq$~99\% probability of detection of an absorption line
    at the redshift of a prior FUV \ovi\ or \hi\ absorption line, according to the $\Delta C \geq 6.6$ criterion. ID is system's identification number as in Tables~\ref{tab:OVIIOVIXMM} through \ref{tab:OVIIIHIChandra}. The additional entry with $\star$ has \cmin/\dof\ slightly larger than the threshold, and it is further discussed because of its prior detections (see Sec.~\ref{sec:3c273ovii}). }
    \label{tab:detections}
\end{table}

Table~\ref{tab:detectionsNED} reports the results of a search for galaxies near the possible
absorption line systems of Table~\ref{tab:detections}, to aid in the interpretation of the origin
of the putative absorption. The search was conducted with the NASA Extragalactic Database (NED) 
within a plane--of--sky radius of 1~Mpc at the redshift of the absorber, and for the standard
flat Planck cosmology with $H_0=67.8$~\kmsMpc\ \citep[e.g.][]{Planck2015-cosmology,planck2020}. In the 
search we allowed a
$\Delta z = \pm 0.0034$ from the nominal redshift of each absorber, which corresponds to approximately
a peculiar velocity of $\pm$~1,000 km/s relative to the Hubble velocity of the absorber. The redshift range
 also permits an investigation of the presence of galaxies along the sightline.

Of the \nDet\ possible absorption line
detections, 14 have at least one galaxy within this volume of the absorber. Most of these galaxies are
at a large impact parameter ($\geq 250$~kpc) from the line of sight to the quasar. 
Only the
systems associated with the $z=0.166$ and $0.363$ PKS~0405 quasar, and the $z=0.0638$ MR~2251 absorber, have a galaxy at an impact parameter $\leq 250$~kpc that makes the absorption associated with the  circum--galactic medium  (CGM) of an individual galaxy a plausible explanation. Moreover, the MR~2251 system is likely an intrinsic absorption system, given that the absorption redshift is consistent with that of the quasar (see Sec.~\ref{sec:mr2251} below). For a majority of possible absorption line systems, we do not
find galaxies at a small impact parameter, and therefore a more likely origin 
for these possible absorption lines is from the WHIM medium associated with larger--scale structures, such as filaments. In \cite{ahoranta2021} we conducted a detailed analysis of the line-of-sight galaxies towards the Ton~S180 absorber at $z=0.062$, one of the systems in this sample, where the closest line-of-sight galaxy has an impact parameter of 0.29~Mpc. Based on an analysis of the virial radius of the line-of-sight galaxies and their properties, we concluded that the CGM is unlikely to be responsible for the possible absorption in Ton~S180. A similar analysis was conducted for \es\ absorber at $z \simeq0.188$, where we could conclude that the possible absorption is unlikely to be associated with the CGM \citep{spence2023}.  

This preliminary analysis indicates that most of the possible absorption line systems
discussed in this section are unlikely to be associated with the warm CGM of individual galaxies. For these systems we do not have available information on the presence of large--scale
filaments that could aid in supporting the filamentary origin of the putative absorption.
A more detailed analysis of the possible origin of the absorption will be provided in a follow--up paper, where the cosmological interpretation of these absorption line systems is presented.

\subsubsection{{3C~273}, \ovii\ at \ovi\ prior, system 21 ($z=0.09018$)}
\label{sec:3c273ovii}
This fit has  ${\Delta C=78.3}$ for \cmin/\dof=100.60/47.~\footnote{The value of \cmin/\dof=2.1 for this fit is just above the arbitrary range of \cmin/\dof=2 chosen for the discussion of possible detections. We chose to provide a discussion of this system, given its large $\Delta C$ value and the prior history with this absorption system.}
The line, which is the strongest detection
in the sample, appears at wavelengths $\lambda=23.5479$, 
which is similar to Galactic \oi\ ($\lambda=23.506$), which is in fact visible in the spectrum (see Fig.~\ref{fig:detections}) and is the primary contributor to the poor \cmin\ \gof\ statistic. 
\cite{ahoranta2020} has already discussed the difficulties in disentangling the
Galactic \oi\ absorption from possible redshifted \ovii\ absorption for this system, and
we do not further attempt to determine its Galactic versus WHIM origin.
We simply 
note that
\cite{ahoranta2020} reported possible detection of \oviii\ at the redshift of system 21, so it is
also possible that part of the \ovii\ absorption is in fact due to the WHIM. The associated \oviii\ line is marginally
significant in the \chandra\ data (system 77, $\Delta C=3.9$) and in the \xmm\ data (system 21, $\Delta C = 1.2$).

\subsubsection{{3C~273}, \ovii\ at \ovi\ priors, system 23, $z=0.1466$}
This fit has ${\Delta C=17.9}$ for \cmin/\dof=91.01/79.
The redshift of the source is $z=0.158$, so if this is an intrinsic line
it would imply a line--of--sight velocity towards the observer of $v=-3,300$~\kms, which appears quite
large to be intrinsic to the source. This redshift was not examined by \cite{ahoranta2020}, because the \ovi\ absorption did not have a reported $b$ parameter. 
The spectra show that the possible absorption is predominantly seen in RGS2,
with no evidence of absorption in RGS1.
The line is not present in the \chandra\ data (system 79, best--fit 
$\tau_0<0$).

\subsubsection{Ton~S180, \ovii\ at \ovi\ prior, system 308, $z=0.0456$}
\label{sec:tons180}
This fit has
${\Delta C=14.4}$ for  \cmin/\dof=39.89/45. This is a confirmation of the previous detection of this line by \cite{ahoranta2021}.
The \chandra\ data do not have a significant detection because of the lower S/N of the spectrum, as previously
noted also by \cite{ahoranta2021}.

\subsubsection{PG~1211, \ovii\ at \ovi\ prior, system 180, $z=0.05117$}
This fit has
${\Delta C=13.5}$ for \cmin/\dof=45.23/34. The line is located at $\lambda=22.705$, which is near the Galactic \oiv\ line at 
$\lambda=22.74$~\AA. Moreover, the line falls on a region of reduced efficiency, which
might not be perfectly calibrated, casting doubts on the reality of this detection.
The 
\chandra\ data do not show any absorption.

\subsubsection{1ES~1553+113, \ovii\ at \ovi\ prior, systems 4 and 5, $z=0.18759$ and $z=0.18775$}
The two fits have ${\Delta C=7.8, 7.5}$ and \cmin/\dof= 102.1/88, 102.3/88, respectively. The two systems are indistinguishable
at the resolution of the \xmm\ data, and only one is shown in Fig.~\ref{fig:detections}. This is the same system that was tentatively identified by \cite{spence2023}, and this re--analysis finds similar results. The \chandra\ data, as already noted in \cite{spence2023}, has substantially lower S/N for this source, and they show no significant features at these wavelengths ($\Delta C=0.6$, and 1.0 respectively).

\subsubsection{PKS~0405-123, \ovii\ at \ovi\ prior, system 261, $z=0.40890$}
\label{sec:pks0405XMM}
This fit has $ \Delta C=7.0$, \cmin/\dof=110.16/86.
The \chandra\ data, which has somewhat larger S/N than the \xmm\ data (See Sec.~\ref{sec:pks0405Chandra} below), does not have any evidence for absorption at those
wavelengths ($\Delta C=0)$.

\subsubsection{PKS~0405-123, \ovii\ at \ovi\ prior, system 45 (\chandra), $z=0.16566$}
\label{sec:pks0405Chandra}
This fit has ${\Delta C=7.2}$, \cmin/\dof=  45.81/38. The lower--exposure  \xmm\ spectrum 
show marginal absorption at these wavelengths ($\Delta C=1.6$).
This system is part of a group of 5 \ovi\ priors (systems 45--49) that span the redshift range $0.16566-0.16712$,
with \chandra\ showing marginal absorption also for those adjacent systems ($\Delta C =7.2-4.3$)
with significant overlap at the \chandra\ resolution.


\subsubsection{NGC~7469, \oviii\ at \ovi\ priors, systems 121 through 123, $z=0.00962-0.0115$}
The three fits have respectively ${\Delta C=45.8, 53.8, 53.4}$ and \cmin/\dof=84.33/74, 94.79/74, and
85.80/74.
Also, there is an \hi\ BLA  prior, system 46 ($z=0.00981$), with $\Delta C=51.6$, that coincides with one of the three systems. Only system 123 is shown in Fig.~\ref{fig:detections}.

NGC~7469 is a well--known Seyfert~1 galaxy at $z=0.0164$, and the redshifted \oviii\ Ly~$\alpha$
line falls near $\lambda= 19.2$~\AA. The \xmm\ X--ray spectra have been previously analyzed by several
groups \citep[e.g.][]{blustin2003,scott2005,behar2017,grafton2020}. The earlier analyses identified a putative \oviii\ Ly~$\alpha$ \emph{emission} line intrinsic to the galaxy at an observed wavelength $\lambda \simeq 19.3$~\AA. Moreover, the same analyses already detected at high significance the  possible \oviii\ Ly~$\alpha$ \emph{absorption} line we also detect, which
they attributed to an unidentified warm absorber associated with the galaxy at a peculiar velocity of $\sim -1,000$~km~s$^{-1}$ towards the observer.
The photoionization modelling of the warm absorber by \cite{grafton2020} does not take into account the
\ovi\ and the \hi\ BLA priors we have used for our search. It is therefore possible to speculate that the \oviii\ absorption line is in fact associated with a genuine line--of--sight WHIM absorber, rather than
an intrinsic absorber, although the intrinsic origin with peculiar velocity along the sightline cannot be
discarded.

\subsubsection{1ES~1553+113,  \oviii\ at \ovi\ prior, system 8, $z=0.31130$}
This fit has
${\Delta C=17.08}$ and \cmin/\dof=113.12/91.  This redshift was not searched 
by \cite{spence2023}, because it only had the $\lambda$=1032~\AA\ \ovi\ 
absorption line detected, and it was not identified by \cite{nicastro2018} or \cite{gatuzz2023}
in their serendipitous searches. The putative absorption is in a region of
 reduced efficiency for both RGS1 and RGS2, as can be seen from Fig.~\ref{fig:detections}. The 
 lower--resolution Chandra data do not
show any absorption at this redshift and there is no associated 
\ovii\ at this redshift.

\subsubsection{1ES~1553+113,  \oviii\ at \ovi\ prior, system 9, $z=0.37868$}
The  fit has $\Delta C=15.34$ and
 \cmin/\dof=113.07/88.  It was not 
observed by \cite{spence2023}, because it only had the $\lambda$=1032~\AA\ \ovi\ 
absorption line detected, and  it was not identified by \cite{nicastro2018} or \cite{gatuzz2023}
in their serendipitous searches. The absorption feature is driven
by the RGS2 data, in a region of reduced efficiency, where the possibility of miscalibration cannot be ruled out (see Fig.~\ref{fig:detections2}). The Chandra data do not
show any absorption at this redshift and there is no associated 
\ovii\ at this redshift.

\subsubsection{Mkn~421, \oviii\ at \ovi\ prior, system 114, $z=0.010$}
\label{sec:mkn421}
This fit has $\Delta C=9.7$ and \cmin/\dof=95.39/57.
The corresponding \chandra\ data, system 81, could 
not be fit to the usual 1~\AA\ band around the line, because
the continuum has an unusual shape,  likely because 
the spectrum is the result of averaging different states of the source. 
Our analysis of the \chandra\ data does not indicate presence of absorption ($\Delta C=0.3$).

\subsubsection{1ES~1553+113,  \oviii\ at \ovi\ prior, system 6, $z=0.1898$}
This fit has ${\Delta C=8.2}$ and \cmin/\dof=25.62/27. This line is at $\lambda=22.5594$~\AA, with two Galactic \ov\ and \oiv\ lines approximately 0.1~\AA\ away on either side. In \cite{spence2023}, this line had a lower significance of $\Delta C= 2.3$, and it was not reported as significant.

\subsubsection{PKS~0405-123, \oviii\ at \ovi\ prior, system 259, $z=0.36329$}
This fit has {$\Delta C=8.0$} and \cmin/\dof=110.13/94. The \chandra\ data do not have an absorption feature at that redshift. See Sec.~\ref{sec:pks0405Chandra} for related comments on this source.

\subsubsection{PKS~2155-304, \oviii\ at \ovi\ prior, system 267, $z=0.05707$}
\label{sec:pks2155o8o6}
This fit has $ \Delta C=7.4$ and \cmin/\dof=76.79/59, and it is based on the RGS~1 data alone.
This absorption line feature is at $\lambda=20.0420$ falls near the $z=0.0543$ \oviii\ absorption line
with a tentative serendipitous detection by \cite{fang2002}, and with
a rich history of follow--up studies (see Sec.~\ref{sec:previousDetections}). Additional discussion is deferred to that section. There is also an \oviii\ line at an \hi\ FUV prior that falls at
virtually the same wavelength, and which is reported below in Sec.~\ref{sec:pks2155o8hi}.

\subsubsection{1ES~1028+511, \oviii\ at \ovi\ prior, system 3, $z=0.33735$}
This fit has $\Delta C=6.6$ and \cmin/\dof=92.62/96.
The shorter \chandra\ data do not have any features at this redshift.
The redshift of this source is $z=0.3604$, and therefore if this
feature was to be intrinsic to the source, it would feature
a rather larger peculiar velocity of $\sim-6,000$~\kms, which appears unlikely.

\subsubsection{3C~273, \oviii\ at \ovi\ prior (\chandra), system 77, $z=0.09018$}
\label{sec:3c273oviii}
This fit has $\Delta C=6.6$ and \cmin/\dof=29.87/38. The \xmm\ data have
marginal evidence for absorption at this wavelength ($\Delta C=1.1$). This system was studied in detail by \cite{ahoranta2020}.
This system is further discussed in Sec.~\ref{sec:previousDetections}.

\subsubsection{PG~0804+761, \ovii\ at \hi\ prior, system 52, $z=0.0502$}
This fit has {$\Delta C=7.4$} and \cmin/\dof=46.31/37.
This is a short \xmm\ exposure, and there are no \chandra\ data available for this sources.

\subsubsection{S50716+714, \ovii\ at \hi\ prior, system 159, $z=0.0883$}
This fit has {$\Delta C=8.0$} and \cmin/\dof=48.91/39.
There are no available \chandra\ data for this source.

\subsubsection{PG~1116+215, \ovii\ at \hi\ prior, system 75, $z=0.0838$}
This fit has {$\Delta C=8.2$} and \cmin/\dof=46.62/32, and it falls near a detector region with reduced sensitivity, whose calibration might affect this detection.
The \chandra\ data, which was used in \cite{bonamente2016}, does not have absorption at this redshift
. This source is further discussed in Sec.~\ref{sec:previousDetections}.

\subsubsection{H1821+643, \ovii\ at \hi\ prior, system 16 (\chandra), $z=0.19817$}
\label{sec:h1821}
This fit has {$\Delta C=6.9$} and \cmin/\dof=30.74/38. The substantially shorter \xmm\ data do not have absorption at this redshift.
The \chandra\ observations of H1821+643 were studied in detail by \cite{kovacs2019}, who stacked the data at \hi\ FUV priors by \cite{tripp1998} to provide a detection of \ovii\ from the stacked spectrum.
Their FUV priors did {not} include the  same $z=0.19817$ system investigated in this paper, but it included an \hi\ prior at $z=0.19905$, for a $\Delta \lambda=19$~m\AA\ that falls within the resolution of the instrument. 

\subsubsection{MR~2251-178, \ovii\ at \hi\ prior, systems 28 and 29 (\chandra), $z=0.0633, 0.0638$}
The two fits have respectively {$\Delta C=6.8, 8.0$} and  \cmin/\dof= 50.52/39, 48.05/38. The redshift of the source is $z=0.064$, so this is probably intrinsic
absorption. Only system 29 is illustrated in Fig.~\ref{fig:detections3}. These systems
have similar redshift to those discussed in Sec.~\ref{sec:mr2251} below, where additional information for this absorber is provided.

\subsubsection{PG~1116+215, \ovii\ at \hi\ prior, system 38 (\chandra), $z=0.13373$}
This fit has {$\Delta C=7.0$} and  \cmin/\dof=41.71/34.
The \chandra\ analysis of \cite{bonamente2016} showed a marginal absorption feature from the longer
observation available for this source. The \xmm\ data has marginal absorption at that redshift as well (system 79, $\Delta C=5.4$) which provides additional support for this possible absorption line detection.

\subsubsection{MR~2251-178, \oviii\ at \hi\ prior, systems 33 through 36 ($z=0.0619-0.0638$)}
\label{sec:mr2251}
These fits have
${\Delta C =16.53, 38.9, 45.7, 45.5}$, \cmin/\dof= 62.95/47, 50.87/49, 44.88/49, 45.72/49.
MR~2251-178 is a quasar at $z=0.064$ whose X--ray emission with \rosat\ and \rxte\ was previously studied respectively by \cite{komossa2001} and \cite{arevalo2008}. Interestingly, the 
low--resolution \rosat\ data analyzed by \cite{komossa2001} did indicate the presence of intrinsic \ovii\ and \oviii\ absorption. At a nominal difference of just $\Delta z \leq 0.002$, it is reasonable
to speculate that the detected absorption is intrinsic to the source, implying peculiar sight--line
negative velocities of $ v \geq - 600$~\kms, which are common for quasars and AGNs.
 Moreover, Table~\ref{tab:detectionsNED} indicates that there are several
galaxies along the sight--line with a small impact parameter from these absorbers. It is therefore also possible that this X--ray
absorption is associated with an individual galaxy rather being intrinsic to the quasar, or
associated with the intervening WHIM.

\subsubsection{PKS~2155-304, \oviii\ at \hi\ prior, system 129, $z=0.05708$}
\label{sec:pks2155o8hi}
This fit has $\Delta C=7.0$ and \cmin/\dof=51.87/51, and it is based on the RGS~1 data alone. This feature has virtually the same redshift as system 267 (\oviii\ at \ovi\ prior) discussed above in Sec.~\ref{sec:pks2155o8o6}, and therefore this is virtually the same fit.  The slightly different best-fit statistics and degrees of freedom
relative to system 267 are the result of   different
choices made in exclusion of bad pixels,  as a test of the effect of bad pixel exclusion in the detection of faint features (the two fits, in fact, lead to virtually the same $\Delta C$ statistic).
The corresponding \chandra\ data do not have any features at this redshift.  Additional discussion
of this source is provided in Sec.~\ref{sec:pks2155}.

\subsubsection{PKS~0405-123, \oviii\ at \hi\ prior, system 121, $z=0.19456$}
The fit has ${\Delta C=13.4}$ and \cmin/\dof=47.40/42.
The source is at $z=0.574$, and the line is at $\lambda=22.6489$~\AA, with the nearest
Galactic line being \oiv\ at $\lambda=22.74$~\AA\ which should not affect significantly
this line. There is only marginal \ovii\ absorption associated with this redshift ($\Delta C=2.0$). 
Fig.~\ref{fig:detections3} suggests that a slight blueshift of the X--ray absorber relative to the FUV prior would make this feature more significant.

\subsubsection{NGC~7469, \oviii\ at \hi\ prior, system 46, $z=0.00981$}
The fit has a $\Delta C=51.6$, with \cmin/\dof=85.67/65.
Given the low redshift, this is probably Galactic absorption with a peculiar velocity of $v\simeq 3,000$~\kms. There are no \chandra\ data for this source.

\subsubsection{H1821+643, \oviii\ at \hi\ prior, system 13 (\chandra), $z=0.0678$}
The fit has {$\Delta C=7.8$} and \cmin/\dof=39.35/38. The substantially shorter \xmm\ data do not have absorption at this redshift.

\begin{table*}
    \centering
    \begin{tabular}{lcccccccc}
    \hline
    \hline
    Name & $z$  & RA & Dec. & (Type) & $z$ & \multicolumn{2}{c}{Separation} \\
         & (QSO) & \multicolumn{2}{c}{(J2000)}  & &  (Abs.) & (arcmin) & (Mpc) \\              
    \hline
1es1553 &  0.4140  & 15 55 43.04 & 11 11 24.4 &  & 0.18775 & & \\
WISEA J155552.00+111536.2 & & 238.966697 & 11.260063 & G &  0.189390 &   4.743 & 0.92 \\
\hline
tons180 &  0.0620  & 0 57 20.0 & 22 22 59.0 & & 0.04560 & & &  \\
(YWP2010) J014.283-22.310 & &14.283000 &  -22.310000 & G &  0.046025  & 5.198 & 0.29\\
2MASX J00570854-2218292   & & 14.285500 & -22.308222  & G &  0.045620  & 5.216 & 0.29\\
WISEA J005757.52-221638.8 & & 14.489674 & -22.277464  & G &  0.045960  & 10.744 & 0.59\\
WISEA J005729.58-223514.6 & & 14.373571 & -22.587381  & G &  0.042930  & 12.461 & 0.69\\
\hline
3c273 &  0.1583 & 12 29 6.7 & 2 3 8.7 &  & 0.14660 & & \\
WISEA J122924.11+020812.1 & & 187.350482 & 2.136719 & G & 0.146568 & 6.672 & 1.0\\
SDSS J122923.28+020822.6  & & 187.347019 & 2.139620 & G & 0.146810 & 6.674 & 1.0\\
\hline
3c273 &  0.1583 & 12 29 6.7 & 2 3 8.7  & & 0.09018 & & \\
WISEA J122851.89+020602.9 & & 187.216214 & 2.100827 & G & 0.090018 & 4.704 & 0.49\\
\hline
pks0405 & 0.5740 & 4 7 48.43 & 12 11 36.7 & & 0.16566 & & \\
WISEA J040751.19-121137.1    & & 61.963324 & -12.193659 & G  & 0.167200  & 0.684 & 0.12\\
GALEXMSC J040754.92-120912.2 & & 61.979167 & -12.153028 & G  & 0.163200  & 2.917 & 0.51\\
\hline
pks2155 & 0.1165 & 21 58 52.07 & 30 13 32.1 & & 0.05707 & & \\
(YWP2010) J329.673-30.325 &  &  329.673000 & -30.325000 &G & 0.057196  & 6.387 & 0.44\\
2PIGG SGPGAL B-0.54199-0.53344 & & 329.670000 &-30.324000 & G & 0.057039     & 6.389 & 0.44\\
2MASX J21584077-3019271        & & 329.670007 &-30.324251 &G  & 0.056723     & 6.403 & 0.44\\
WISEA J215823.82-301931.6   & &   329.599262& -30.325463 &G   & 0.054081     & 8.555 & 0.59\\
(SPS98) 215638.5-303057.4   & &  329.883974 &-30.275962 &G    & 0.057226     & 9.163 & 0.63\\
WISEA J215857.20-300245.9   & &  329.738335 &-30.046086 &G    & 0.057100     & 10.826 & 0.74\\
WISEA J215856.61-300229.3   & &  329.735889 &-30.041486 &G    & 0.057290     & 11.089 & 0.76\\
WISEA J215949.70-301619.9   & &  329.957120 &-30.272195 &G    & 0.056300     & 12.752 & 0.86\\
WISEA J215752.77-301931.5   & &   329.469890 &-30.325432 &G    & 0.056800     & 14.141 & 0.97\\
\hline
pks0405 &  0.5740 & 4 7 48.43 & 12 11 36.7 & & 0.36329 & & \\
PKS 0405-12:(EY94) 177 & & 61.941083 & -12.185750 & G  &  0.36140 & 0.777 & 0.24 \\
\hline
mrk421 &  0.0300 & 11 4 27.3 & 38 12 32.0 &  & 0.01009 & & \\
WISEA J110130.42+374716.8 & & 165.376690 & 37.788071 & G & 0.011727 & 43.035 & 0.55\\
\hline
1es1553 &  0.4140 & 15 55 43.04 & 11 11 24.4 & & 0.31130 & & \\
WISEA J155527.54+111126.0 &  & 238.864761 & 11.190568 & G & 0.314670 &  3.791 & 1.0\\
\hline
ngc7469 & 0.0163 & 23 3 15.6 & 8 52 26.0 & & 0.01153 & & \\
MRK 0524 &  & 345.317621 & 9.599036 & G  & 0.014834  & 52.542 & 0.77\\
\hline
s50716 &  0.2315 & 7 21 53.45 & 71 20 36.4 & & 0.08834 & & \\
WISEA J072116.68+711607.0 & &110.319504 & 71.268624 & G & 0.088000  & 5.372 & 0.55\\
WISEA J072142.61+711118.6 & & 110.427548 & 71.188522 & G & 0.088430  & 9.336 & 0.95\\
\hline
pg1116 &  0.1763 & 11 19 8.6 & 21 19 18.0 & & 0.08382 & & \\
WISEA J111929.52+211455.6 & & 169.873015 & 21.248780 & G & 0.083460 & 6.548 & 0.63\\
WISEA J111919.66+211112.6 & & 169.831927 & 21.186834 & G & 0.083050 & 8.491 & 0.83\\
\hline
pg1116 &  0.1763 & 11 19 8.6 & 21 19 18.0 & & 0.13373 & & \\
WISEA J111908.84+212255.4 &  & 169.787169 &  21.382054 & G & 0.134110 & 3.624 & 0.53\\
\hline
mr2251 &   0.0640 & 22 54 5.88 & 17 34 55.3 & & 0.06381 & \\
MR 2251-178               & &     343.524522 &-17.582035 &G &      0.063980     &0.004 & 0.0\\
WISEA J225405.76-173416.2 & &     343.524036 &-17.571178 &G &      0.066090     &0.652 & 0.05\\
WISEA J225400.80-173703.7 & &     343.503344 &-17.617701 &G &      0.064400     &2.461 & 0.19\\
WISEA J225356.19-173727.9 & &     343.484157 &-17.624434 &G &      0.066833     &3.438 & 0.26\\
APMUKS(BJ) B225118.97-175400.8 & &343.495599 &-17.633724 &G &      0.063257     &3.517 & 0.27\\
ABELL S1071               & &     343.491563 &-17.633507 &GClstr&  0.064200     &3.620 & 0.28\\
WISEA J225353.74-173810.3 & &     343.473929 &-17.636203 &G  &     0.066800     &4.354 & 0.33\\
WISEA J225344.12-173534.6 & &     343.433863 &-17.592956 &G  &     0.066603     &5.230 & 0.40\\
WISEA J225416.24-174046.2 & &     343.567678 &-17.679511 &G  &     0.064465     &6.347 & 0.48\\
WISEA J225418.29-174059.0 & &     343.576217 &-17.683058 &G  &     0.065889     &6.743 & 0.51\\
WISEA J225342.09-173017.2 & &     343.425416 &-17.504786 &G  &     0.065700     &7.326 & 0.56\\
WISEA J225325.25-173532.1 &  &    343.355210 &-17.592259 &G &     0.065800     &9.707  & 0.74\\
\hline
    \end{tabular}
    \caption{List of NED galaxy--type objects along the sight--line to the possible absorption line systems of Table~\ref{tab:detections}. The search was done in a cone of $\pm$~1~Mpc in the plane of the sky at the
    resdhift of the absorber, and within $\Delta z=0.0034$ of the nominal absorption redshift, approximately corresponding to a peculiar velocity of $\pm 1,000$~km/s relative to the absorption redshift. For systems with substantial redshift overlap with each other, such as some for \es\ and for MR2251, only one list is reported.}
    \label{tab:detectionsNED}
\end{table*}

\def\figSize{2.1}
\begin{figure*}
    \centering
    \includegraphics[width=\figSize in]{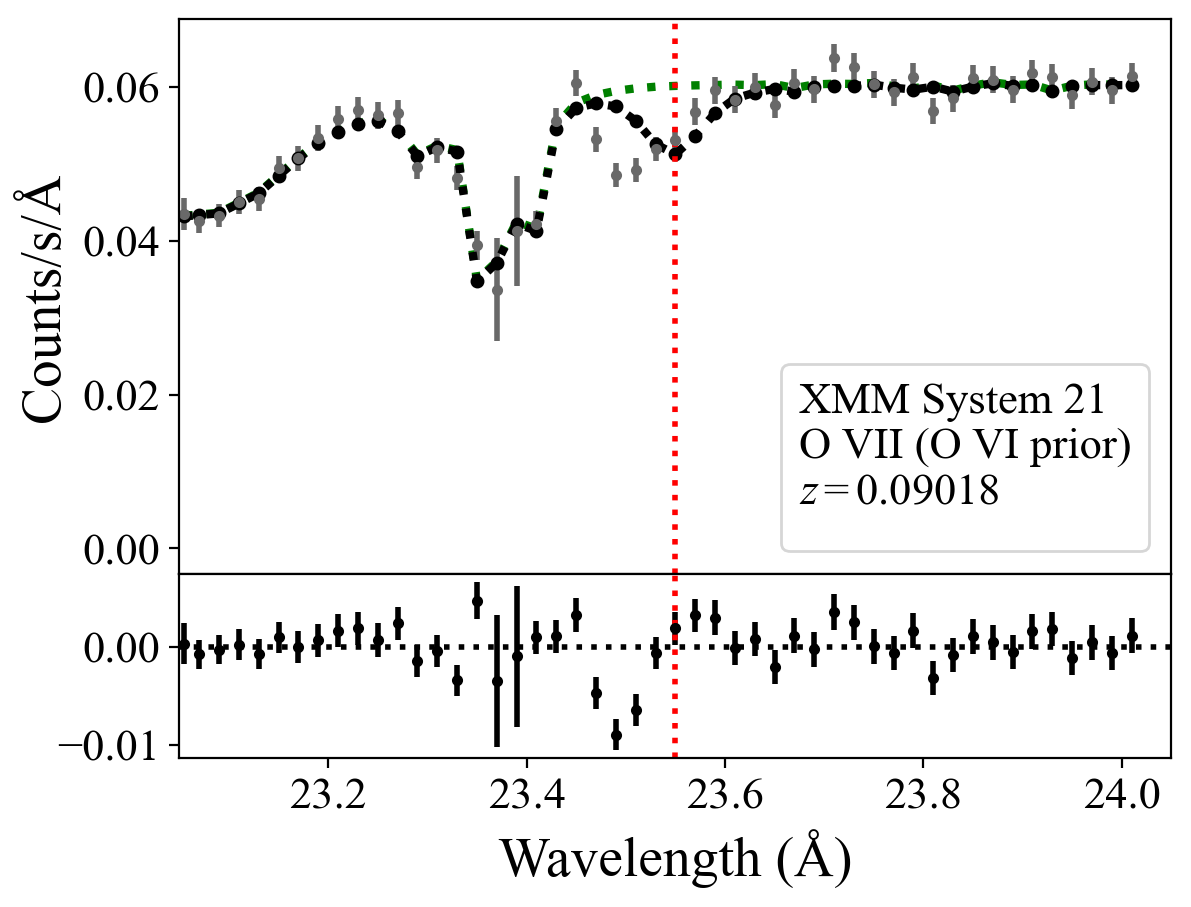}
    \includegraphics[width=\figSize in]{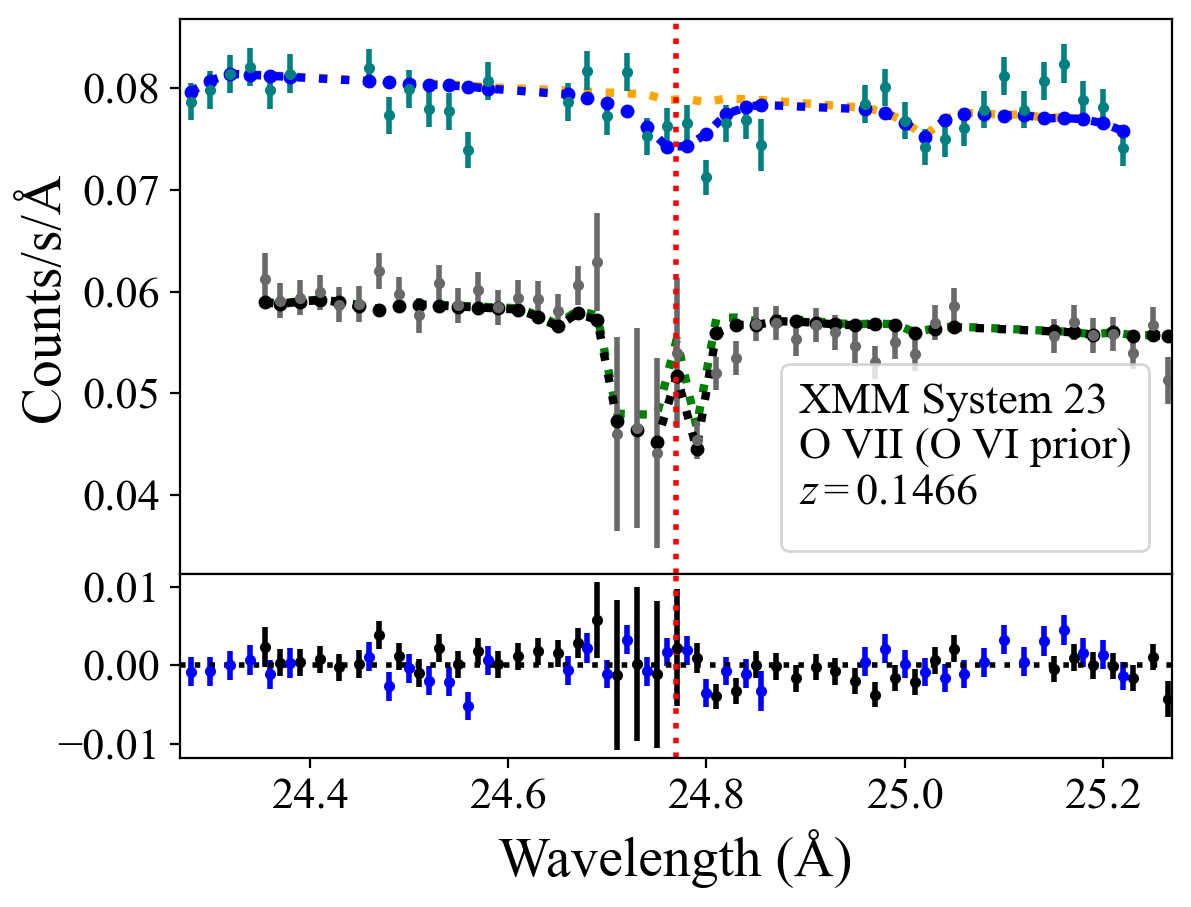}
    \includegraphics[width=\figSize in]{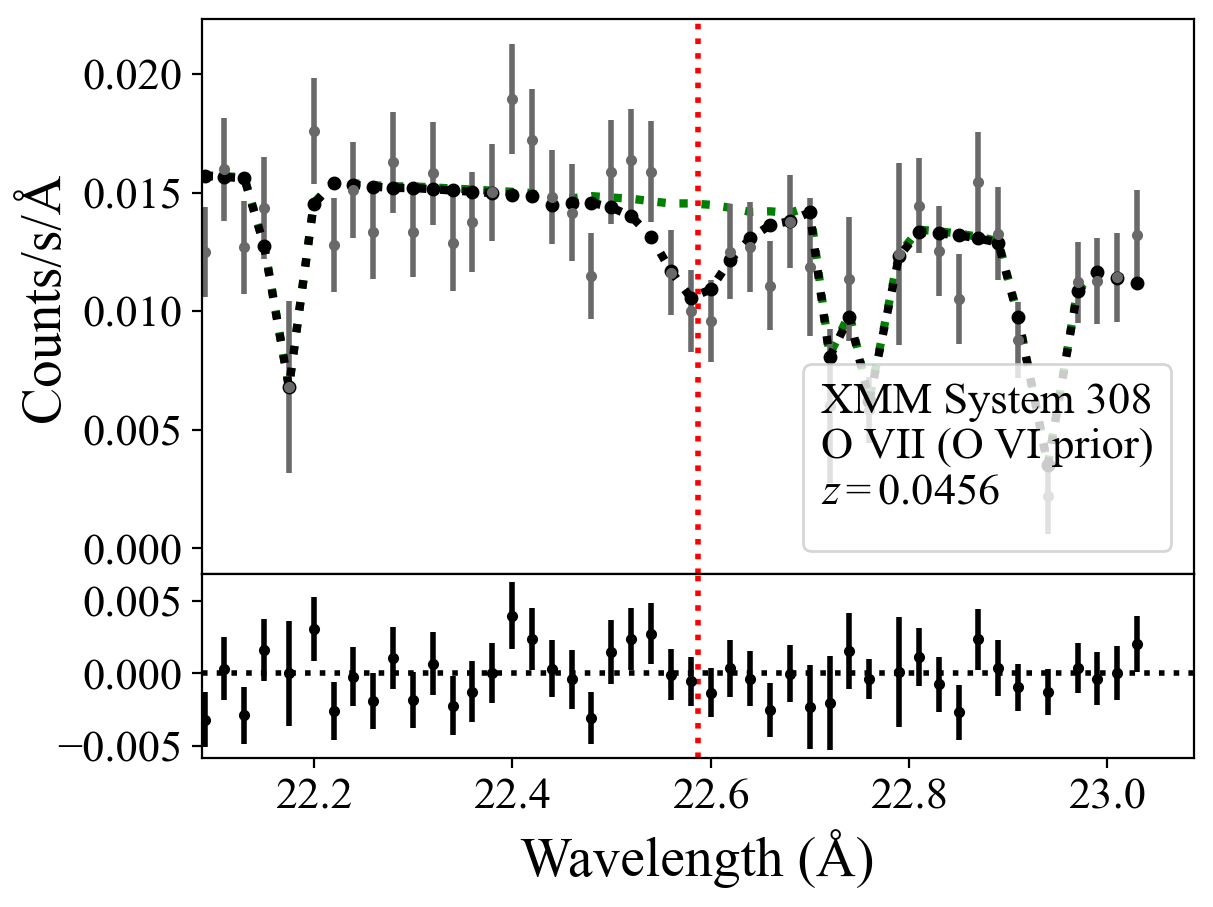}
    \includegraphics[width=\figSize in]{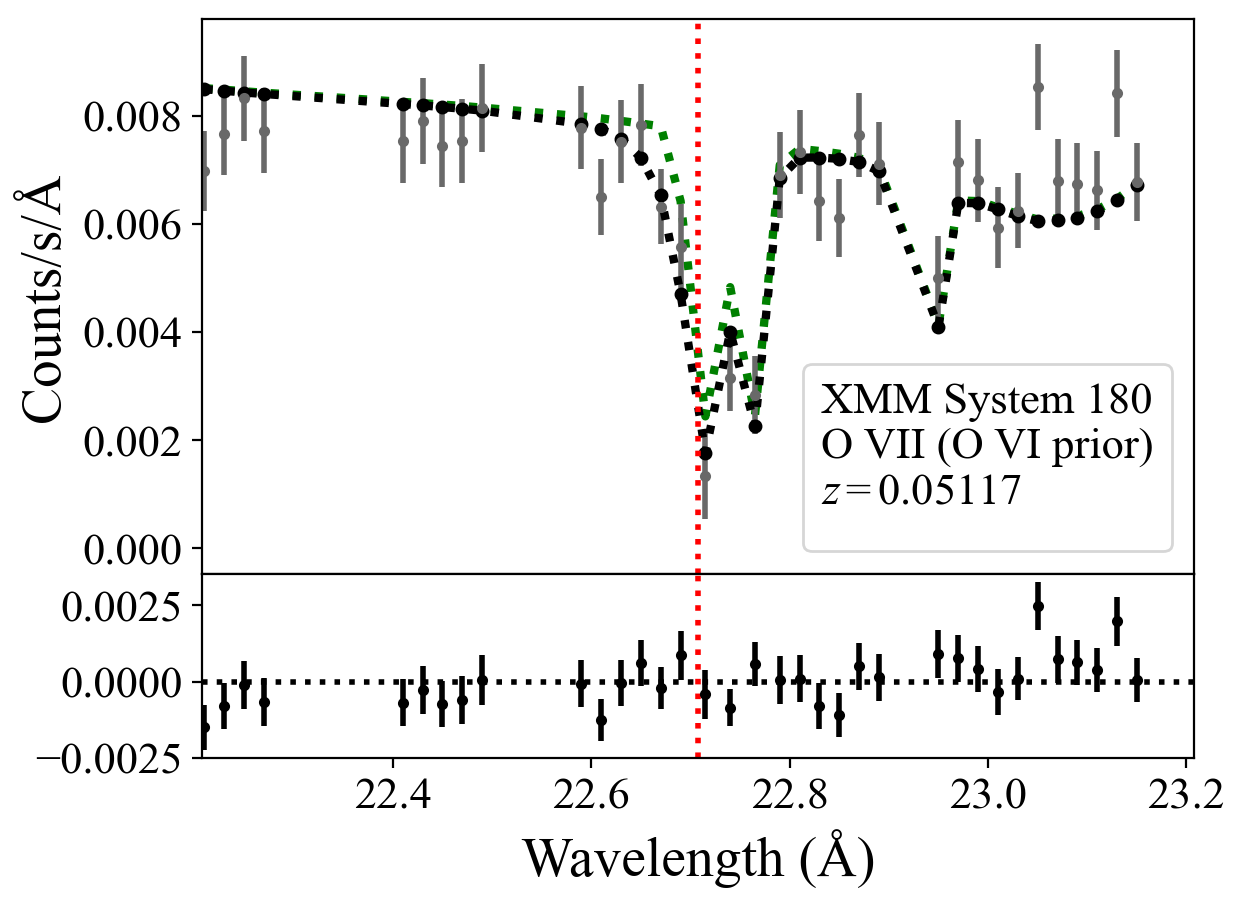}
     \includegraphics[width=\figSize in]{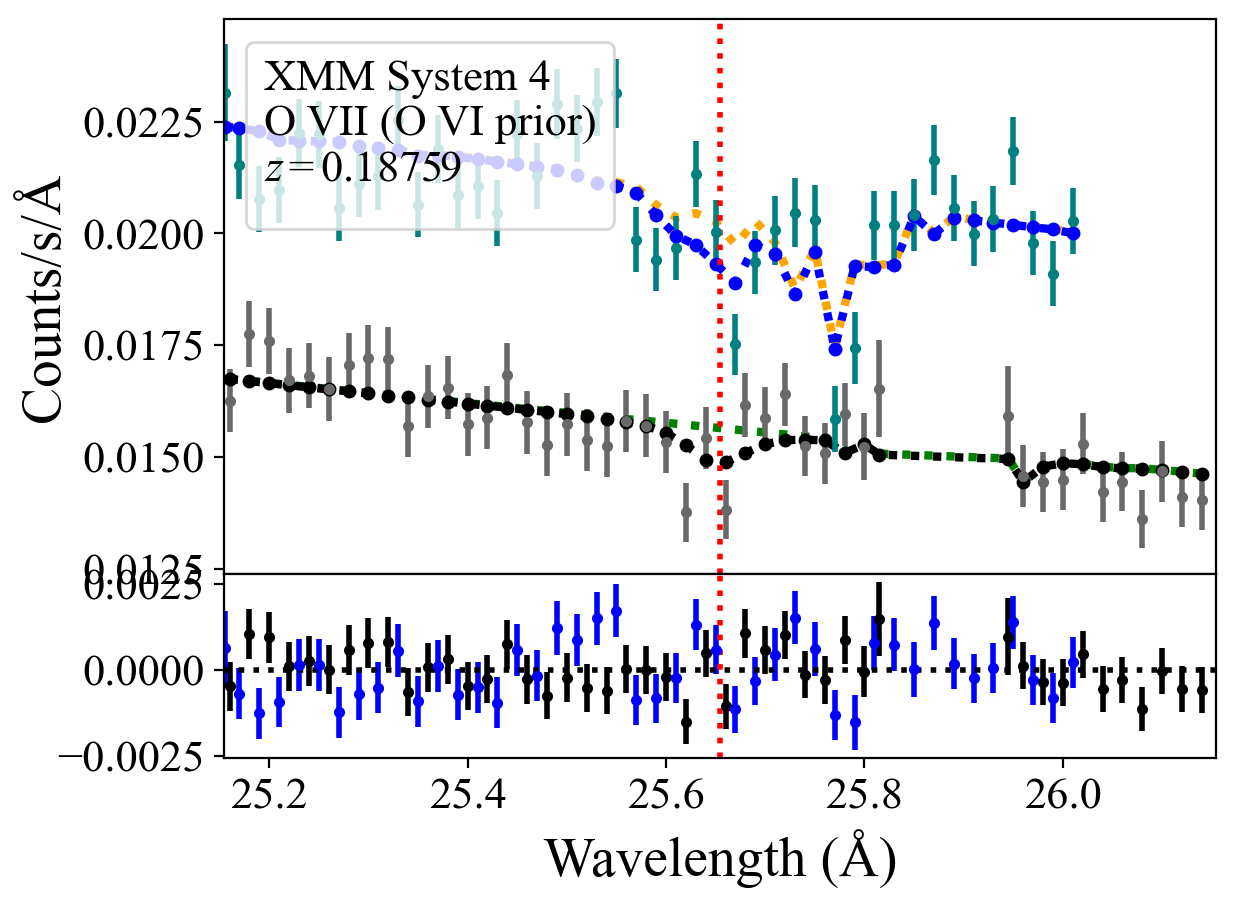}
     \includegraphics[width=\figSize in]{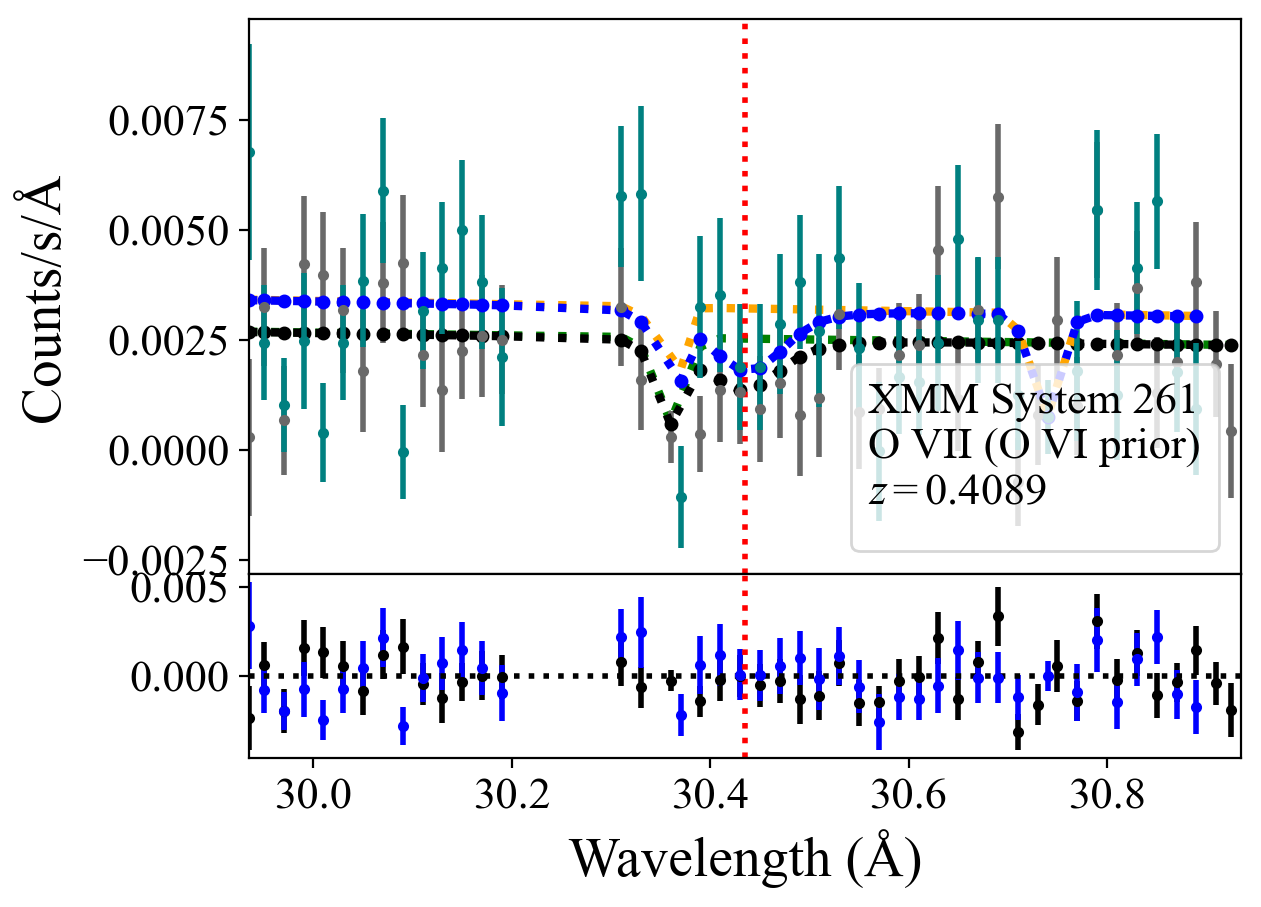}
      \includegraphics[width=\figSize in]{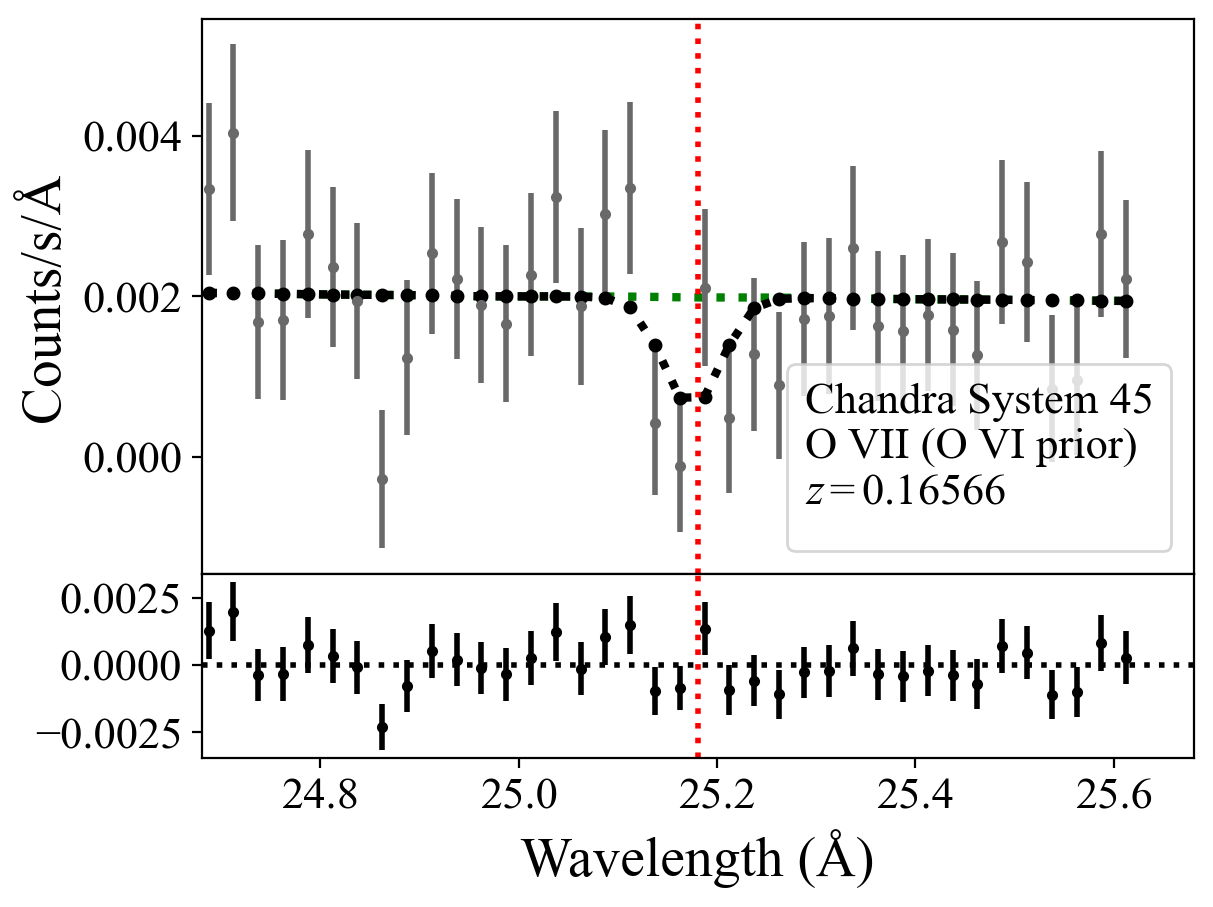}
      \includegraphics[width=\figSize in]{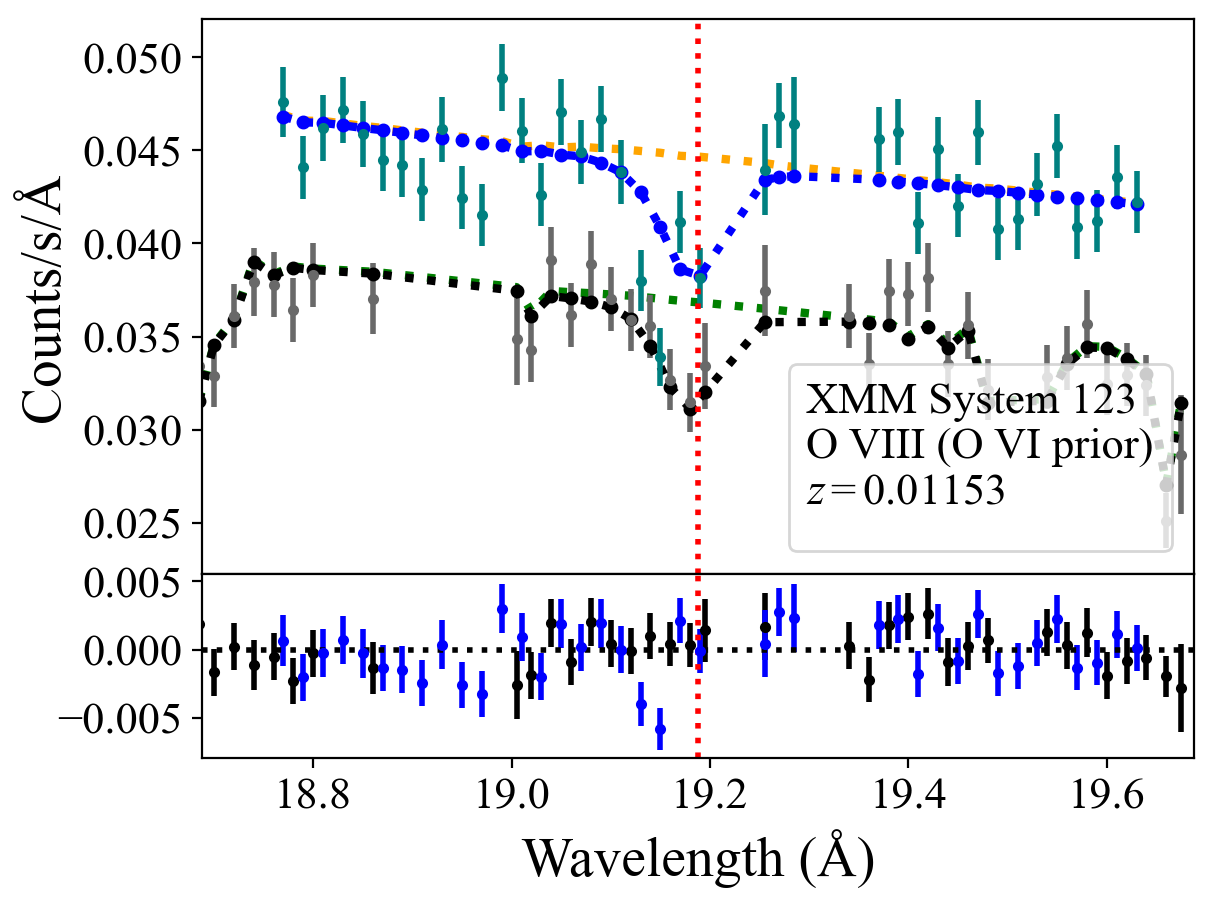}
      \includegraphics[width=\figSize in]{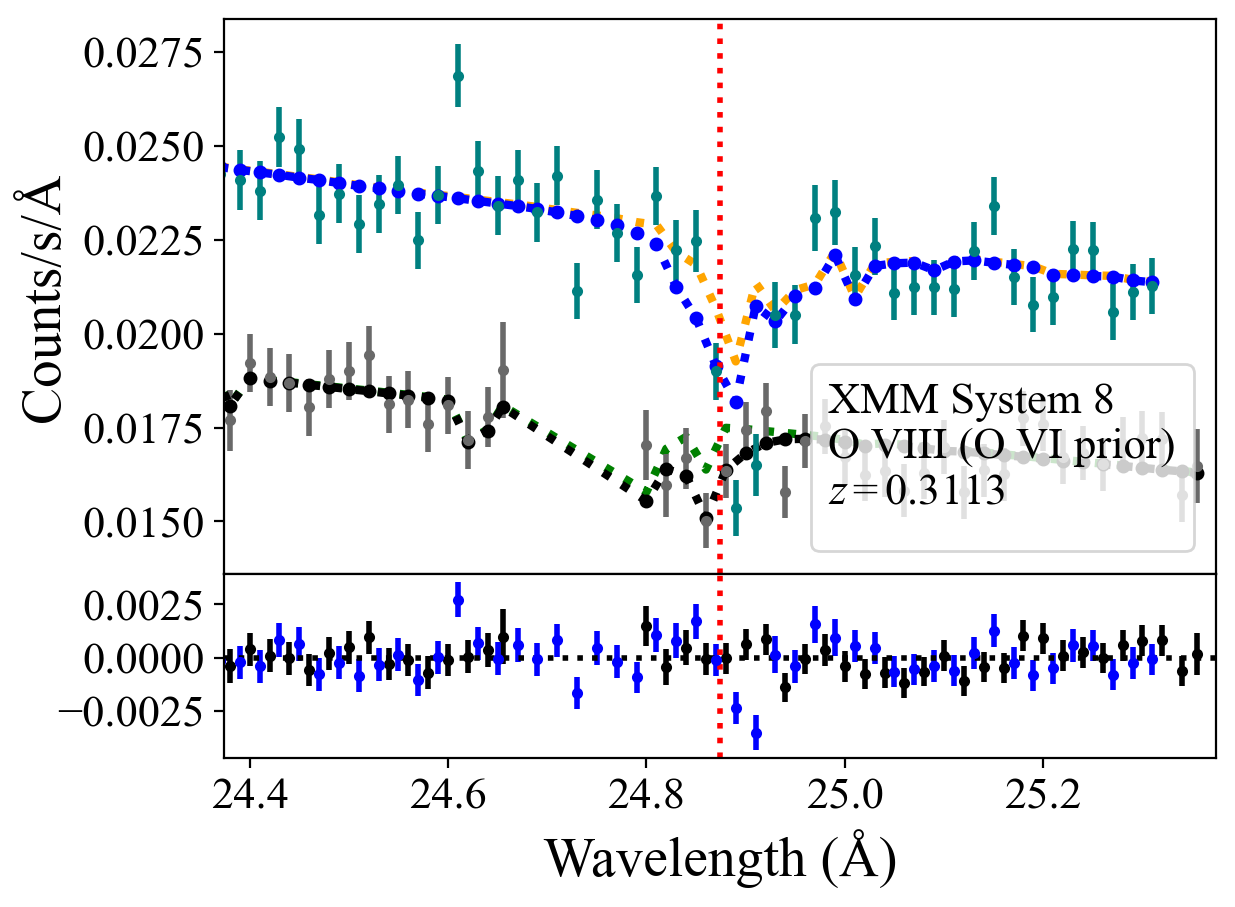}
    \caption{Spectra for systems with possible WHIM detections discussed in Sec.~\ref{sec:detections}. For each spectrum, the bottom panel contains the residuals (data minus model) in the same units.}
    \label{fig:detections}
\end{figure*}

\begin{figure*}
    \centering
     \includegraphics[width=\figSize in]{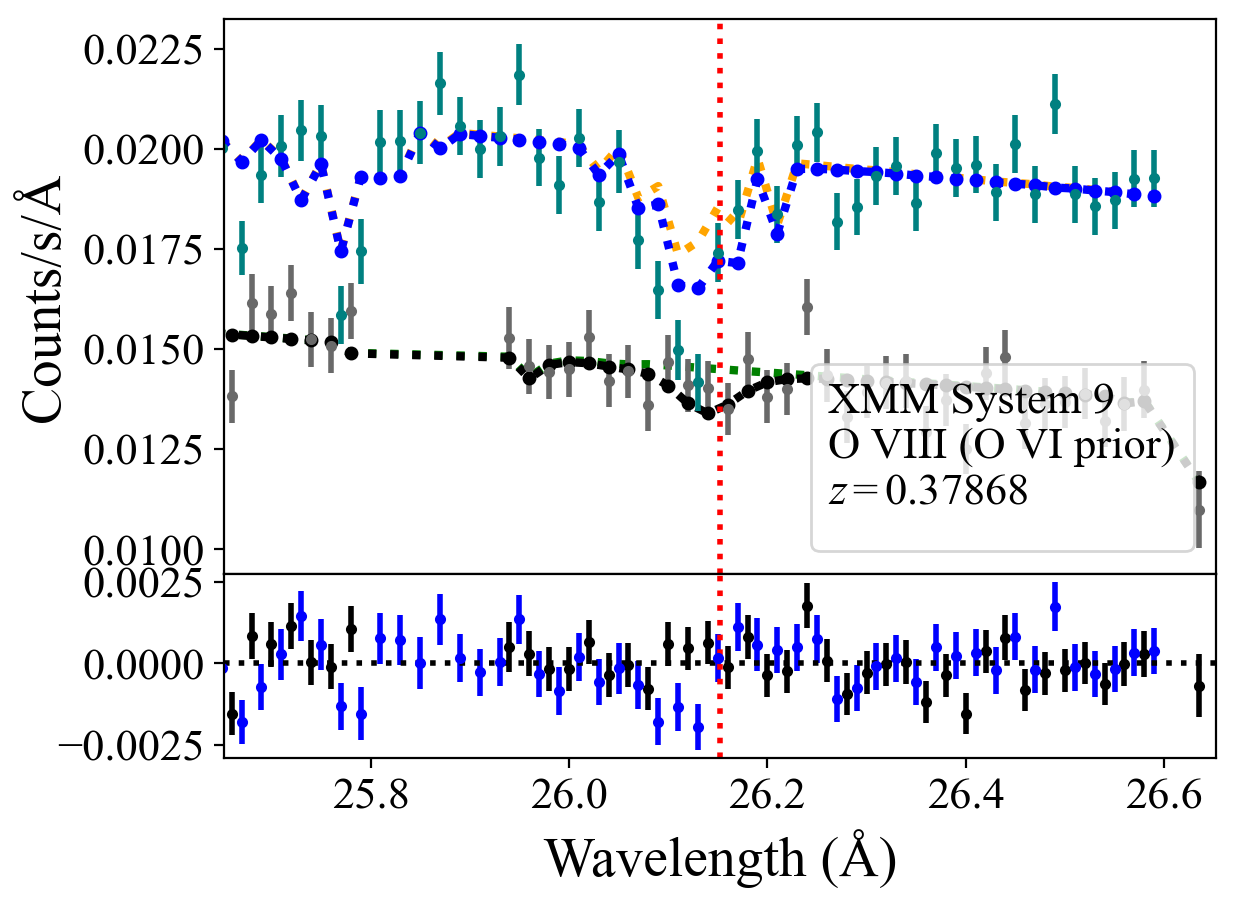}
    \includegraphics[width=\figSize in]{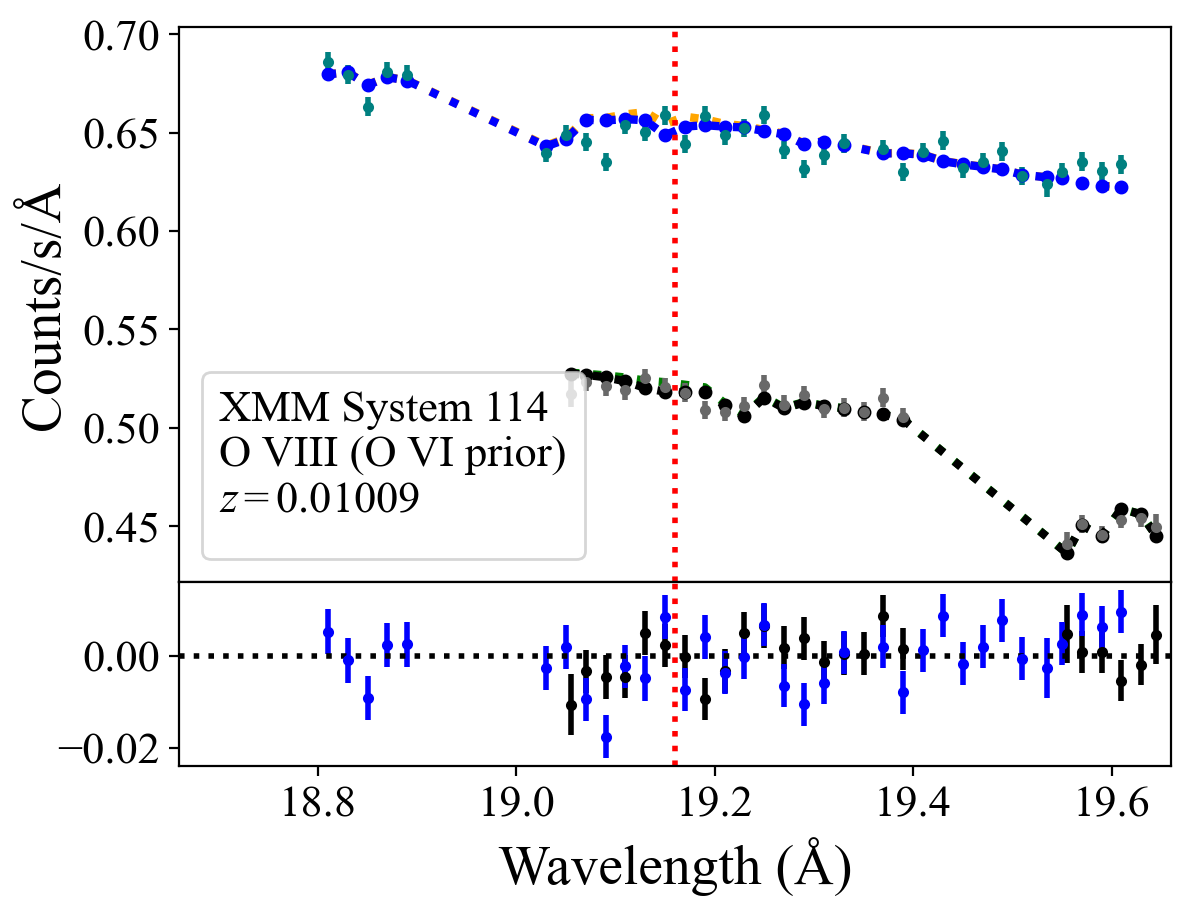}
     \includegraphics[width=\figSize in]{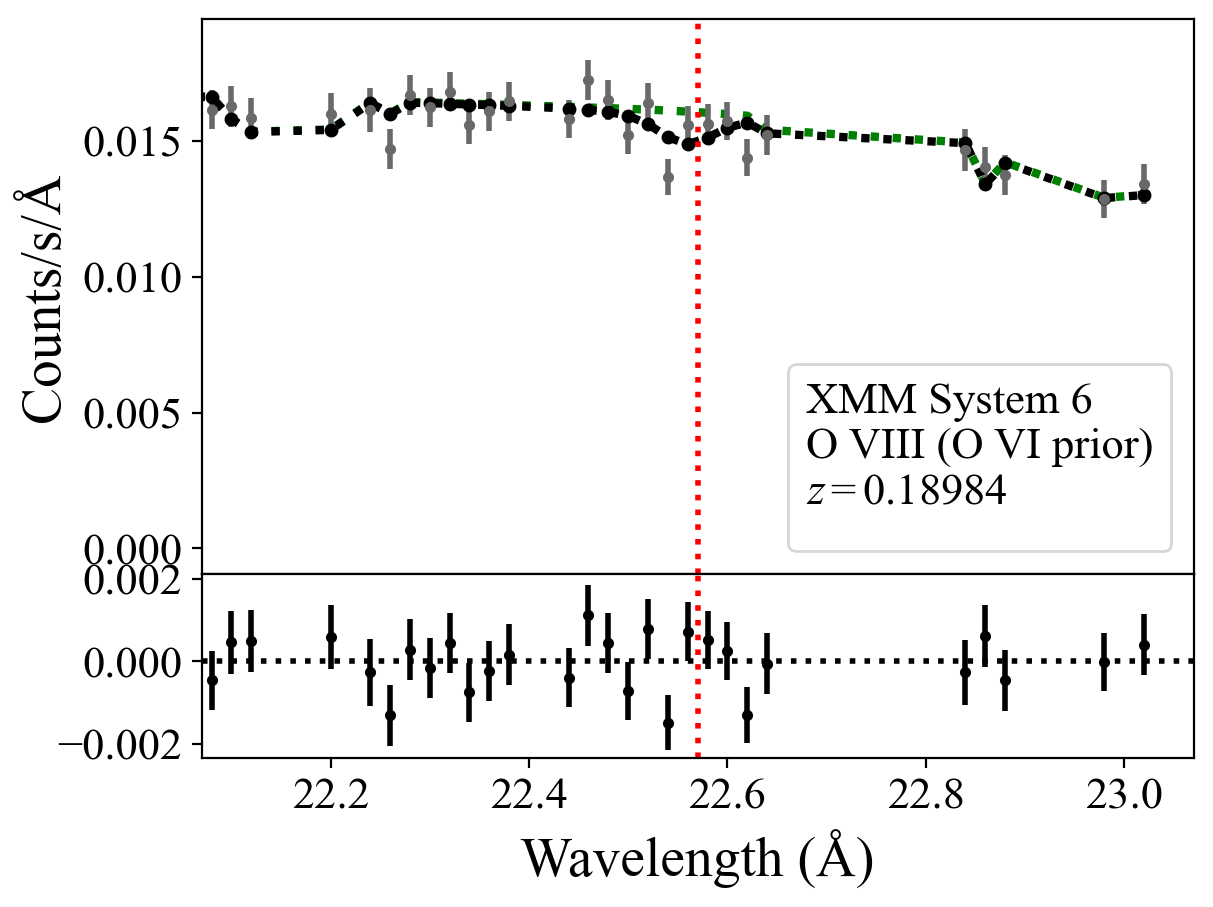}
      \includegraphics[width=\figSize in]{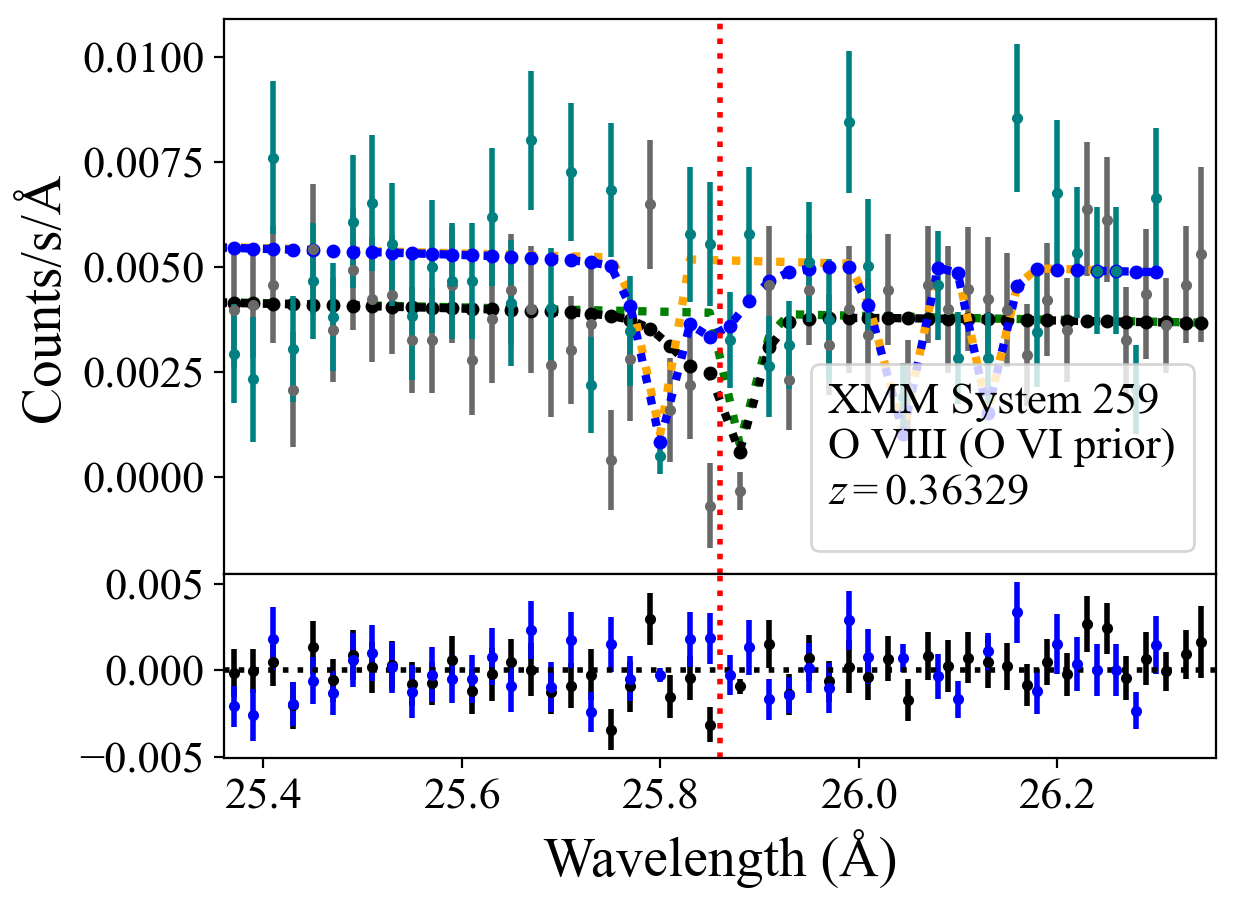}
      \includegraphics[width=\figSize in]{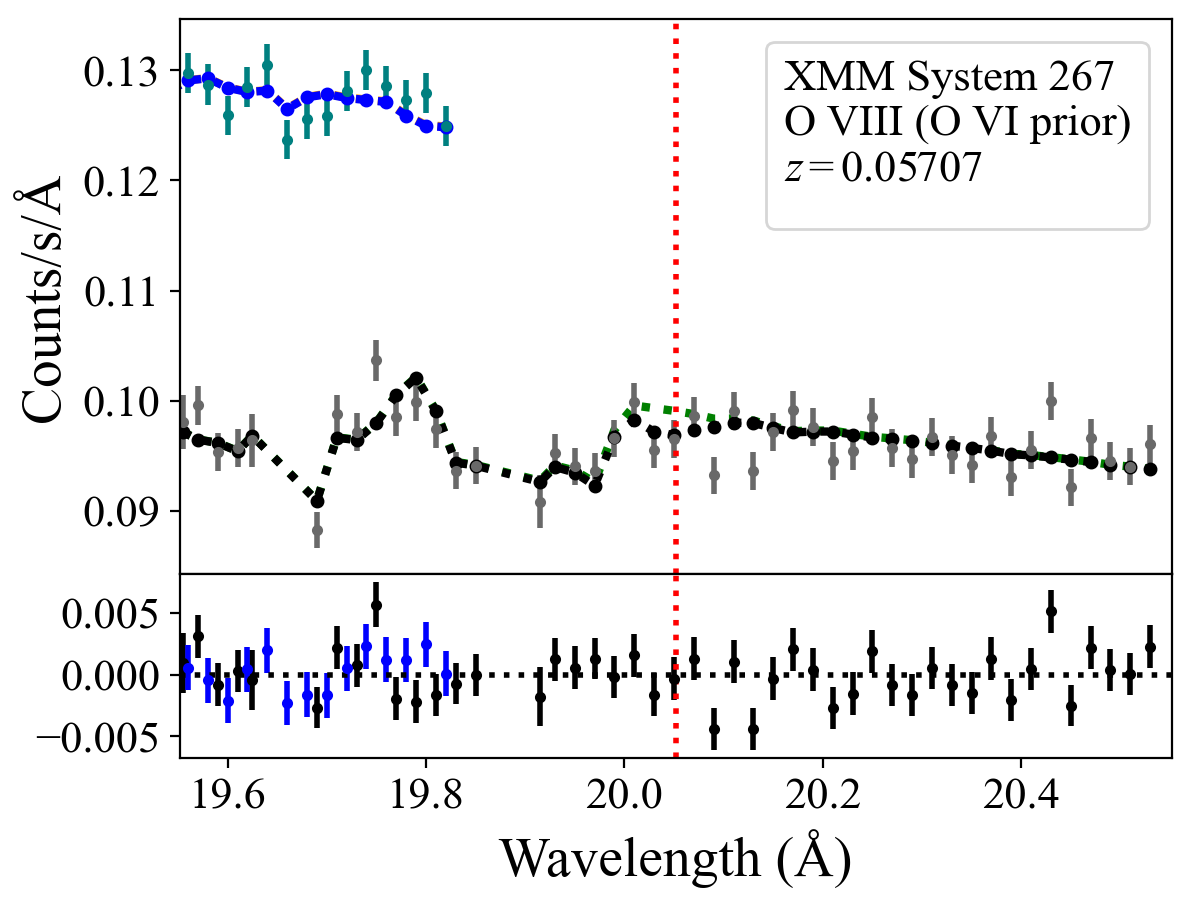}
      \includegraphics[width=\figSize in]{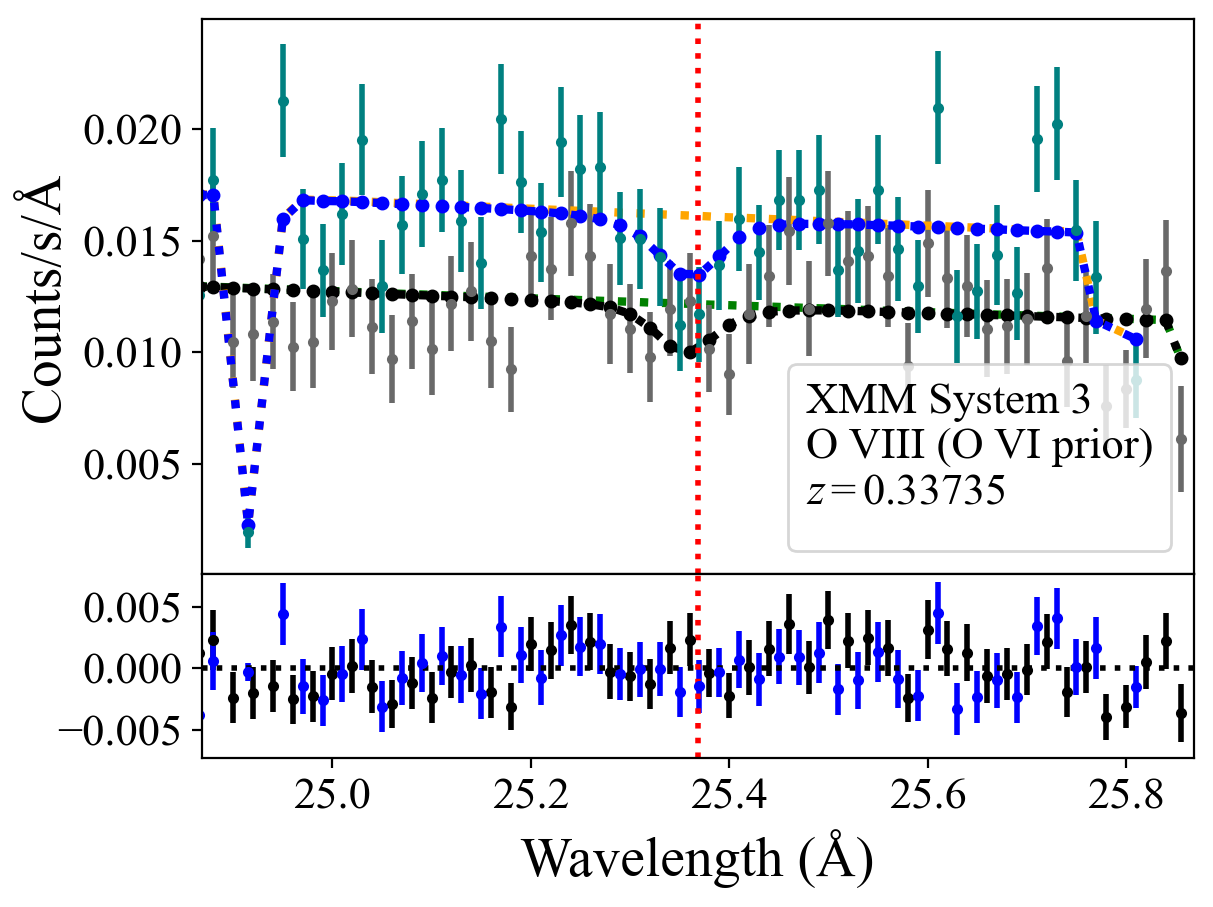}
      \includegraphics[width=\figSize in]{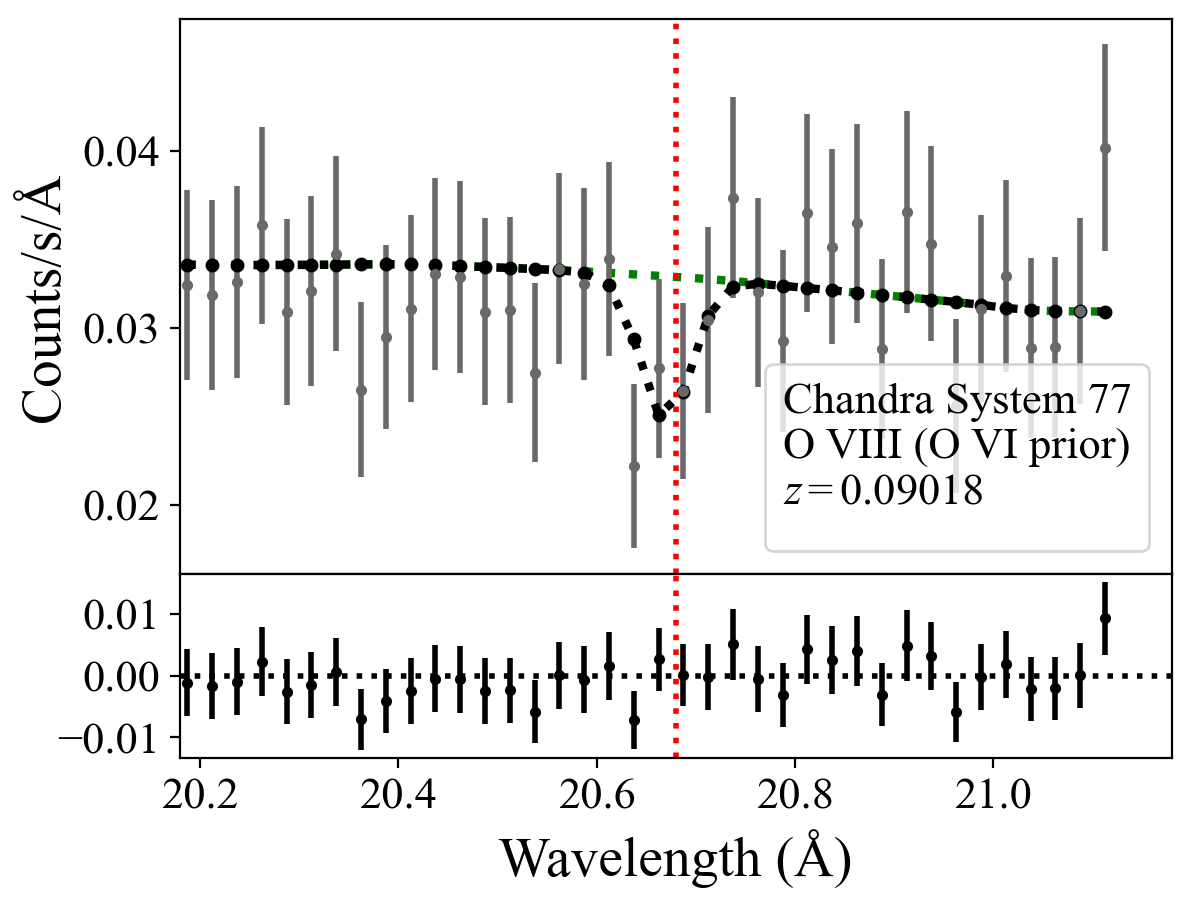}
          \includegraphics[width=\figSize in]{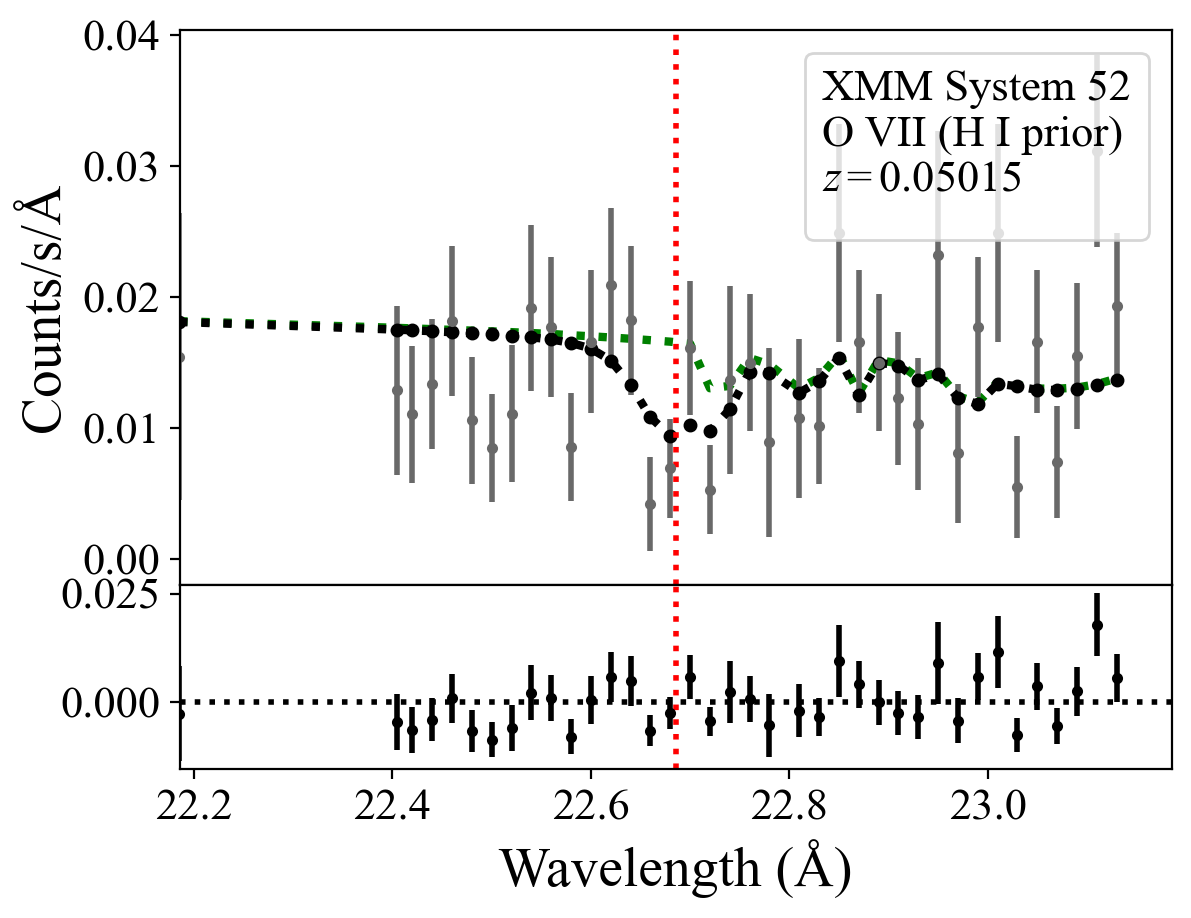}
     \includegraphics[width=\figSize in]{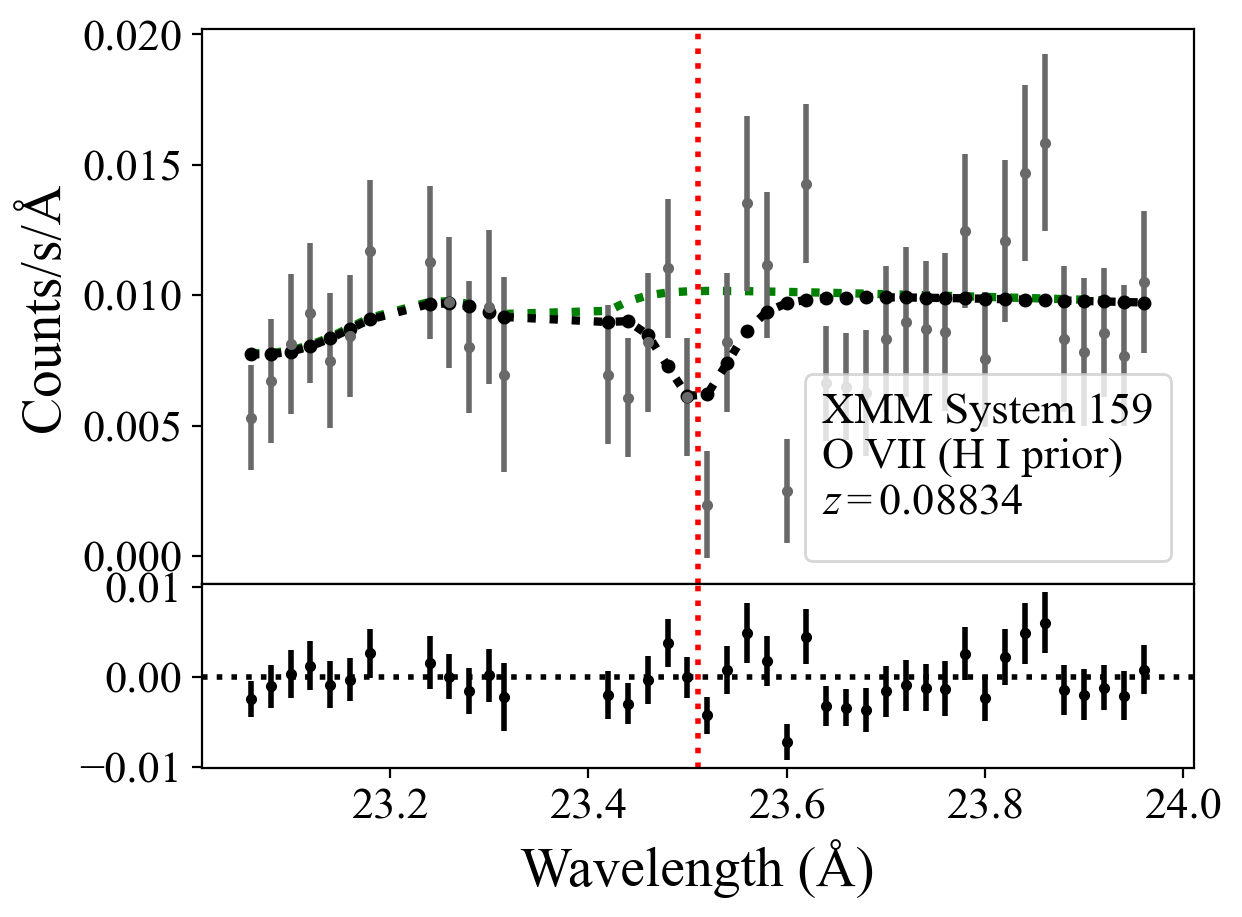}
    \caption{Spectra for systems with possible WHIM detections (cont'd).}
    \label{fig:detections2}
\end{figure*}

\begin{figure*}
    \centering  
    \includegraphics[width=\figSize in]{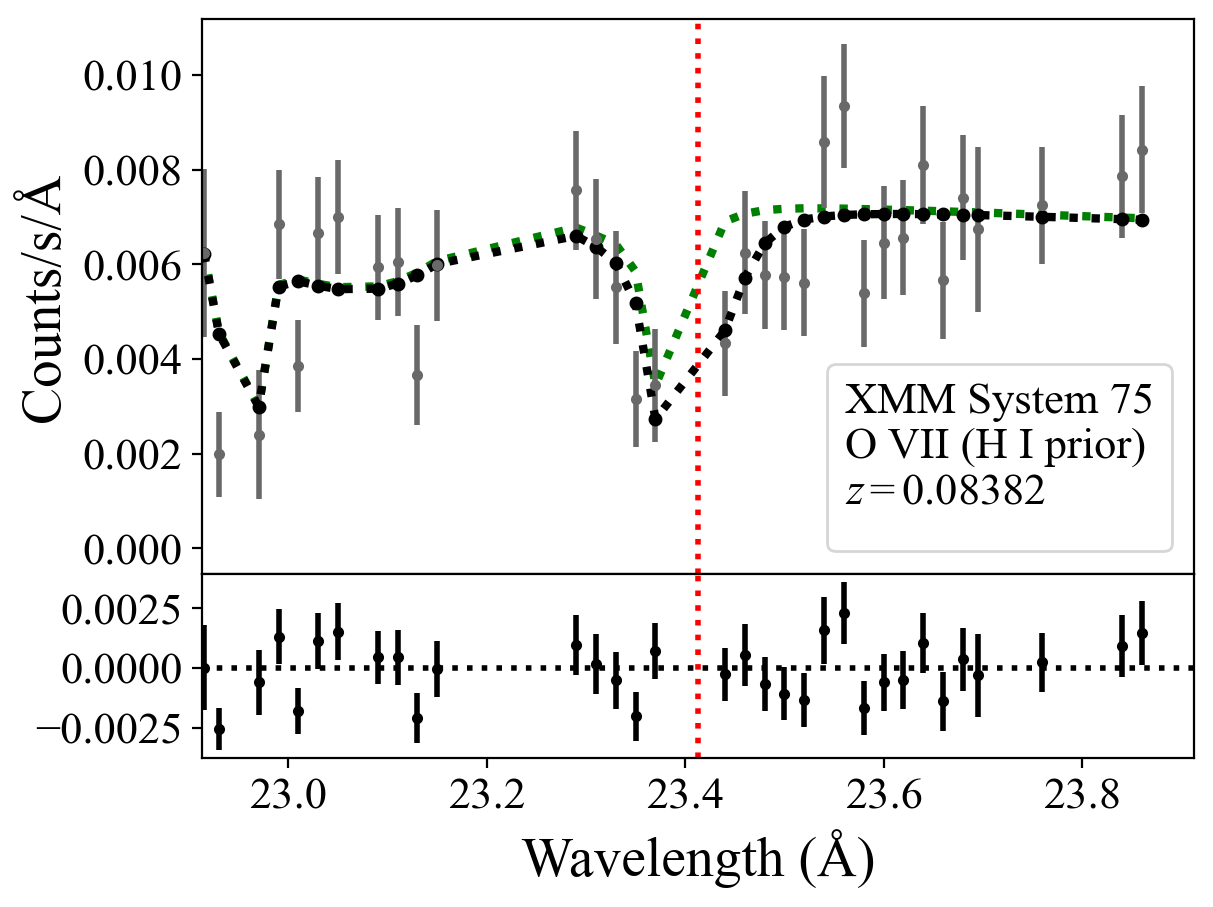}
      \includegraphics[width=\figSize in]{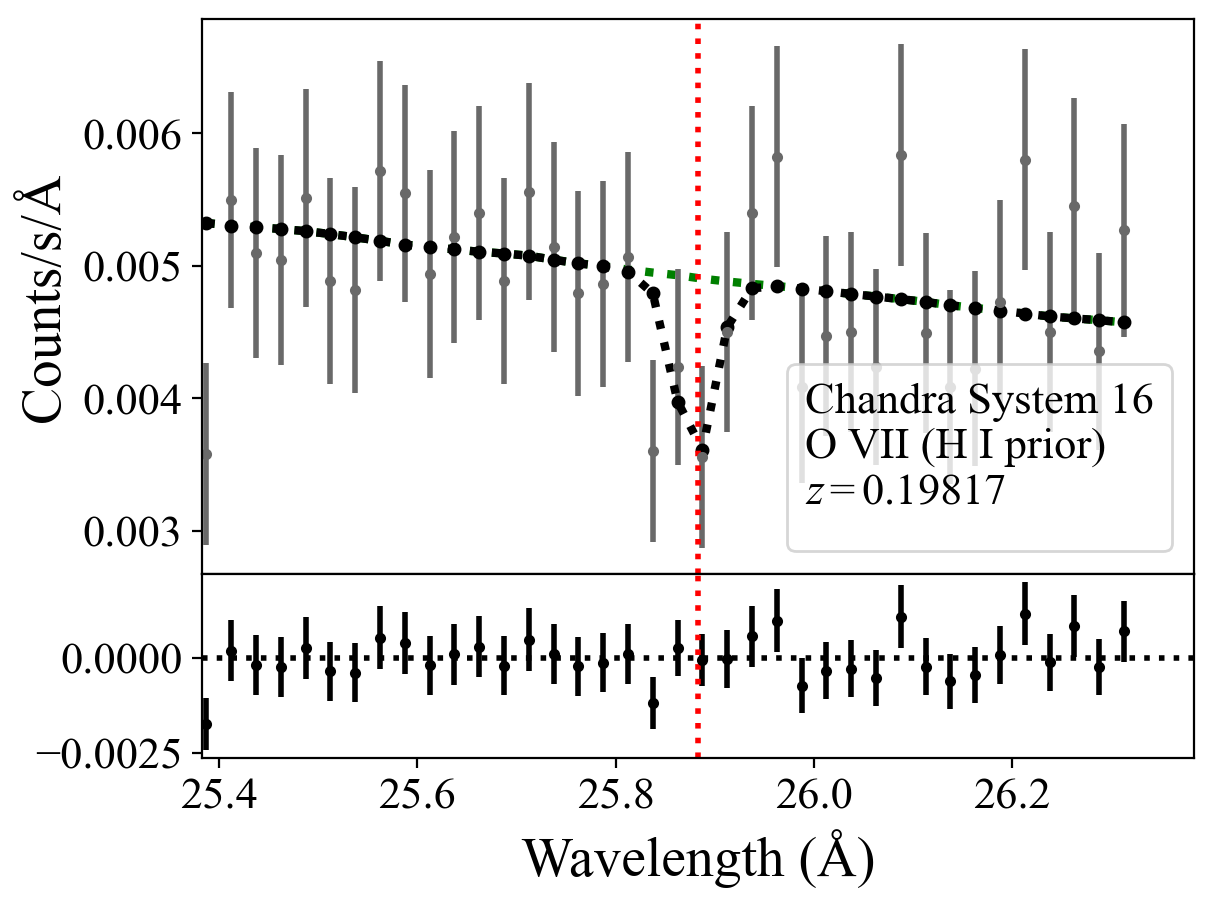}
          \includegraphics[width=\figSize in]{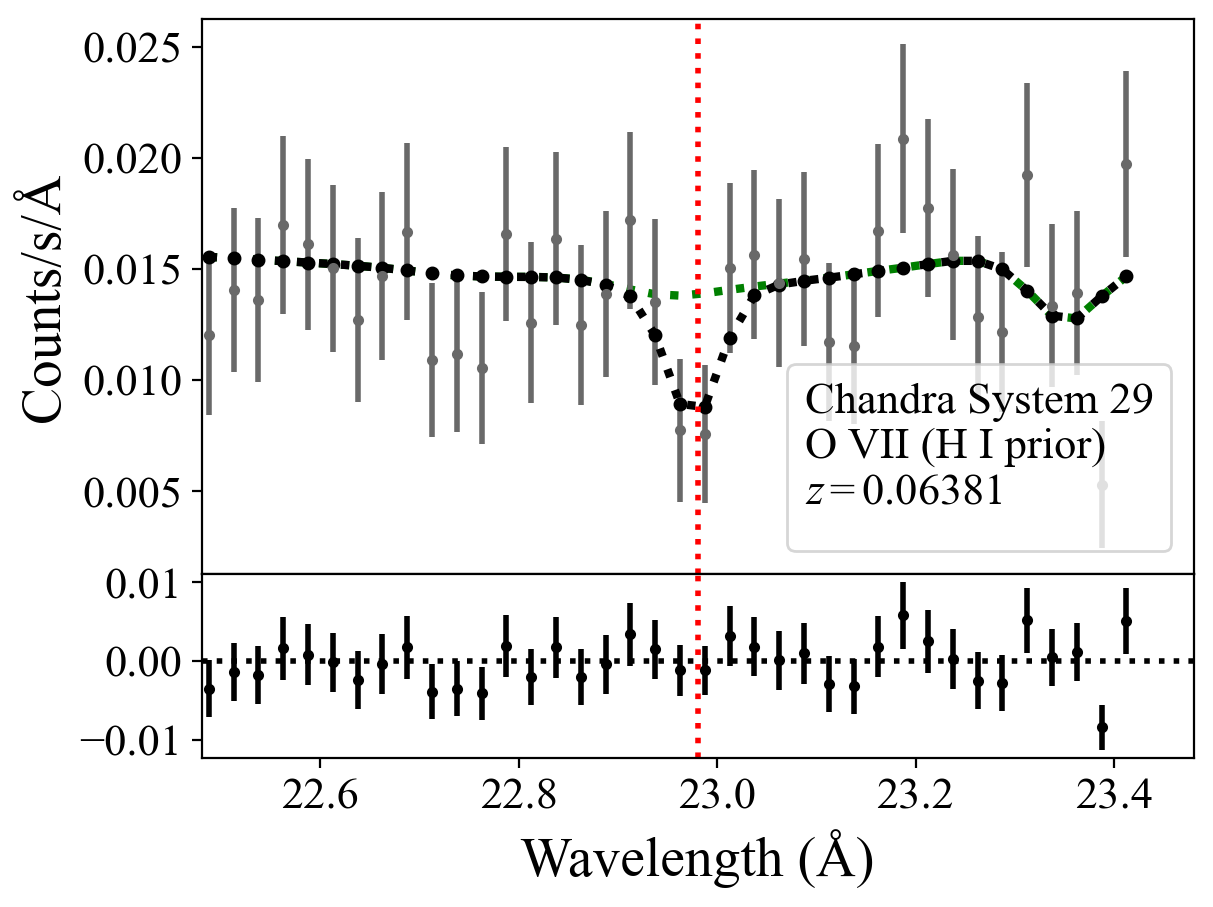} 
           \includegraphics[width=\figSize in]{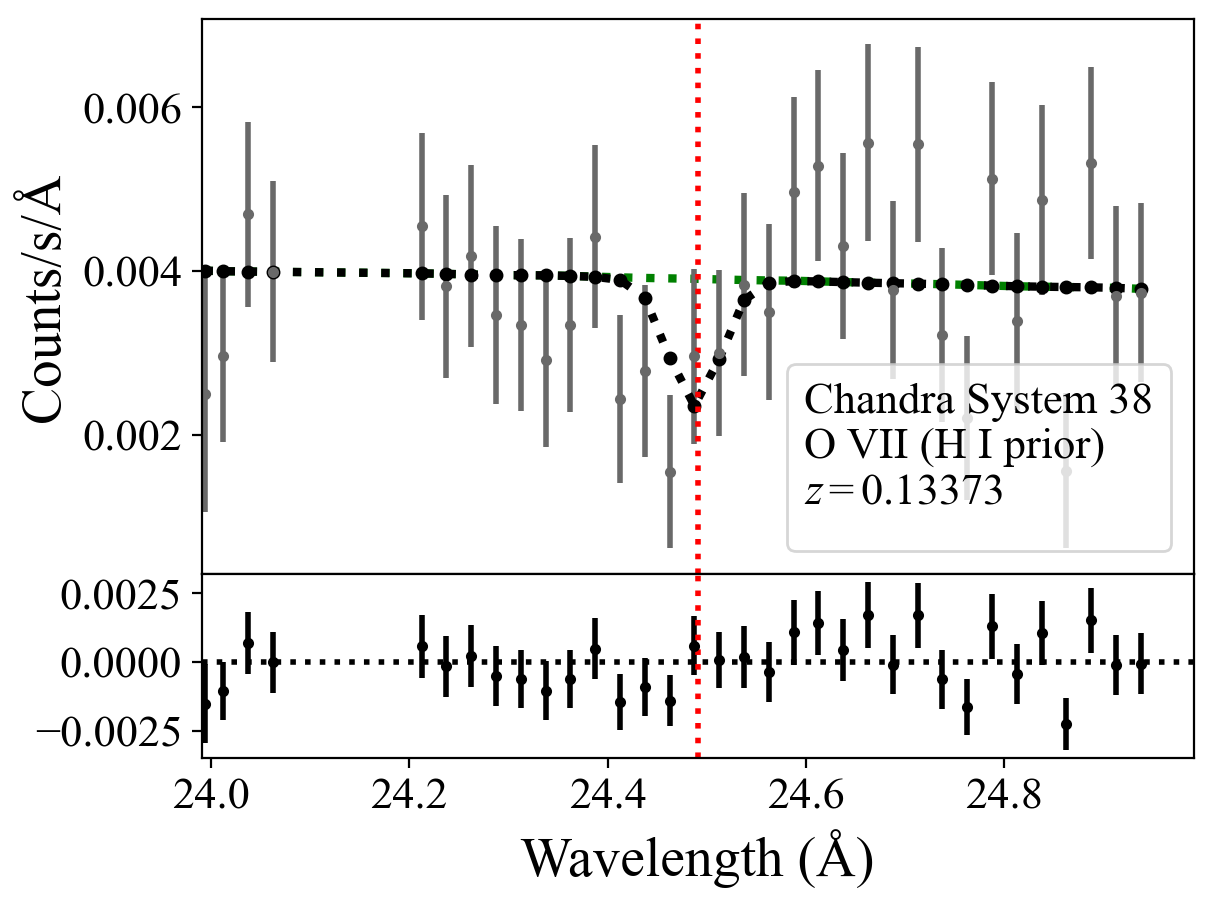}
           \includegraphics[width=\figSize in]{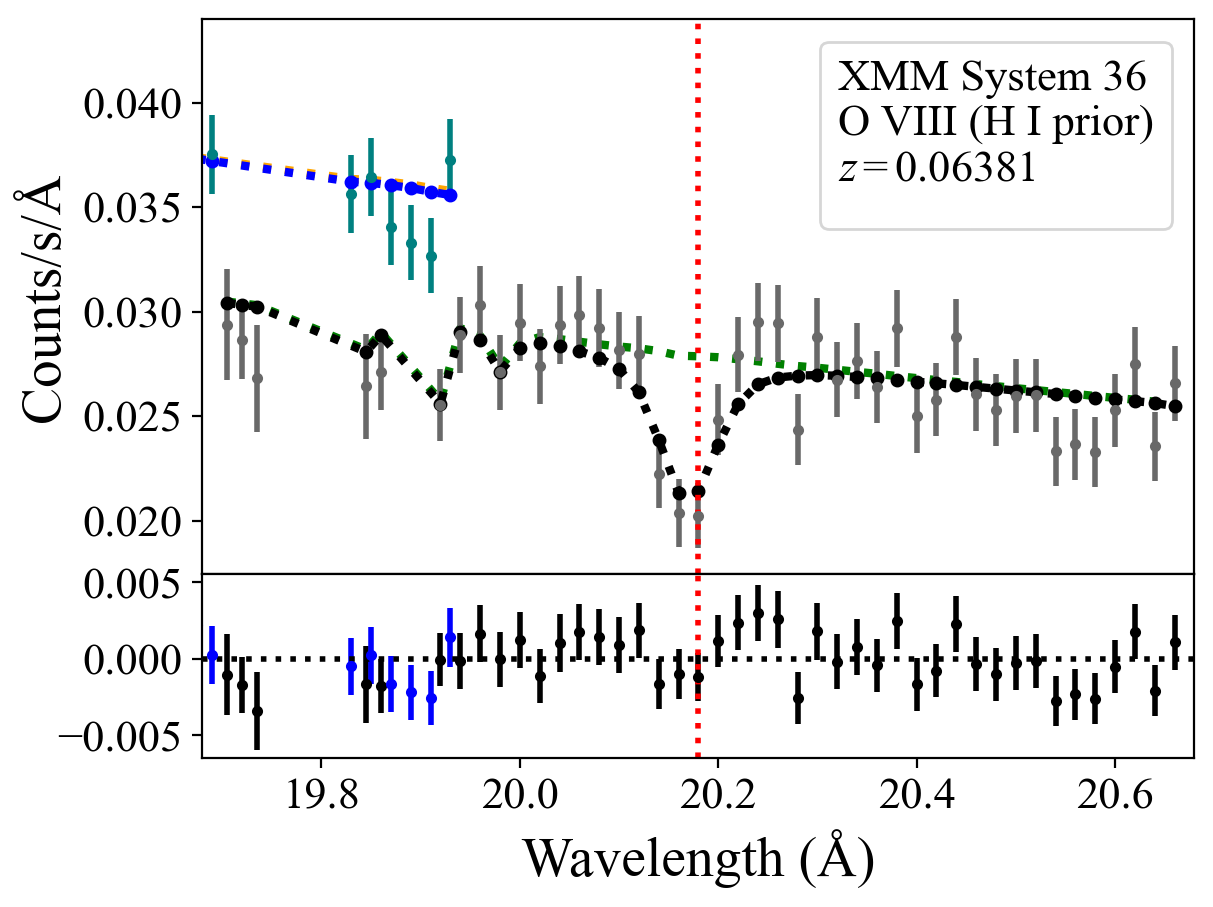}
           \includegraphics[width=\figSize in]{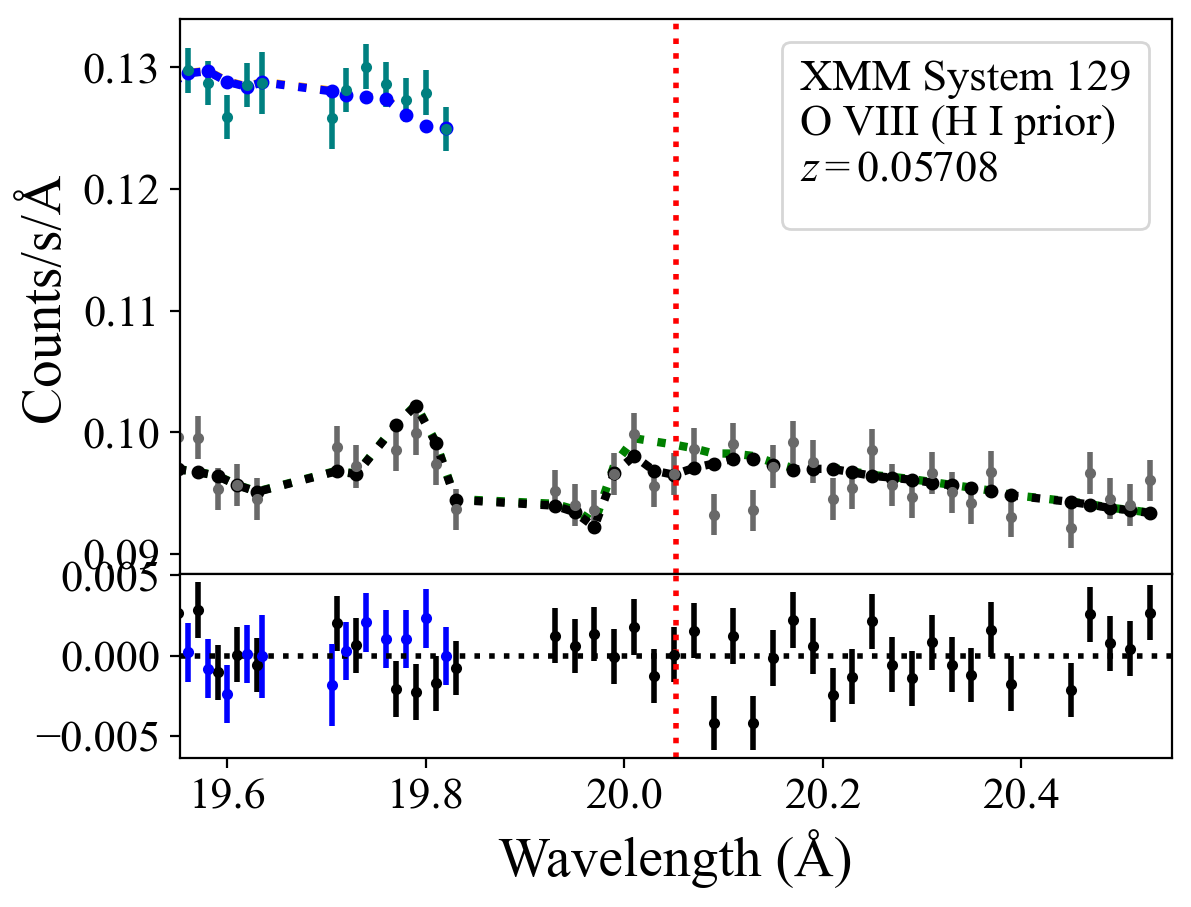}
    \includegraphics[width=\figSize in]{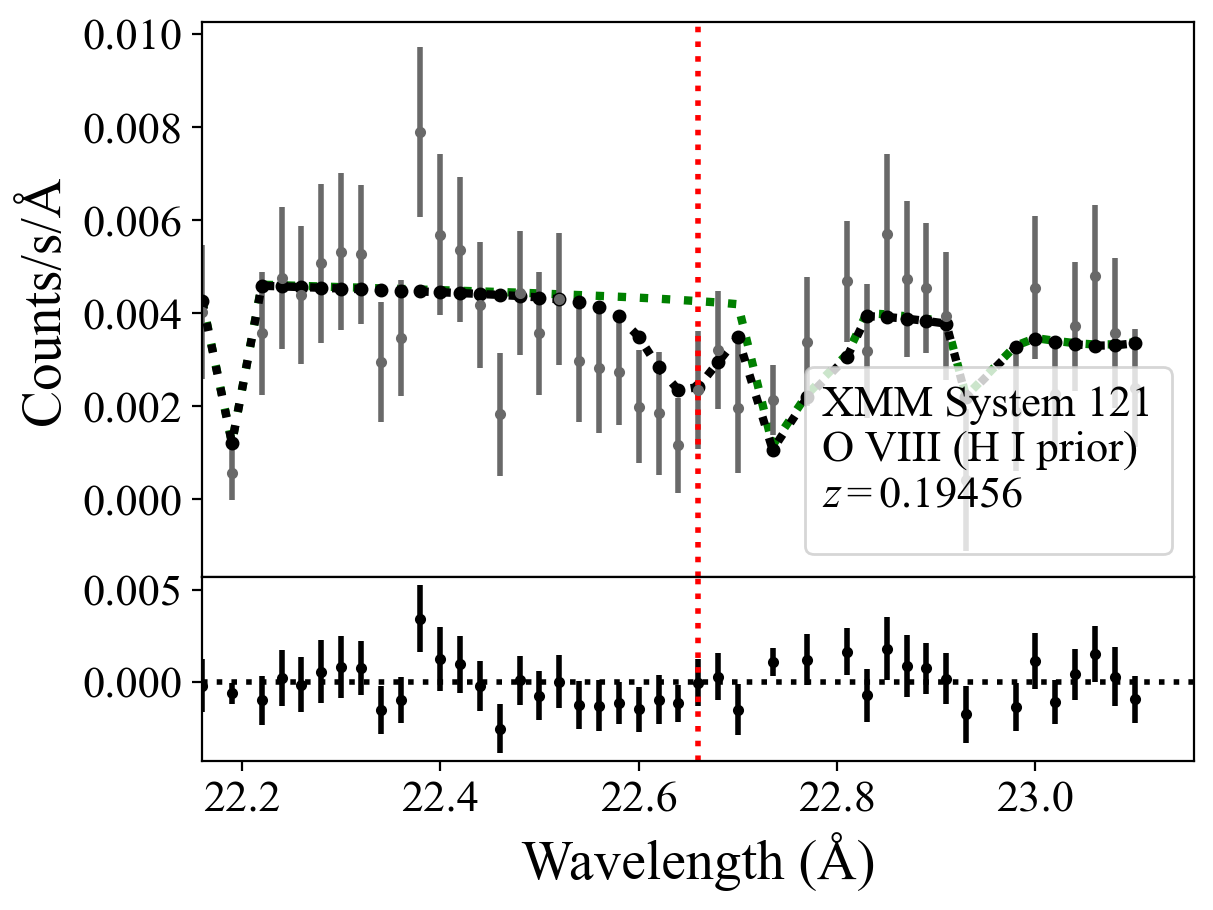}
\includegraphics[width=\figSize in]{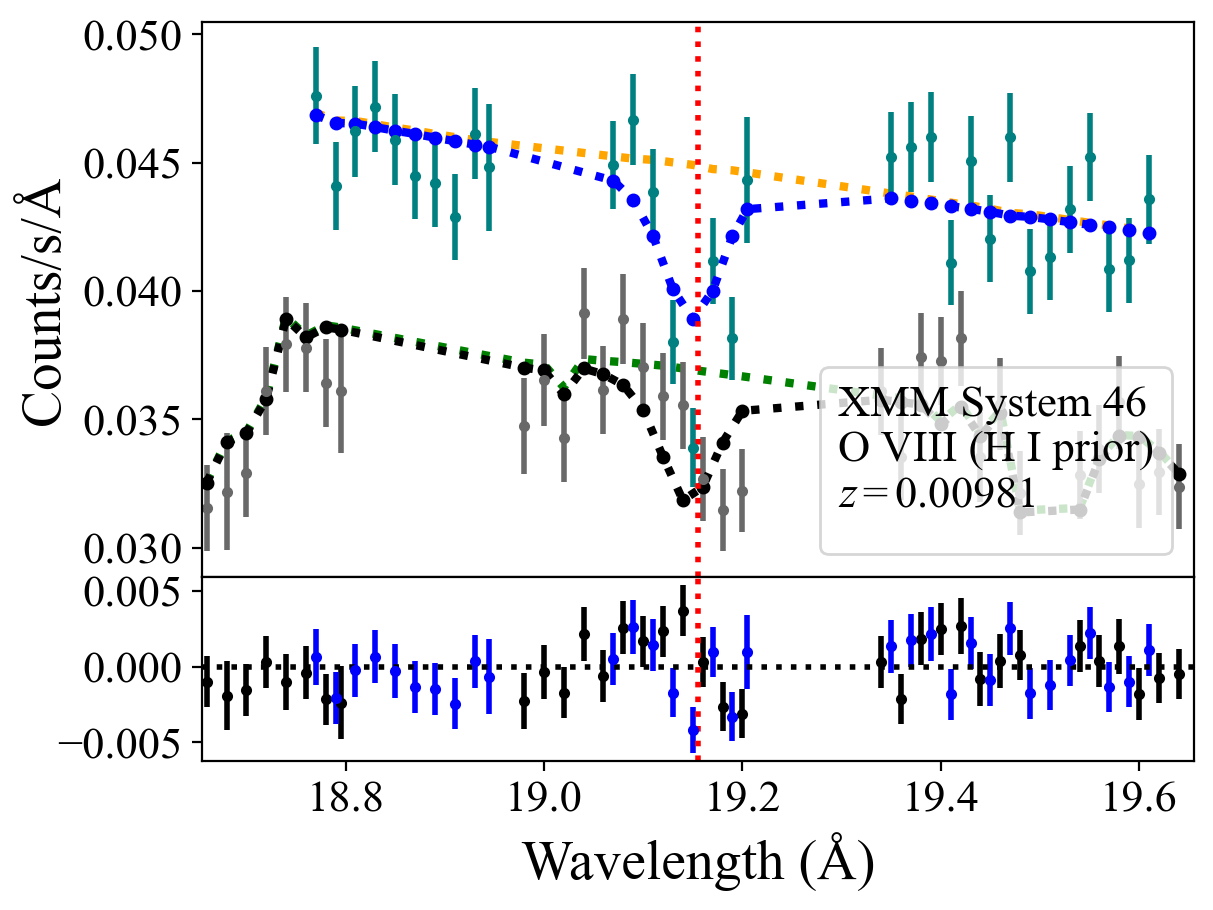}
      \includegraphics[width=\figSize in]{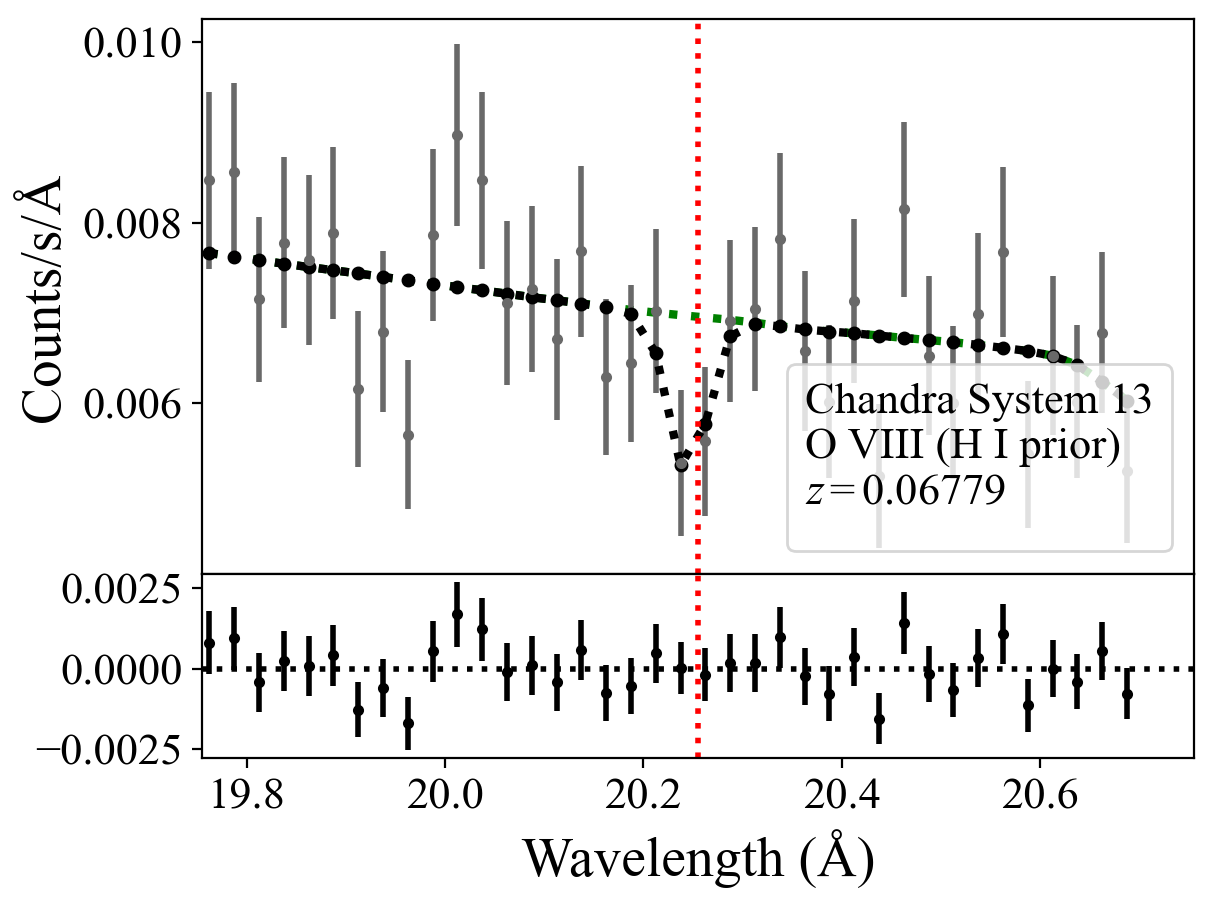}
\caption{Spectra for systems with possible WHIM detections (cont'd).}
    \label{fig:detections3}
\end{figure*}

\section{Discussion}
\label{sec:discussion}

\subsection{Sources and redshifts with previously reported detections}
\label{sec:previousDetections}

Among the sources analyzed for this work, there are a few with previously reported
possible X--ray absorption systems associated with the WHIM. A list of these systems
is provided in Table~\ref{tab:previousDetections}. Comments on source with FUV priors
used in previous analyses are provided in this section.

\begin{table*}
    \centering
    \begin{tabular}{l|p{0.5cm}p{1.0cm}p{1cm}p{1.5cm}p{2.8cm}p{4.5cm}}
    \hline
    \hline
       Source  & Species & Redshift & \multicolumn{2}{c}{FUV Prior} & References & Notes\\
               &         &          &  \multicolumn{2}{c}{\hrulefill} & & \\
               &         &          & ID & $\Delta C$ (X; C) & & \\
       \hline
        1ES~1553+113 & \ovii\ & 0.4335, 0.3551 &  \multicolumn{2}{c}{\nodata} & \cite{nicastro2018}, \cite{gatuzz2023} & No FUV priors were used for those analyses. \\
                     & \ovii & 0.1878 & \#4(OVI) & (7.8; 0.6) &  \cite{spence2023} & Reported as a marginal detection.\\
                     &  & & \#5(OVI) & (7.5; 0.6) & & Nearly same redshift as \#4.\\
        3C~273 & \oviii & 0.090 & \#21(OVI) & (1.2; 6.6)& \cite{ahoranta2020} & Redshifted \ovii\ coincides with Galactic \oi.\\
        H~1821+643 & \ovii\ & (Stacked) &  \multicolumn{2}{c}{\nodata} & \cite{kovacs2019} & Several FUV priors were stacked\\
        H~2356-309 & \ovii & 0.03 & \multicolumn{2}{c}{\nodata} &\cite{fang2010, buote2009, gatuzz2023, zappacosta2010} &No FUV priors were used\\
        Mrk~421 & \ovii & 0.011 &  \#231(OVI)& ($\nodata$;9.9$^{(*)}$)  & \cite{nicastro2005,rasmussen2007,yao2012, gatuzz2023} & \#231 uses \ovi\ prior at 0.0100, not used in searches. \xmmshort\ redshifted \ovii\ unavailable (21.5-21.9~\AA\ ignored)\\
		& \ovii & 0.027 &  \multicolumn{2}{c}{\nodata} & & No FUV prior used in this search, disagreement among authors.\\
        PG~1116+215 & \oviii & 0.093 & \#76(HI) & (0.2; 0.5)  & \cite{bonamente2016, bonamente2019b, gatuzz2023} & \xmmshort\ has nearby detector gap.\\
           &  &  & \#77(HI) & (0.2;0.6) & & Nearly same redshift as \#76. \\
        PKS~0558-123 & \ovii & 0.117 &  \multicolumn{2}{c}{\nodata} & \cite{nicastro2010, gatuzz2013} & 
        No FUV prior was used.\\
        PKS~2155-304 & \oviii & 0.0543 & & & \cite{fang2005,fang2007, fang2002, yao2009,cagnoni2004} & No FUV priors used in the search; detection is contested.\\
             & \oviii & 0.054 & \#266(OVI) & (0; 0.73) & \cite{nevalainen2019} & Disagreement among instruments in N19.\\
                     &        & 0.057 & \#267(OVI) & (7.4; 0.1) &  &  \\
                     &        &       &  \#129(HI) & (7.0,0) & & \\
        Ton~S180 & \ovii & 0.04579 & \#308(OVI)  &  (14.4; 2.1$^{(*)}$)& \cite{ahoranta2021} & XMM detection confirmed, Chandra had low S/N, as originally noted.\\
         \hline
         \hline
    \end{tabular}
    
    Notes: $(*)$ Model has emission line feature with negative $\tau_0$.
    \caption{Analysis of selected sources with prior reported detection of possible WHIM X--ray absorption lines. The $\Delta C$ statistics are reported for \xmm\ (X) and \chandra\ (C), respectively.}
    \label{tab:previousDetections}
\end{table*}

\subsubsection{1ES~1553+113} The \cite{spence2023} analysis is virtually identical to that provided in this work, and the putative \ovii\ corresponding to an \ovi\ absorber at $z \simeq 0.188$
is found with similar significance as the earlier publication. 
Although our work is limited to absorption lines at FUV redshift priors, we comment that this source was also analyzed by \cite{nicastro2018, nicastro2018b} and by \cite{gatuzz2023} for serendipitous lines without FUV priors. The $z\simeq0.434$ \ovii\ line reported by \cite{nicastro2018}  was shown by \cite{johnson2019} to be likely associated with the 
intra--group medium where the \es\ is located, rather than the WHIM. \cite{gatuzz2023} did not find
evidence of absorption at other redshifts. Our analysis did not investigate these X--ray serendipitous redshifts.

\subsubsection{3C~273} The putative FUV--prior \oviii\ detected by \cite{ahoranta2020} at $z=0.090$ is tentatively confirmed in these observations 
(see Sec.~\ref{sec:3c273oviii}). 
Note that the \cite{ahoranta2020} results were based on a modelling of all WHIM species via a collisional ionization equilibrium 
(\texttt{slab}) model for both the \chandra\ and \xmm\ data, which included the putative
detection of \neix\ that we do not investigate in this paper. 
 The stronger detection of this absorption system with \chandra, compared to \xmmshort, is consistent with the \cite{ahoranta2020} 
analysis (see Table~4 and Figure~4 of that paper). Also, there is evidence for \ovii\ at this redshift, see Sec.~\ref{sec:3c273ovii}. However, the \cite{ahoranta2020} analysis concluded that it is not possible to distinguish possible \ovii\ at this redshift from Galactic \oi\ absorption.

\subsubsection{Mrk~421} The putative \ovii\ absorption line at $z=0.011$ tentatively detected
by \cite{nicastro2005} falls close to a known RGS 
detector artifact near $21.8$~\AA\ that was excluded in our analysis. This is consistent with the non--detections reported by \cite{rasmussen2007} and \cite{yao2012}, and by the analysis of \cite{gatuzz2023} which found no significant evidence of absorption.
In Sec.~\ref{sec:mkn421} we briefly discussed a possible \oviii\ detection at $z=0.010$, which
however may have been affected by averaging of different states of the source, and therefore
should not be regarded as conclusive.

\subsubsection{PG~1116+215} 
\cite{bonamente2016} had reported a tentative detection of \oviii\ near $z=0.091$ from the 
analysis of the deeper \chandra\ data available for this source.
The current \chandra\ analysis is based on all available observations, whereas the original report of an \oviii\ line was based only on the earlier
observation. The later observations, which we studied in  \cite{bonamente2019b}, had significantly higher background, and it was determined that those data could not detect the line. The addition of the \chandra\ datasets are a possible reason for 
the failure to confirm the \cite{bonamente2016} putative detection in this analysis.
The \xmm\ observations have an RGS gap near the putative line, and do not provide useful data. The
analysis by \cite{gatuzz2023} did not find evidence of the reported \oviii\ absorption line.

\subsubsection{PKS~2155-304} 
\label{sec:pks2155}
\cite{fang2002, fang2005, fang2007}  reported a possible $z=0.0543$ \oviii\ line, which would fall between the two \ovi\ priors investigated in this paper (at $z=0.054$ and 0.057), which where investigated by \cite{nevalainen2019}. 
\cite{yao2009, cagnoni2004} did not confirm those results on the \oviii\ line. \cite{nevalainen2019} did not find evidence of any \ovii\ lines at the FUV prior redshifts, and found marginal evidence of a possible \oviii\ line at one of the two redshifts ($z=0.054$), but with inconsistent results among the instruments used. We notice that some of the earlier detection
of the possible $z=0.0543$ line were condicted with \chandra\ LETG/ACIS combination, whereas we only use the LETG/HRC combination for this study.

Our analysis does not detect \oviii\ absorption at the lower of the two redshift ($z=0.054$), but it finds marginal evidence for absorption at $z=0.057$ (see Sec.~\ref{sec:pks2155o8o6} and \ref{sec:pks2155o8hi}). This  finding is in disagreement with the previous analysis by \cite{nevalainen2019}, where evidence of absorption at $z=0.054$ was reported.

Moreover, at those redshifts, any 
\ovii\ absorption would fall near an inefficiency in the 
\xmmshort\ detectors at 22.7-22.8~\AA\ that makes the assessment very uncertain.
Specifically, system 266 ($z=0.054$, $\lambda=22.7675$~\AA) 
has a very poor fit if the data in the inefficiency are included,
and that fit should not be considered useful. And system 267
($z=0.057$, $\lambda=22.8327$~\AA), with the 22.7-22.8~\AA\  data
removed from the fit, does not find any evidence for \ovii\ absorption (best--fit $\tau_0$ is negative).

\subsubsection{Ton~S180} The tentative \xmm\ detection of \ovii\ corresponding to an \ovi\ prior by \cite{ahoranta2021} is confirmed by this
re--analysis (see system 308, Sec.~\ref{sec:tons180}).

\subsubsection{H1821+643} This quasar was the subject of the \cite{kovacs2019} analysis, who used a larger number of \hi\ FUV priors from \cite{tripp1998}
to provide a tentative detection of \ovii\ from the stacking of the spectra.
Our tentative detection of \ovii\ at a redshift that is similar to
one of those used in their search (see Sec.~\ref{sec:h1821}) lends positive support for the presence of
WHIM along the sightline to this source.

\subsection{Systematics}
\label{sec:systematics}
The large number of sources and of absorption--line systems probed in this
work required certain simplifying assumptions to carry out a 
homogeneous analysis. Such assumptions may have limited or biased
our results for certain sources. Accordingly, we address the major sources of systematic error introduced by our analysis in this section.

\subsubsection{Fixed redshift}

By design, this study makes no attempt to optimize the search for possible X--ray absorption lines at prior FUV redshifts, e.g., by adjusting the redshift, as was done for example for the case of PG~1116+215 \citep{bonamente2016}. Given the binning and the resolution of the data, a 20~m\AA\ uncertainty in the centroid of an absorption line corresponds to a redshift uncertainty of approximately $\Delta z \simeq 0.001$, or a peculiar velocity of the putative
absorber of $v \simeq 300$~\kms. Larger uncertainties in the line centroids are not included directly in this study,  as
their account would require another order--of--magnitude larger computational effort that goes beyond the scopes of this paper,  but they are considered as systematic errors that will be used when making cosmological ineferences from these data.

For example, this analysis of the quasar  PG~1116+215 
does not find evidence for absorption at $z=0.0928$ at an \hi\ FUV prior (see Tab.~\ref{tab:OVIIIHIChandra}), which was however detected by varying redshift to $z=0.0911$ by \cite{bonamente2016} upon visual inspection of the spectra. The choice to not vary the FUV redshift is 
motivated by the goal of  providing
a uniform analysis that does not involve peculiar velocities between the FUV and the X--ray absorber. Such peculiar velocities would complicate
not just the analysis, but also the interpretation and the statistical
significance of detection, which would then require considerations of
the effective number of redshift trials, as discussed in Sec.~\ref{sec:sample}. 

 To illustrate and quantify the effects of this systematic, we repeated our analysis with $\Delta \lambda=\pm20$~m\AA\ and 
$\Delta \lambda=\pm40$~m\AA\ shifts to the line center of two representative spectra: the high S/N \xmmshort\ spectrum
for system 114, \oviii/\ovi\ 
corresponding to the possible $z=0.01$ absorption line for Mrk~421; and to the low S/N \xmmshort\ spectrum for system 179, \oviii/\ovi\ which features a non-detection. These shifts correspond to approximately a redshift uncertainty of 
$\Delta z=\pm0.001$ and $\Delta z=\pm0.002$.

Results of the fits are provided in Table~\ref{tab:shift}. For system 114, three of the four fits with a shifted line centroids give equivalent results as the nominal fit. The $\Delta \lambda=-40$~m\AA\ fit, marked with a star in the table, gives a substantially strong line detection ($\Delta C=15.3$ versus the nominal $\Delta C=9.0$). 
This difference may be due to two reasons.
First, the new line centroid is closer to a gap in the spectral coverage, which is visible in Fig.~\ref{fig:detections2}, thereby
making the continuum around the line less accurately determined. Second, the higher significance may be due to
the random fluctuations in flux of this relatively high S/N source, which results in a strong detection when the centroid corresponds to a datapoint with low flux. The latter reason also highlights the possible perils of adjusting the absorption redshift to maximize the significance of detection of a line \citep[e.g.][]{bonamente2016}.
For system 179, all fits with shifted line centroids give equivalent results to the nominal fit, 
a result which is consistent with
the lower S/N of this spectrum. 
We conclude that the systematic error associated with a fixed redshift in the search
may be significant for some of the strongest sources, but not significant for the majority of the other sources.

\begin{table}
    \centering
    \begin{tabular}{l|cll}
    \hline
    \hline
    \multicolumn{4}{c}{\xmmshort\ \oviii/\ovi\ System 114} \\
    \hline
    Statistic & Nom. $\lambda$ & $\Delta \lambda=+0.02$~\AA\ & $\Delta \lambda=-0.02$~\AA \\
                   & \multicolumn{3}{c}{\hrulefill} \\
    $\Delta C$    & 9.0  & 9.2 & 10.9 \\
    $\tau_0$   &  0.136$\pm$0.048  & 0.138$\pm$0.048 & 0.150$\pm$0.048 \\
    \hline
                &         & $\Delta \lambda=+40$~m\AA\ & $\Delta \lambda=-40$~m\AA \\
                & \multicolumn{3}{c}{\hrulefill} \\
     $\Delta C$ &  (same)   & 7.6 & 15.3$^{\star}$ \\
       $\tau_0$  &  (same)  & 0.125$\pm$0.048 & 0.191$\pm$0.050$^{\star}$\\
       \hline
       \multicolumn{4}{c}{\xmmshort\ \oviii/\ovi\ System 179} \\
    \hline
    Statistic & Nom. $\lambda$ & $\Delta \lambda=+0.02$~\AA\ & $\Delta \lambda=-0.02$~\AA \\
                   & \multicolumn{3}{c}{\hrulefill} \\
    $\Delta C$    & 1.0  & 0.3 & 1.0\\
    $\tau_0$   &  0.48$\pm$0.54  & 0.23$\pm$0.48 & 0.48$\pm$0.54 \\
    \hline
                &         & $\Delta \lambda=+40$~m\AA\ & $\Delta \lambda=-40$~m\AA \\
                & \multicolumn{3}{c}{\hrulefill} \\
     $\Delta C$ &  (same)   & 0.0 & 1.0 \\
       $\tau_0$  &  (same)  & 0.00$\pm$0.44 & 0.47$\pm$0.54\\
    \hline
    \hline
    \end{tabular}
    \caption{Comparison of results between the nominal fits for two spectra, and those with an absorption
    line centroid shift. $\Delta C$ is the usual statistic that is used for the significance of detection, and $\tau_0$
    the optical depth at line center, both also reported in Table~\ref{tab:OVIIIOVIXMM} for the  fits with the nominal
    wavelength.}
    \label{tab:shift}
\end{table}

\subsubsection{Source variability}
Source variability is another issue that may affect the ability of 
our method of analysis to provide accurate results for certain sources.
The sample considered in this study includes X--ray blazars such as Mrk~421 and PKS~2155-304 and AGNs such as 3C~273
that are known to be highly time--variable \citep[e.g.,][and references therein]{gupta2020, soldi2008}, and other quasars can also be variable
over several time scales \citep[e.g.][]{middei2017}.
Variability might be a potential issue, in particular, for Mrk~421, whereas we could not fit the
spectrum in the usual 1~\AA\ baseline, possibly due to difficulties in averaging X--ray spectra at different epochs. The choice of a 
relatively short wavelength range for the fit is in fact designed to minimize this possible issue, but more accurate models might be required for
certain bright sources \citep[as was done, e.g., in][]{spence2023}.

For possible absorption lines that are detected according to a large value of the $\Delta C$ statistic (see Sec.~\ref{sec:detections} and Fig.~\ref{fig:detections}), a reanalysis that addresses source variability across the observations, 
peculiar velocities and other systematics, may be warranted. Such analysis would involve the fit of individual spectra for each epoch with 
a more complex model (e.g, a spline as in \citealt{spence2023}), which was not feasible 
to conduct for all sources in this paper in a uniform way.
Issues involving source variability and their relationship to WHIM absorption line detections were also investigated in previous analyses of some of the sources in this sample \citep[e.g.][]{nevalainen2019}.

\subsubsection{Intrinsic scatter}

Calibration uncertainties in the \xmm\ and \chandra\ data
\citep[e.g.][]{kaastra2018, spence2023} and inadequacies of the simple power--law model for the continuum may give rise to larger--than--statistical
fluctuations of the data relative to the best--fit model, as noted in Sec.~\ref{sec:dataAnalysis}. 
The presence of calibration uncertainties leading to intrinsic scatter 
is in fact suggested by the presence of a small but significant number of fits, especially in the \xmm\
data, as summarized in Tab.~\ref{tab:statistics}. 
Such sources of intrinsic scatter can be accounted in the Poisson regression, for example following the methods presented in \cite{bonamente2023} and \cite{bonamente2024}. 
In the presence
of intrinsic scatter, we have shown that the parent distribution of the \cmin\ statistic is modified in such a way as to make even substantially larger--than--one reduced \cmin\ statistics formally acceptable \citep[for details, see][]{bonamente2024}. This was the reason to list best--fit values for regressions in Tables~\ref{tab:OVIIOVIXMM} 
through \ref{tab:OVIIIHIChandra} even when there were large values of the \cmin\ statistic.

Another effect of the intrinsic scatter is that of lowering the level of significance of the possible detections presented in Sec.~\ref{sec:detections}. The exact effect of systematic errors on the $\Delta C$ statistic for the Poisson regression is not yet known exactly \citep[as noted, e.g., in][]{bonamente2024}, and 
therefore a detailed analysis of the 
significance of detection in the presence of intrinsic scatter is not feasible.
We  limit ourselves to point out that the $p$--values for the detection of the nested absorption--line model component in Sec.~\ref{sec:detections} will become lower in the presence of intrinsic scatter.
This effect could be at play for a number of systems discussed in Sec.~\ref{sec:detections}, whereas
the presence of systematic errors would lower the significance of detection and make some
of the $\Delta C \geq 6.6$ systems in fact \emph{less} statistically significant than the nominal
99~\% confidence.

\section{Conclusions}
\label{sec:conclusions}

This study was designed with the goal of analyzing, in a uniform way, the largest possible number of X--ray sources and absorption line systems in search of the missing baryons in the WHIM. With a total of \nSources\ extragalactic sources with X--ray data, we have uniformly studied the presence of possible \ovii\ and \oviii\ absorption in \nsystems\ absorption line systems at FUV priors provided by the \fuse\ and \hst\ analyses of \cite{tilton2012} and \cite{danforth2016}. We have opted for a simple power--law model of the continuum in the vicinity of the putative absorption lines, instead of a more physically motivated model such as the one used by \cite{gatuzz2023}. With this choice, we were able
to identify \nDet\ possible X--ray absorption lines and  set simple upper limits to the non--detection of \ovii\ and \oviii\ for all \nsystems\ systems. These measurements will be used to constrain the cosmological density of these ions, following the methods presented in \cite{spence2023}. Cosmological constraints will be presented in detail in a companion paper.

The search for X--ray absorption lines presented in this paper highlights the challenges of achieving significant detections with the current generation of X--ray grating spectrometers. Of the
\nsystems\ \ovii\ and \oviii\ X--ray systems with FUV priors analyzed, only 
\nDet\ have a statistically significant negative deviation from the continuum model 
according to the $\Delta C$ statistic, and  just 
a handful of these (Sec.~\ref{sec:detections}) have overall convincing evidence for the presence of possible \ovii\ or \oviii\ absorption at the FUV redshift priors. 
The main challenges towards a successful detection of X--ray absorption lines 
are associated with the presence of sources of systematic errors that are
likely to lower the formal significance of detection of these fluctuations, as discussed in Sec.~\ref{sec:systematics}.
The results of this search are therefore qualitatively consistent with previous searches, including the most comprehensive serendipitous search available to date \citep{gatuzz2023}, in finding it difficult to use
\xmm\ and 
\chandra\ grating data to detect the X--ray WHIM. 

Even amidst these substantial challenges, however, the search at fixed FUV redshift priors has shown that it is in fact possible to identify likely X--ray absorption lines with the \xmm\ and \chandra\ grating spectrometers. Specifically,
our analysis has confirmed certain previous detections, and found a few additional possible
absorption line systems that previous searches had been unable to identify. In total we have
identified \nDet\ systems with evidence of absorption of either \ovii\ or \oviii\ at 
a formal $\geq 99$~\% confidence. Given the presence of substantial sources of systematic error, a more detailed analysis of many of these systems
is needed, before making a conclusive determination on the nature of those fluctuations. 
Likewise, stacking of the spectra at the wavelength of the redshifted \ovii\ and \oviii\ lines (e.g.,
as in \citealt{kovacs2019})
might shed additional light on the overall presence of absorbing gas in the inter--galactic medium. 
Such analyses, however, go beyond the scopes of this paper, and they are deferred to future publications. 

Many of the sources in this sample have either a shallow exposure, or an intrinsically faint flux which, combined with the limited
resolution of the two instruments, make the resulting data not sufficiently sensitive to the detection of the type of column densities that are typically present in the WHIM \citep[e.g.][]{wijers2019,tuominen2023}. The main reason to perform this analysis for all the source in Table~\ref{tab:sample} is to set upper limits to the non--detection of \ovii\ and \oviii\ for a redshift path of $\Delta z \simeq 10$ that is cosmologically significant, instead of using a substantially smaller path for only the brightest sources.
Both likely positive detections and upper limits will be used to
constrain the cosmological density of X--ray absorbing baryons following the methods discussed in \cite{spence2023}, to be presented in a companion paper. 

\section*{Acknowledgments}
DS and MB acknowledge support from NASA 2ADAP2018 program `Closing the Gap on the Missing Baryons at Low Redshift with multi--wavelength observations of the Warm--Hot Intergalactic Medium' awarded to the University of Alabama in Huntsville. TT acknowledges the support of the Academy of Finland grant no. 339127.

 This research has made use of the NASA/IPAC Extragalactic Database, which is funded by the National Aeronautics and Space Administration and operated by the California Institute of Technology.

\section*{Data Availability Statement}
All data contained in Tables~\ref{tab:oviPaper}--\ref{tab:hiPaper} and Tables~\ref{tab:OVIIOVIXMM}--\ref{tab:OVIIIHIChandra} are provided in full length in the on--line version of the paper, and also in
machine--readable format. 


\bibliographystyle{mnras}

\begin{thebibliography}{}
\makeatletter
\relax
\def\mn@urlcharsother{\let\do\@makeother \do\$\do\&\do\#\do\^\do\_\do\%\do\~}
\def\mn@doi{\begingroup\mn@urlcharsother \@ifnextchar [ {\mn@doi@}
  {\mn@doi@[]}}
\def\mn@doi@[#1]#2{\def\@tempa{#1}\ifx\@tempa\@empty \href
  {http://dx.doi.org/#2} {doi:#2}\else \href {http://dx.doi.org/#2} {#1}\fi
  \endgroup}
\def\mn@eprint#1#2{\mn@eprint@#1:#2::\@nil}
\def\mn@eprint@arXiv#1{\href {http://arxiv.org/abs/#1} {{\tt arXiv:#1}}}
\def\mn@eprint@dblp#1{\href {http://dblp.uni-trier.de/rec/bibtex/#1.xml}
  {dblp:#1}}
\def\mn@eprint@#1:#2:#3:#4\@nil{\def\@tempa {#1}\def\@tempb {#2}\def\@tempc
  {#3}\ifx \@tempc \@empty \let \@tempc \@tempb \let \@tempb \@tempa \fi \ifx
  \@tempb \@empty \def\@tempb {arXiv}\fi \@ifundefined
  {mn@eprint@\@tempb}{\@tempb:\@tempc}{\expandafter \expandafter \csname
  mn@eprint@\@tempb\endcsname \expandafter{\@tempc}}}

\bibitem[\protect\citeauthoryear{{Ahoranta} et~al.,}{{Ahoranta}
  et~al.}{2020}]{ahoranta2020}
{Ahoranta} J.,  et~al., 2020, \mn@doi [\aap] {10.1051/0004-6361/201935846},
  \href {https://ui.adsabs.harvard.edu/abs/2020A&A...634A.106A} {634, A106}

\bibitem[\protect\citeauthoryear{{Ahoranta}, {Finoguenov}, {Bonamente},
  {Tilton}, {Wijers}, {Muzahid}  \& {Schaye}}{{Ahoranta}
  et~al.}{2021}]{ahoranta2021}
{Ahoranta} J.,  {Finoguenov} A.,  {Bonamente} M.,  {Tilton} E.,  {Wijers} N.,
  {Muzahid} S.,   {Schaye} J.,  2021, \mn@doi [\aap]
  {10.1051/0004-6361/202038021}, \href
  {https://ui.adsabs.harvard.edu/abs/2021A&A...656A.107A} {656, A107}

\bibitem[\protect\citeauthoryear{{Anders} \& {Grevesse}}{{Anders} \&
  {Grevesse}}{1989}]{anders1989}
{Anders} E.,  {Grevesse} N.,  1989, \gca, 53, 197

\bibitem[\protect\citeauthoryear{{Ar{\'e}valo}, {Uttley}, {Kaspi}, {Breedt},
  {Lira}  \& {McHardy}}{{Ar{\'e}valo} et~al.}{2008}]{arevalo2008}
{Ar{\'e}valo} P.,  {Uttley} P.,  {Kaspi} S.,  {Breedt} E.,  {Lira} P.,
  {McHardy} I.~M.,  2008, \mn@doi [\mnras] {10.1111/j.1365-2966.2008.13719.x},
  \href {https://ui.adsabs.harvard.edu/abs/2008MNRAS.389.1479A} {389, 1479}

\bibitem[\protect\citeauthoryear{{Asplund}, {Grevesse}, {Sauval}  \&
  {Scott}}{{Asplund} et~al.}{2009}]{asplund2009}
{Asplund} M.,  {Grevesse} N.,  {Sauval} A.~J.,   {Scott} P.,  2009, \mn@doi
  [\araa] {10.1146/annurev.astro.46.060407.145222}, \href
  {http://adsabs.harvard.edu/abs/2009ARA%26A..47..481A} {47, 481}

\bibitem[\protect\citeauthoryear{{Behar} et~al.,}{{Behar}
  et~al.}{2017}]{behar2017}
{Behar} E.,  et~al., 2017, \mn@doi [\aap] {10.1051/0004-6361/201629943}, \href
  {https://ui.adsabs.harvard.edu/abs/2017A&A...601A..17B} {601, A17}

\bibitem[\protect\citeauthoryear{{Bertone}, {Schaye}  \& {Dolag}}{{Bertone}
  et~al.}{2008}]{bertone2008}
{Bertone} S.,  {Schaye} J.,   {Dolag} K.,  2008, \mn@doi [\ssr]
  {10.1007/s11214-008-9318-3}, \href
  {http://adsabs.harvard.edu/abs/2008SSRv..134..295B} {134, 295}

\bibitem[\protect\citeauthoryear{{Blustin} et~al.,}{{Blustin}
  et~al.}{2003}]{blustin2003}
{Blustin} A.~J.,  et~al., 2003, \mn@doi [\aap] {10.1051/0004-6361:20030236},
  \href {https://ui.adsabs.harvard.edu/abs/2003A&A...403..481B} {403, 481}

\bibitem[\protect\citeauthoryear{Bonamente}{Bonamente}{2019}]{bonamente2019}
Bonamente M.,  2019, \mn@doi [Journal of Applied Statistics]
  {10.1080/02664763.2018.1531976}, 46, 1129

\bibitem[\protect\citeauthoryear{Bonamente}{Bonamente}{2020}]{bonamente2020}
Bonamente M.,  2020, \mn@doi [Journal of Applied Statistics]
  {10.1080/02664763.2019.1704703}, 47, 2044

\bibitem[\protect\citeauthoryear{{Bonamente}}{{Bonamente}}{2023}]{bonamente2023}
{Bonamente} M.,  2023, \mn@doi [\mnras] {10.1093/mnras/stad463}, \href
  {https://ui.adsabs.harvard.edu/abs/2023MNRAS.522.1987B} {522, 1987}

\bibitem[\protect\citeauthoryear{{Bonamente}, {Nevalainen}, {Tilton},
  {Liivam{\"a}gi}, {Tempel}, {Hein{\"a}m{\"a}ki}  \& {Fang}}{{Bonamente}
  et~al.}{2016}]{bonamente2016}
{Bonamente} M.,  {Nevalainen} J.,  {Tilton} E.,  {Liivam{\"a}gi} J.,  {Tempel}
  E.,  {Hein{\"a}m{\"a}ki} P.,   {Fang} T.,  2016, \mn@doi [\mnras]
  {10.1093/mnras/stw285}, \href
  {http://adsabs.harvard.edu/abs/2016MNRAS.457.4236B} {457, 4236}

\bibitem[\protect\citeauthoryear{Bonamente, Ahoranta, Nevalainen  \&
  Holt}{Bonamente et~al.}{2019}]{bonamente2019b}
Bonamente M.,  Ahoranta J.,  Nevalainen J.,   Holt P.,  2019, \mn@doi [Research
  Notes of the {AAS}] {10.3847/2515-5172/ab2132}, 3, 75

\bibitem[\protect\citeauthoryear{{Bonamente}, {Chen}  \&
  {Zimmerman}}{{Bonamente} et~al.}{2024}]{bonamente2024}
{Bonamente} M.,  {Chen} Y.,   {Zimmerman} D.,  2024, \apj\ in press

\bibitem[\protect\citeauthoryear{{Buote}, {Zappacosta}, {Fang}, {Humphrey},
  {Gastaldello}  \& {Tagliaferri}}{{Buote} et~al.}{2009}]{buote2009}
{Buote} D.~A.,  {Zappacosta} L.,  {Fang} T.,  {Humphrey} P.~J.,  {Gastaldello}
  F.,   {Tagliaferri} G.,  2009, \mn@doi [\apj] {10.1088/0004-637X/695/2/1351},
  \href {http://adsabs.harvard.edu/abs/2009ApJ...695.1351B} {695, 1351}

\bibitem[\protect\citeauthoryear{{Cagnoni}, {Nicastro}, {Maraschi}, {Treves}
  \& {Tavecchio}}{{Cagnoni} et~al.}{2004}]{cagnoni2004}
{Cagnoni} I.,  {Nicastro} F.,  {Maraschi} L.,  {Treves} A.,   {Tavecchio} F.,
  2004, \mn@doi [\apj] {10.1086/381698}, \href
  {http://adsabs.harvard.edu/abs/2004ApJ...603..449C} {603, 449}

\bibitem[\protect\citeauthoryear{{Cash}}{{Cash}}{1976}]{cash1976}
{Cash} W.,  1976, \aap, \href
  {http://adsabs.harvard.edu/abs/1976A%26A....52..307C} {52, 307}

\bibitem[\protect\citeauthoryear{{Cash}}{{Cash}}{1979}]{cash1979}
{Cash} W.,  1979, \apj, 228, 939

\bibitem[\protect\citeauthoryear{{Cautun}, {van de Weygaert}, {Jones}  \&
  {Frenk}}{{Cautun} et~al.}{2014}]{cautun2014}
{Cautun} M.,  {van de Weygaert} R.,  {Jones} B. J.~T.,   {Frenk} C.~S.,  2014,
  \mn@doi [\mnras] {10.1093/mnras/stu768}, \href
  {https://ui.adsabs.harvard.edu/abs/2014MNRAS.441.2923C} {441, 2923}

\bibitem[\protect\citeauthoryear{{Cen} \& {Ostriker}}{{Cen} \&
  {Ostriker}}{1999}]{cen1999}
{Cen} R.,  {Ostriker} J.~P.,  1999, \mn@doi [\apj] {10.1086/306949}, \href
  {http://adsabs.harvard.edu/cgi-bin/nph-bib_query?bibcode=1999ApJ...514....1C&db_key=AST}
  {514, 1}

\bibitem[\protect\citeauthoryear{{Danforth} et~al.,}{{Danforth}
  et~al.}{2016}]{danforth2016}
{Danforth} C.~W.,  et~al., 2016, \mn@doi [\apj] {10.3847/0004-637X/817/2/111},
  \href {http://adsabs.harvard.edu/abs/2016ApJ...817..111D} {817, 111}

\bibitem[\protect\citeauthoryear{{Das}, {Mathur}, {Gupta}, {Nicastro}  \&
  {Krongold}}{{Das} et~al.}{2019}]{das2019}
{Das} S.,  {Mathur} S.,  {Gupta} A.,  {Nicastro} F.,   {Krongold} Y.,  2019,
  \mn@doi [\apj] {10.3847/1538-4357/ab5846}, \href
  {https://ui.adsabs.harvard.edu/abs/2019ApJ...887..257D} {887, 257}

\bibitem[\protect\citeauthoryear{{Dav{\'e}} et~al.,}{{Dav{\'e}}
  et~al.}{2001}]{dave2001}
{Dav{\'e}} R.,  et~al., 2001, \mn@doi [\apj] {10.1086/320548}, \href
  {http://adsabs.harvard.edu/cgi-bin/nph-bib_query?bibcode=2001ApJ...552..473D&db_key=AST}
  {552, 473}

\bibitem[\protect\citeauthoryear{{Detmers, R. G.}, {Kaastra, J. S.},
  {Costantini, E.}, {Verbunt, F.}, {Cappi, M.}  \& {de Vries, C.}}{{Detmers, R.
  G.} et~al.}{2010}]{detmers2010}
{Detmers, R. G.} {Kaastra, J. S.} {Costantini, E.} {Verbunt, F.} {Cappi, M.}
  {de Vries, C.} 2010, \mn@doi [A&A] {10.1051/0004-6361/200913879}, 516, A61

\bibitem[\protect\citeauthoryear{{Draine}}{{Draine}}{2011}]{draine2011}
{Draine} B.~T.,  2011, {Physics of the Interstellar and Intergalactic Medium}

\bibitem[\protect\citeauthoryear{{Fang}, {Marshall}, {Lee}, {Davis}  \&
  {Canizares}}{{Fang} et~al.}{2002}]{fang2002}
{Fang} T.,  {Marshall} H.~L.,  {Lee} J.~C.,  {Davis} D.~S.,   {Canizares}
  C.~R.,  2002, \mn@doi [\apjl] {10.1086/341665}, \href
  {http://adsabs.harvard.edu/abs/2002ApJ...572L.127F} {572, L127}

\bibitem[\protect\citeauthoryear{{Fang}, {Croft}, {Sanders}, {Houck},
  {Dav{\'e}}, {Katz}, {Weinberg}  \& {Hernquist}}{{Fang}
  et~al.}{2005}]{fang2005}
{Fang} T.,  {Croft} R.~A.~C.,  {Sanders} W.~T.,  {Houck} J.,  {Dav{\'e}} R.,
  {Katz} N.,  {Weinberg} D.~H.,   {Hernquist} L.,  2005, \mn@doi [\apj]
  {10.1086/428656}, \href {http://adsabs.harvard.edu/abs/2005ApJ...623..612F}
  {623, 612}

\bibitem[\protect\citeauthoryear{{Fang}, {Canizares}  \& {Yao}}{{Fang}
  et~al.}{2007}]{fang2007}
{Fang} T.,  {Canizares} C.~R.,   {Yao} Y.,  2007, \mn@doi [\apj]
  {10.1086/522560}, \href {http://adsabs.harvard.edu/abs/2007ApJ...670..992F}
  {670, 992}

\bibitem[\protect\citeauthoryear{{Fang}, {Buote}, {Humphrey}, {Canizares},
  {Zappacosta}, {Maiolino}, {Tagliaferri}  \& {Gastaldello}}{{Fang}
  et~al.}{2010}]{fang2010}
{Fang} T.,  {Buote} D.~A.,  {Humphrey} P.~J.,  {Canizares} C.~R.,  {Zappacosta}
  L.,  {Maiolino} R.,  {Tagliaferri} G.,   {Gastaldello} F.,  2010, \mn@doi
  [\apj] {10.1088/0004-637X/714/2/1715}, \href
  {http://adsabs.harvard.edu/abs/2010ApJ...714.1715F} {714, 1715}

\bibitem[\protect\citeauthoryear{{Gatuzz} et~al.,}{{Gatuzz}
  et~al.}{2013}]{gatuzz2013}
{Gatuzz} E.,  et~al., 2013, \mn@doi [\apj] {10.1088/0004-637X/768/1/60}, \href
  {http://adsabs.harvard.edu/abs/2013ApJ...768...60G} {768, 60}

\bibitem[\protect\citeauthoryear{{Gatuzz}, {Garc{\'{\i}}a}, {Kallman},
  {Mendoza}  \& {Gorczyca}}{{Gatuzz} et~al.}{2015}]{gatuzz2015}
{Gatuzz} E.,  {Garc{\'{\i}}a} J.,  {Kallman} T.~R.,  {Mendoza} C.,   {Gorczyca}
  T.~W.,  2015, \mn@doi [\apj] {10.1088/0004-637X/800/1/29}, \href
  {http://adsabs.harvard.edu/abs/2015ApJ...800...29G} {800, 29}

\bibitem[\protect\citeauthoryear{{Gatuzz}, {Garc{\'\i}a}, {Churazov}  \&
  {Kallman}}{{Gatuzz} et~al.}{2023}]{gatuzz2023}
{Gatuzz} E.,  {Garc{\'\i}a} J.~A.,  {Churazov} E.,   {Kallman} T.~R.,  2023,
  \mn@doi [\mnras] {10.1093/mnras/stad698}, \href
  {https://ui.adsabs.harvard.edu/abs/2023MNRAS.521.3098G} {521, 3098}

\bibitem[\protect\citeauthoryear{{Grafton-Waters} et~al.,}{{Grafton-Waters}
  et~al.}{2020}]{grafton2020}
{Grafton-Waters} S.,  et~al., 2020, \mn@doi [\aap]
  {10.1051/0004-6361/201935815}, \href
  {https://ui.adsabs.harvard.edu/abs/2020A&A...633A..62G} {633, A62}

\bibitem[\protect\citeauthoryear{Gupta}{Gupta}{2020}]{gupta2020}
Gupta A.~C.,  2020, \mn@doi [Galaxies] {10.3390/galaxies8030064}, 8

\bibitem[\protect\citeauthoryear{{Johnson} et~al.,}{{Johnson}
  et~al.}{2019}]{johnson2019}
{Johnson} S.~D.,  et~al., 2019, \mn@doi [\apjl] {10.3847/2041-8213/ab479a},
  \href {https://ui.adsabs.harvard.edu/abs/2019ApJ...884L..31J} {884, L31}

\bibitem[\protect\citeauthoryear{{Kaastra}}{{Kaastra}}{2017}]{kaastra2017}
{Kaastra} J.~S.,  2017, \mn@doi [Astronomy and Astrophysics]
  {10.1051/0004-6361/201629319}, \href
  {http://adsabs.harvard.edu/abs/2017A%26A...605A..51K} {605, A51}

\bibitem[\protect\citeauthoryear{{Kaastra}, {Mewe}  \&
  {Nieuwenhuijzen}}{{Kaastra} et~al.}{1996}]{kaastra1996}
{Kaastra} J.~S.,  {Mewe} R.,   {Nieuwenhuijzen} H.,  1996, in {Yamashita} K.,
  {Watanabe} T.,  eds, UV and X-ray Spectroscopy of Astrophysical and
  Laboratory Plasmas. pp 411--414

\bibitem[\protect\citeauthoryear{{Kaastra}, {de~Vries}  \& {den
  Herder}}{{Kaastra} et~al.}{2018}]{kaastra2018}
{Kaastra} J.~S.,  {de~Vries} C.,   {den Herder} J.,  2018

\bibitem[\protect\citeauthoryear{{Kirkman}, {Tytler}, {Suzuki}, {O'Meara}  \&
  {Lubin}}{{Kirkman} et~al.}{2003}]{kirkman2003}
{Kirkman} D.,  {Tytler} D.,  {Suzuki} N.,  {O'Meara} J.~M.,   {Lubin} D.,
  2003, \mn@doi [\apjs] {10.1086/378152}, \href
  {http://adsabs.harvard.edu/abs/2003ApJS..149....1K} {149, 1}

\bibitem[\protect\citeauthoryear{{Komossa}, {Gliozzi}  \&
  {Papadakis}}{{Komossa} et~al.}{2001}]{komossa2001}
{Komossa} S.,  {Gliozzi} M.,   {Papadakis} I.,  2001, \mn@doi [Astronomical and
  Astrophysical Transactions] {10.1080/10556790108229722}, \href
  {https://ui.adsabs.harvard.edu/abs/2001A&AT...20..329K} {20, 329}

\bibitem[\protect\citeauthoryear{{Kov{\'a}cs}, {Bogd{\'a}n}, {Smith}, {Kraft}
  \& {Forman}}{{Kov{\'a}cs} et~al.}{2019}]{kovacs2019}
{Kov{\'a}cs} O.~E.,  {Bogd{\'a}n} {\'A}.,  {Smith} R.~K.,  {Kraft} R.~P.,
  {Forman} W.~R.,  2019, \mn@doi [\apj] {10.3847/1538-4357/aaef78}, \href
  {http://adsabs.harvard.edu/abs/2019ApJ...872...83K} {872, 83}

\bibitem[\protect\citeauthoryear{Li, Chen, Meng, Kashyap  \& Bonamente}{Li
  et~al.}{2024}]{li2024}
Li X.,  Chen Y.,  Meng X.,  Kashyap V.,   Bonamente M.,  2024, to be submitted

\bibitem[\protect\citeauthoryear{{Martizzi} et~al.,}{{Martizzi}
  et~al.}{2019}]{martizzi2019}
{Martizzi} D.,  et~al., 2019, \mn@doi [\mnras] {10.1093/mnras/stz1106}, \href
  {https://ui.adsabs.harvard.edu/abs/2019MNRAS.486.3766M} {486, 3766}

\bibitem[\protect\citeauthoryear{{Mazzotta}, {Mazzitelli}, {Colafrancesco}  \&
  {Vittorio}}{{Mazzotta} et~al.}{1998}]{mazzotta1998}
{Mazzotta} P.,  {Mazzitelli} G.,  {Colafrancesco} S.,   {Vittorio} N.,  1998,
  \mn@doi [\aaps] {10.1051/aas:1998330}, \href
  {http://adsabs.harvard.edu/abs/1998A%26AS..133..403M} {133, 403}

\bibitem[\protect\citeauthoryear{{Middei, R.}, {Vagnetti, F.}, {Bianchi, S.},
  {La Franca, F.}, {Paolillo, M.}  \& {Ursini, F.}}{{Middei, R.}
  et~al.}{2017}]{middei2017}
{Middei, R.} {Vagnetti, F.} {Bianchi, S.} {La Franca, F.} {Paolillo, M.}
  {Ursini, F.} 2017, \mn@doi [A\&A] {10.1051/0004-6361/201629940}, 599, A82

\bibitem[\protect\citeauthoryear{{Nevalainen} et~al.,}{{Nevalainen}
  et~al.}{2019}]{nevalainen2019}
{Nevalainen} J.,  et~al., 2019, \mn@doi [\aap] {10.1051/0004-6361/201833109},
  \href {http://adsabs.harvard.edu/abs/2019A%26A...621A..88N} {621, A88}

\bibitem[\protect\citeauthoryear{{Nicastro}}{{Nicastro}}{2018}]{nicastro2018b}
{Nicastro} F.,  2018, arXiv e-prints, \href
  {http://adsabs.harvard.edu/abs/2018arXiv181103498N} {}

\bibitem[\protect\citeauthoryear{{Nicastro} et~al.,}{{Nicastro}
  et~al.}{2005}]{nicastro2005}
{Nicastro} F.,  et~al., 2005, \mn@doi [\apj] {10.1086/431270}, \href
  {http://adsabs.harvard.edu/abs/2005ApJ...629..700N} {629, 700}

\bibitem[\protect\citeauthoryear{Nicastro, Krongold, Fields, Conciatore,
  Zappacosta, Elvis, Mathur  \& Papadakis}{Nicastro
  et~al.}{2010}]{nicastro2010}
Nicastro F.,  Krongold Y.,  Fields D.,  Conciatore M.~L.,  Zappacosta L.,
  Elvis M.,  Mathur S.,   Papadakis I.,  2010, \mn@doi [The Astrophysical
  Journal] {10.1088/0004-637X/715/2/854}, 715, 854

\bibitem[\protect\citeauthoryear{{Nicastro} et~al.,}{{Nicastro}
  et~al.}{2013}]{nicastro2013}
{Nicastro} F.,  et~al., 2013, \mn@doi [\apj] {10.1088/0004-637X/769/2/90},
  \href {http://adsabs.harvard.edu/abs/2013ApJ...769...90N} {769, 90}

\bibitem[\protect\citeauthoryear{{Nicastro}, {Senatore}, {Gupta}, {Guainazzi},
  {Mathur}, {Krongold}, {Elvis}  \& {Piro}}{{Nicastro}
  et~al.}{2016}]{nicastro2016}
{Nicastro} F.,  {Senatore} F.,  {Gupta} A.,  {Guainazzi} M.,  {Mathur} S.,
  {Krongold} Y.,  {Elvis} M.,   {Piro} L.,  2016, \mn@doi [\mnras]
  {10.1093/mnras/stv2923}, \href
  {http://adsabs.harvard.edu/abs/2016MNRAS.457..676N} {457, 676}

\bibitem[\protect\citeauthoryear{{Nicastro} et~al.,}{{Nicastro}
  et~al.}{2018}]{nicastro2018}
{Nicastro} F.,  et~al., 2018, \mn@doi [Nature] {10.1038/s41586-018-0204-1},
  \href {https://ui.adsabs.harvard.edu/#abs/2018Natur.558..406N} {558, 406}

\bibitem[\protect\citeauthoryear{{Planck Collaboration} et~al.,}{{Planck
  Collaboration} et~al.}{2015}]{Planck2015-cosmology}
{Planck Collaboration} et~al., 2015, preprint, \href
  {http://adsabs.harvard.edu/abs/2015arXiv150201589P} {} (\mn@eprint {arXiv}
  {1502.01589})

\bibitem[\protect\citeauthoryear{{Planck Collaboration} et~al.,}{{Planck
  Collaboration} et~al.}{2020}]{planck2020}
{Planck Collaboration} et~al., 2020, \mn@doi [\aap]
  {10.1051/0004-6361/201833910}, \href
  {https://ui.adsabs.harvard.edu/abs/2020A&A...641A...6P} {641, A6}

\bibitem[\protect\citeauthoryear{{Protassov}, {van Dyk}, {Connors}, {Kashyap}
  \& {Siemiginowska}}{{Protassov} et~al.}{2002}]{protassov2002}
{Protassov} R.,  {van Dyk} D.~A.,  {Connors} A.,  {Kashyap} V.~L.,
  {Siemiginowska} A.,  2002, \mn@doi [\apj] {10.1086/339856}, \href
  {https://ui.adsabs.harvard.edu/abs/2002ApJ...571..545P} {571, 545}

\bibitem[\protect\citeauthoryear{{Rasmussen}, {Kahn}, {Paerels}, {Herder},
  {Kaastra}  \& {de Vries}}{{Rasmussen} et~al.}{2007}]{rasmussen2007}
{Rasmussen} A.~P.,  {Kahn} S.~M.,  {Paerels} F.,  {Herder} J.~W.~d.,  {Kaastra}
  J.,   {de Vries} C.,  2007, \mn@doi [\apj] {10.1086/509865}, \href
  {http://adsabs.harvard.edu/abs/2007ApJ...656..129R} {656, 129}

\bibitem[\protect\citeauthoryear{{Rauch}}{{Rauch}}{1998}]{rauch1998}
{Rauch} M.,  1998, \mn@doi [\araa] {10.1146/annurev.astro.36.1.267}, \href
  {https://ui.adsabs.harvard.edu/abs/1998ARA&A..36..267R} {36, 267}

\bibitem[\protect\citeauthoryear{{Ren}, {Fang}  \& {Buote}}{{Ren}
  et~al.}{2014}]{ren2014}
{Ren} B.,  {Fang} T.,   {Buote} D.~A.,  2014, \mn@doi [\apjl]
  {10.1088/2041-8205/782/1/L6}, \href
  {http://adsabs.harvard.edu/abs/2014ApJ...782L...6R} {782, L6}

\bibitem[\protect\citeauthoryear{{Scott} et~al.,}{{Scott}
  et~al.}{2005}]{scott2005}
{Scott} J.~E.,  et~al., 2005, \mn@doi [\apj] {10.1086/496911}, \href
  {https://ui.adsabs.harvard.edu/abs/2005ApJ...634..193S} {634, 193}

\bibitem[\protect\citeauthoryear{{Soldi, S.} et~al.,}{{Soldi, S.}
  et~al.}{2008}]{soldi2008}
{Soldi, S.} et~al., 2008, \mn@doi [A\&A] {10.1051/0004-6361:200809947}, 486,
  411

\bibitem[\protect\citeauthoryear{{Spence}, {Bonamente}, {Nevalainen},
  {Tuominen}, {Ahoranta}, {de Plaa}, {Liu}  \& {Wijers}}{{Spence}
  et~al.}{2023}]{spence2023}
{Spence} D.,  {Bonamente} M.,  {Nevalainen} J.,  {Tuominen} T.,  {Ahoranta} J.,
   {de Plaa} J.,  {Liu} W.,   {Wijers} N.,  2023, \mn@doi [\mnras]
  {10.1093/mnras/stad1345}, \href
  {https://ui.adsabs.harvard.edu/abs/2023MNRAS.523.2329S} {523, 2329}

\bibitem[\protect\citeauthoryear{{Tilton}, {Danforth}, {Shull}  \&
  {Ross}}{{Tilton} et~al.}{2012}]{tilton2012}
{Tilton} E.~M.,  {Danforth} C.~W.,  {Shull} J.~M.,   {Ross} T.~L.,  2012,
  \mn@doi [\apj] {10.1088/0004-637X/759/2/112}, \href
  {http://adsabs.harvard.edu/abs/2012ApJ...759..112T} {759, 112}

\bibitem[\protect\citeauthoryear{{Tripp}, {Lu}  \& {Savage}}{{Tripp}
  et~al.}{1998}]{tripp1998}
{Tripp} T.~M.,  {Lu} L.,   {Savage} B.~D.,  1998, \mn@doi [\apj]
  {10.1086/306397}, \href {http://adsabs.harvard.edu/abs/1998ApJ...508..200T}
  {508, 200}

\bibitem[\protect\citeauthoryear{{Tuominen} et~al.,}{{Tuominen}
  et~al.}{2021}]{tuominen2021}
{Tuominen} T.,  et~al., 2021, \mn@doi [\aap] {10.1051/0004-6361/202039221},
  \href {https://ui.adsabs.harvard.edu/abs/2021A&A...646A.156T} {646, A156}

\bibitem[\protect\citeauthoryear{{Tuominen}, {Nevalainen}, {Hein{\"a}m{\"a}ki},
  {Tempel}, {Wijers}, {Bonamente}, {Aragon-Calvo}  \& {Finoguenov}}{{Tuominen}
  et~al.}{2023}]{tuominen2023}
{Tuominen} T.,  {Nevalainen} J.,  {Hein{\"a}m{\"a}ki} P.,  {Tempel} E.,
  {Wijers} N.,  {Bonamente} M.,  {Aragon-Calvo} M.~A.,   {Finoguenov} A.,
  2023, \mn@doi [\aap] {10.1051/0004-6361/202244508}, \href
  {https://ui.adsabs.harvard.edu/abs/2023A&A...671A.103T} {671, A103}

\bibitem[\protect\citeauthoryear{{Verner}, {Verner}  \& {Ferland}}{{Verner}
  et~al.}{1996}]{verner1996}
{Verner} D.~A.,  {Verner} E.~M.,   {Ferland} G.~J.,  1996, \mn@doi [Atomic Data
  and Nuclear Data Tables] {10.1006/adnd.1996.0018}, \href
  {http://adsabs.harvard.edu/abs/1996ADNDT..64....1V} {64, 1}

\bibitem[\protect\citeauthoryear{Wasserstein \& Lazar}{Wasserstein \&
  Lazar}{2016}]{asa2016}
Wasserstein R.~L.,  Lazar N.~A.,  2016, \mn@doi [The American Statistician]
  {10.1080/00031305.2016.1154108}, 70, 129

\bibitem[\protect\citeauthoryear{{Weinberg}, {Miralda-Escud{\'e}}, {Hernquist}
  \& {Katz}}{{Weinberg} et~al.}{1997}]{weinberg1997}
{Weinberg} D.~H.,  {Miralda-Escud{\'e}} J.,  {Hernquist} L.,   {Katz} N.,
  1997, \mn@doi [\apj] {10.1086/304893}, \href
  {https://ui.adsabs.harvard.edu/abs/1997ApJ...490..564W} {490, 564}

\bibitem[\protect\citeauthoryear{{Wijers}, {Schaye}, {Oppenheimer}, {Crain}  \&
  {Nicastro}}{{Wijers} et~al.}{2019}]{wijers2019}
{Wijers} N.~A.,  {Schaye} J.,  {Oppenheimer} B.~D.,  {Crain} R.~A.,
  {Nicastro} F.,  2019, \mn@doi [\mnras] {10.1093/mnras/stz1762}, \href
  {https://ui.adsabs.harvard.edu/abs/2019MNRAS.488.2947W} {488, 2947}

\bibitem[\protect\citeauthoryear{{Yao}, {Tripp}, {Wang}, {Danforth},
  {Canizares}, {Shull}, {Marshall}  \& {Song}}{{Yao} et~al.}{2009}]{yao2009}
{Yao} Y.,  {Tripp} T.~M.,  {Wang} Q.~D.,  {Danforth} C.~W.,  {Canizares} C.~R.,
   {Shull} J.~M.,  {Marshall} H.~L.,   {Song} L.,  2009, \mn@doi [\apj]
  {10.1088/0004-637X/697/2/1784}, \href
  {http://adsabs.harvard.edu/abs/2009ApJ...697.1784Y} {697, 1784}

\bibitem[\protect\citeauthoryear{{Yao}, {Shull}, {Wang}  \& {Cash}}{{Yao}
  et~al.}{2012}]{yao2012}
{Yao} Y.,  {Shull} J.~M.,  {Wang} Q.~D.,   {Cash} W.,  2012, \mn@doi [\apj]
  {10.1088/0004-637X/746/2/166}, \href
  {http://adsabs.harvard.edu/abs/2012ApJ...746..166Y} {746, 166}

\bibitem[\protect\citeauthoryear{{Zappacosta}, {Nicastro}, {Maiolino},
  {Tagliaferri}, {Buote}, {Fang}, {Humphrey}  \& {Gastaldello}}{{Zappacosta}
  et~al.}{2010}]{zappacosta2010}
{Zappacosta} L.,  {Nicastro} F.,  {Maiolino} R.,  {Tagliaferri} G.,  {Buote}
  D.~A.,  {Fang} T.,  {Humphrey} P.~J.,   {Gastaldello} F.,  2010, \mn@doi
  [\apj] {10.1088/0004-637X/717/1/74}, \href
  {https://ui.adsabs.harvard.edu/abs/2010ApJ...717...74Z} {717, 74}

\bibitem[\protect\citeauthoryear{{den Herder, J. W.} et~al.,}{{den Herder, J.
  W.} et~al.}{2001}]{denherder2001}
{den Herder, J. W.} et~al., 2001, \mn@doi [A&A] {10.1051/0004-6361:20000058},
  365, L7

\makeatother
\end{thebibliography}
\input{main.bbl}

\appendix

\section{Tests of the \texttt{slab} and \texttt{line} models for the measurement of column densities}
\label{app:NumericalTests}
This Appendix presents numerical simulations aimed at testing the
accuracy of the narrow--band \texttt{line} model to measure the column density of the \ovii\ and \oviii\ ions from the WHIM.

\subsection{Context and atomic data}
The He--like \ovii\ and the H--like \oviii\ ions give rise to a series of resonance lines,
of which the strongest lines occur respectively at $\lambda=21.602$~\AA\ (often referred to as He-$\alpha$) and at
$\lambda=18.969$~\AA\ (Ly-$\alpha$, a doublet with components that are indistinguishable at this resolution).
Weaker absorption lines occur at shorter wavelengths; for example, the He-$\beta$ line is at 18.63~\AA, the He-$\gamma$
at 17.77~\AA, etc., etc., with an \ovii\ absorption edge shortwards of $\sim$17~\AA\ \citep[e.g.][]{verner1996}.
Lines in this series have progressively smaller oscillator strengths, and are detectable in absorption for high
column densities (e.g., $N \geq 10^{17}$~cm$^{-2}$), such as those that occur in the interstellar medium (ISM) in the spectra of 
bright sources \citep[e.g.][]{gatuzz2013, gatuzz2015}.

Typical column densities of \ovii\ and \oviii\ in the WHIM are substantially smaller, e.g., they are typically expected to be 
$\leq 10^{16}$~cm$^{-2}$ \citep[e.g.][]{wijers2019,tuominen2023}. Moreover, 
X--ray fluxes of extragalatic sources such as those used in this paper are lower than those used for the type of ISM studies that can detect high-order lines in the series. In fact, to date there have been only a few possible detections of 
$\alpha$--order lines from \ovii\ or \oviii, and there has been no convincing detection of any high-order lines in any of the extragalactic sources used for WHIM studies (see literature review in Sec.~\ref{sec:introduction} and \ref{sec:searchPrior}).

\subsection{Simulations with the \texttt{slab} model}
\label{sec:slab}
\begin{figure}
    \centering
    \includegraphics[width=\linewidth]{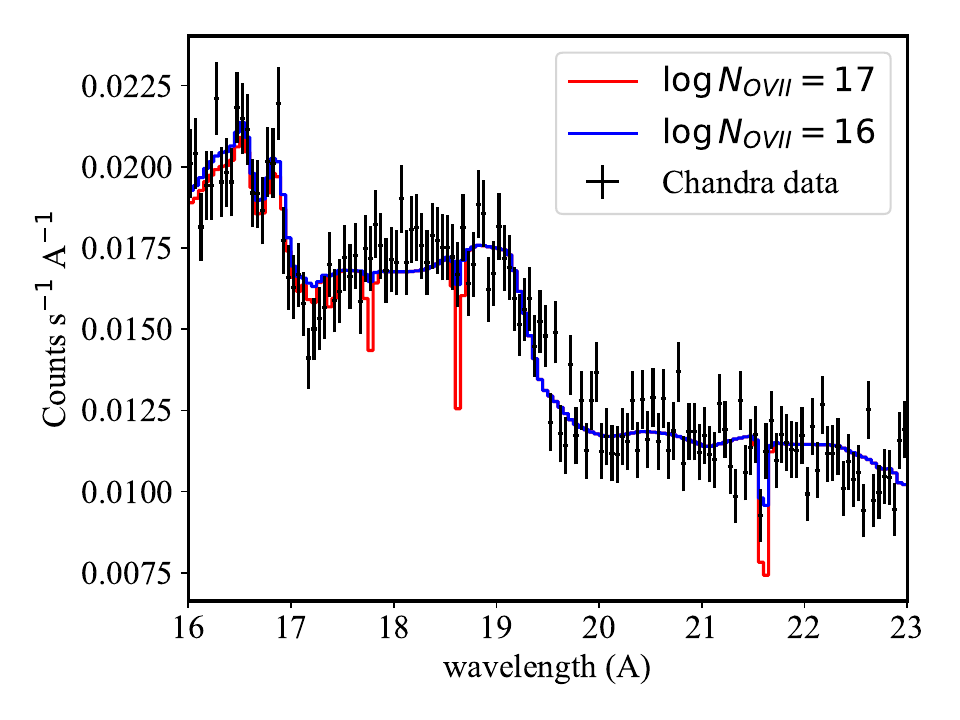}
    \caption{\chandra\ HRC observation of \es, and illustration of the \texttt{slab} models used for the simulations.}
    \label{fig:1ESChandraSim}
\end{figure}

To address the use of narrow--band fits  and the \texttt{line} model (see Sec.~\ref{sec:dataAnalysis}) to measure the 
column density of these ions in the spectra of typical extragalactic sources, 
we simulated 1~Ms \chandra\ HRC LETG spectra of \es, one of the brightest sources in the sample. The simulation 
included a power--law continuum, and \texttt{slab} model with a fixed \ovii\ column density of either $\log N=17$~cm$^{-2}$ or  $\log N=16$~cm$^{-2}$,
representative respectively of a column density that is substantially larger than what is expected in this study, and one that
is similar to the largest columns expected in our sample. For a review of the use of the \texttt{slab} model, see e.g. \cite{spence2023}, where we
used both the \texttt{slab} and the \texttt{line} models. Fig.~\ref{fig:1ESChandraSim} provides one of the simulated spectra, clearly showing that the $\log N=17$~cm$^{-2}$ column does result in significant absorption lines in the series, while the $\log N=16$~cm$^{-2}$
case results in a barely detectable He-$\alpha$ line, with the other lines (He-$\beta$, etc.) being undetectable.

\begin{table}
    \centering
    \begin{tabular}{llllll}
        \hline
    \hline
    \multicolumn{1}{c}{Fit range} & Exposure & \multicolumn{4}{c}{Column densities and errors} \\
    \multicolumn{1}{c}{(A)} & $\log T$(s) & \multicolumn{2}{c}{$\log N_{\text{OVII}}$ (cm$^{-2}$)} & \multicolumn{2}{c}{Average Errors} \\
     & & true & best-fit & & \\
    \hline
    16-23 & 6.0 & 17.0 & 17.00 $\pm$ 0.12 & 0.08  & -0.08 \\ 
20-23 & 6.0 & 17.0 & 16.96 $\pm$ 0.18 & 0.23  & -0.23 \\ 
\hline  
16-23 & 6.0 & 16.0 & 15.98 $\pm$ 0.14 & 0.13  & -0.13 \\ 
20-23 & 6.0 & 16.0 & 16.05 $\pm$ 0.12 & 0.14  & -0.14 \\ 
\hline  
16-23 & 5.0 & 16.0 & 15.86 $\pm$ 0.44 & 0.55  & -0.55 \\ 
20-23 & 5.0 & 16.0 & 16.07 $\pm$ 0.21 & 0.50  & -0.50 \\ 
\hline
\hline
    \end{tabular}
    \caption{Results of the \texttt{pyspex} simulation of the \es\ \chandra\ data with the \texttt{slab} model. Errors in the
    4-th column is the rms scatter of the best--fit, while the `Average Error' columns report the averages of the uncertainties in the 10 realizations.}
    \label{tab:1ESSim}
\end{table}

To address in more detail the ability of the narrow--band fits to recover the parent column density, as done in Sec.~\ref{sec:dataAnalysis} of this paper, we performed two sets of simulations, each with 10~independent realizations.
First, we fit a broad-band (16-23~\AA) spectrum that includes all the lines in the series (at $z=0$),
and a narrow-band (20-23~\AA) range that includes only the He-$\alpha$ line, both with the same \texttt{slab} model as in the
input to the simulations. As illustrated in Table~\ref{tab:1ESSim}, the narrow-band
fits are unbiased in their ability to recover the correct column densities, for both column densities. For the
larger column density, the narrower band has a larger uncertainty, which is expected by the presence of measurable higher-order lines in the series. But for the $\log N=16$~cm$^{-2}$ column, the simulations show that there is no advantage to fitting the broader
band, since the expected higher-order lines fall within the noise of the instrument; in fact, there is even indication that a
narrower band might be a better choice, given the smaller scatter in the recovered column densities, compared with
broad-band fits. We also performed similar simulations for \oviii, showing equivalent results. 
 We conclude that the use of narrow--band fits that only include the primary line in the series (e.g., Ly-$\alpha$
or He-$\alpha$) are unbiased
estimators of the parent column density.

\subsection{Simulations with \texttt{slab} and \texttt{line} models}
\label{sec:slabLine}
A second set of simulations was performed using the same simulation setup as for the earlier simulation 
that make use of the \texttt{slab} model, but using the Gaussian \texttt{line} model
to fit the He~$\alpha$ absorption line at 21.602~\AA\ in a narrow band. The equivalent width 
of the \texttt{line} model is automatically calculated by \texttt{SPEX} via the \texttt{ewa} parameter, 
and it is then converted it to a column density using the curves of growth 
method \citep[COG, e.g.,][]{draine2011}. 
The \texttt{slab} model assumes a velocity dispersion parameter that was converted to the $b$ parameter for the
COG analysis. 

We performed two sets of 10 realizations each for two typical \ovii\ column densities, $\log N=16$~cm$^{-2}$ (corresponding to the
largest WHIM column densities expected) and $\log N=15$~cm$^{-2}$ (a more typical column density expected). For each realization,
we calculate the column density and its rms scatter, and obtained respectively an average $\log N=16.09\pm0.40$
and $\log N=15.07\pm0.53$. The agreement with the parent column density from the \texttt{slab} model indicates that the
narrow--band \texttt{line} method is unbiased, and therefore it is accurate to measure typical WHIM column densities.

\subsection{Constraints on $b$ parameter and other considerations}

At the resolution of the data in this paper, we cannot significantly constrain the $b$ parameter of any putative line, as already indicated in \cite{spence2023}. This is another
reason why the \texttt{line} model is deemed sufficient to recover the column density. 
In fact, when attempting to use the fits in Sec.~\ref{sec:slab} above for $\log N=16$~cm$^{-2}$ and fit them
with a \texttt{slab} model now with a 
 free \texttt{v} parameter ---
corresponding to the rms velocity of the absorbing gas and related to the $b$ parameter --- 
\texttt{spex} typically fails to converge rapidly towards a best--fit, and it leaves the parameter
basically unconstrained when it does converge. 

This is illustrated in
Fig~\ref{fig:1ESChandraSim}, which simulates one of the longest observations, for one of the brightest sources,
and in Fig.~\ref{fig:detections}--\ref{fig:detections3} for the \xmm\ and \chandra\ data:
the resolution of the data is so limited that there are not enough degrees of freedom to constrain the line broadening.  Instead, the $b$ parameter of the line 
will be assumed based on the range of temperatures expected in the WHIM, 
while acknowledging the possible effects of additional non–thermal line broadening,
which determines the degree of saturation of lines,  as a systematic effect in the measurement of the equivalent width, similar to what
was done in a previous study for Ton~S180 \citep{ahoranta2021}. 
Additional details on the COG method, and its use for the measurement and upper limits of column densities, 
will be presented in a subsequent paper that contains the astrophysical interpretation of our results.

It is also useful to point out that, at significantly larger column densities of the type present in some ISM sight--lines (e.g.,
$\log N \geq 17$~cm$^{-2}$), the Lorentzian broadening of the line becomes significant, and therefore a Gaussian fit is less accurate in recovering the column density. 
While the same effect may also be important at lower column densities when the lines are narrow 
\citep[e.g., see discussion in][]{ahoranta2021}, this effect cannot be modeled directly from the data, but can be studied through parallel COG analysis considering different physical scenarios.


\end{document}